\documentclass[aps,pra,longbibliography,reprint,onecolumn,superscriptaddress,floatfix]{revtex4-2}

\usepackage{graphicx}
\usepackage{dcolumn}
\usepackage{bm}

\usepackage[utf8]{inputenc}
\usepackage{ragged2e}

\usepackage{amsmath}
\usepackage{ulem}
\usepackage{layout}
\usepackage{mathtools}
\usepackage[ruled,longend,noend]{algorithm2e}

\usepackage{xcolor}
\usepackage{graphicx}
\usepackage{adjustbox}
\usepackage{amssymb}
\usepackage{siunitx}
\usepackage{stmaryrd}
\usepackage{float}
\usepackage{amsthm}
\usepackage{tikz-cd}
\usepackage{tikz}
\usetikzlibrary{backgrounds}
\usepackage[breaklinks]{hyperref}
\usepackage{bbm}
\hypersetup{
    colorlinks = true,
    linkcolor = violet,
    linkbordercolor = {white}
}
\usepackage{quantikz}
\usetikzlibrary{decorations.markings}
\usetikzlibrary{shapes.geometric}
\pgfdeclarelayer{edgelayer}
\pgfdeclarelayer{nodelayer}
\pgfsetlayers{edgelayer,nodelayer,main}
\tikzstyle{none}=[inner sep=0pt]

\tikzstyle{input}=[circle,fill=blue,draw=black,line width=0.8 pt]
\tikzstyle{output}=[circle,fill=black,draw=black,line width=0.8 pt,minimum size=0.1cm,inner sep=0pt]

\tikzstyle{simple}=[-,draw=black,line width=2.000]
\tikzstyle{arrow}=[-,draw=black,postaction={decorate},decoration={markings,mark=at position .5 with {\arrow{>}}},line width=2.000]
\tikzstyle{tick}=[-,draw=black,postaction={decorate},decoration={markings,mark=at position .5 with {\draw (0,-0.1) -- (0,0.1);}},line width=2.000]
\tikzstyle{z-node}=[fill={rgb,255: red,0; green,192; blue,0}, draw=black, shape=circle,minimum size=0.2cm,inner sep=0pt]
\tikzstyle{x-node}=[fill={rgb,255: red,192; green,0; blue,0}, draw=black, shape=circle]
\tikzstyle{h-box}=[fill=yellow, draw=black, shape=rectangle]
\tikzstyle{pivot}=[fill={rgb,255: red,255; green,128; blue,0}, draw=black, shape=circle]
\tikzstyle{h}=[-, draw=blue, dashed,dash pattern=on 2pt off 1pt]

\theoremstyle{definition}
\newtheorem{theorem}{Theorem}[section]
\newtheorem{corollary}[theorem]{Corollary}
\newtheorem{lemma}[theorem]{Lemma}
\newtheorem{claim}[theorem]{Claim}
\newtheorem{definition}[theorem]{Definition}
\newtheorem{proposition}[theorem]{Proposition}
\newtheorem{conjecture}[theorem]{Conjecture}
\newtheorem{fact}[theorem]{Fact}

\newcommand{\Z}{\mathbb{Z}}

\newcommand{\seq}{\subseteq}

\newcommand{\ketbra}[2]{\ket{#1} \bra{#2}}

\newcommand{\br}[1]{\left( #1 \right)}

\newcommand{\set}[1]{\{ #1 \}} 

\newcommand{\bounds}[2]{\bigg\rvert_{#1}^{#2}}
\newcommand{\boudns}[2]{\bounds} 

\newcommand{\floor}[1]{\left\lfloor #1 \right\rfloor}
\newcommand{\ceil}[1]{\left\lceil #1 \right\rceil}

\newcommand{\g}{\gamma}

\renewcommand{\d}{\delta}
\newcommand{\D}{\Delta}
\newcommand{\e}{\epsilon}

\renewcommand{\th}{\theta}
\newcommand{\Th}{\Theta}
\newcommand{\w}{\omega}
\newcommand{\W}{\Omega}
\renewcommand{\L}{\Lambda}

\newcommand{\CA}{\mathcal{A}}

\newcommand{\CC}{\mathcal{C}}
\newcommand{\CD}{\mathcal{D}}
\newcommand{\CE}{\mathcal{E}}

\newcommand{\CI}{\mathcal{I}}

\newcommand{\CP}{\mathcal{P}}
\newcommand{\CL}{\mathcal{L}}
\newcommand{\CO}{\mathcal{O}}

\newcommand{\CR}{\mathcal{R}}
\newcommand{\CS}{\mathcal{S}}

\newcommand{\CW}{\mathcal{W}}

\begin{document}

\preprint{APS/123-QED}

\title{
Universal Graph Representation of Stabilizer Codes 
}

\author{Andrey Boris Khesin}
\email{khesin@mit.edu}
\altaffiliation{Presently at Department of Computer Science, University of Oxford, Oxford, UK}
\affiliation{Department of Mathematics, Massachusetts Institute of Technology, Cambridge, MA}%
\author{Jonathan Z. Lu}
\email{lujz@mit.edu}
\affiliation{Department of Mathematics, Massachusetts Institute of Technology, Cambridge, MA}%
\author{Peter W. Shor}
\affiliation{Department of Mathematics, Massachusetts Institute of Technology, Cambridge, MA}%

\date{\today}

\begin{abstract}
While stabilizer tableaus have proven useful as a descriptive tool for additive quantum codes, they otherwise offer little guidance for concrete constructions or algorithm analysis.
We introduce a representation of stabilizer codes as graphs with certain structures, and prove via the ZX Calculus that this representation is related to stabilizer tableaus by an efficiently computable bijection.
This gives a new universal recipe for code construction by way of finding graphs with nice properties.
The graph representation gives insight into both code construction and algorithms. 
We construct as examples families of $\llbracket n, \;\Theta(\frac{n}{\log n}), \;\Th(\log n)\rrbracket$ and $\llbracket n, \;\W(n^{4/5}), \;\Th(n^{1/5}) \rrbracket$ codes.
We use graphs in a probabilistic analysis to extend the quantum Gilbert-Varshamov bound into a three-way distance-rate-weight trade-off.
Moreover, code properties such as distance and encoding circuit depth are bounded by simple functions of the graph degree.
We prove that key coding algorithms---distance approximation, minimum weight generator selection, and decoding---are unified as instances of one optimization game on a graph. 
By studying this game, we construct an efficient greedy decoder and prove that it corrects all recoverable errors for all graphs with cycle lengths no shorter than 13 (reducible to 5 with mild extra constraints); these include the above two families.
Our results suggest that graphs are generically useful for the study of stabilizer codes.
\end{abstract}

\maketitle

\tableofcontents

\section{\label{sec:intro}Introduction}
Most quantum algorithms which are believed to provide significant speedups to certain computational problems require both many qubits and error correction to succeed in practice~\cite{t1,t2,t3,t4,t5,t6,t7,nielsen2002quantum}. As advancements in the experimental scaling of quantum computers have steadily marched forward, the problem of designing good quantum codes for the practical implementation of various quantum algorithms has in turn become of more pressing interest~\cite{e1,e2,e3,e4,e5}.

With a few exceptions, stabilizer codes have emerged as one of the most intensely studied families of codes due to their simple description~\cite{ec1,ec2,ec3,shor1995scheme,gottesman2009introduction,nielsen2002quantum,gottesman1997stabilizer}. Roughly speaking, stabilizer codes are the most general quantum analogue of linear codes for classical computing. More precisely, a $\llbracket  n, k\rrbracket$ stabilizer code encodes $k$ logical qubits to $n$ physical qubits by embedding $2^k$ logical degrees of freedom into a subspace defined as the simultaneous $+1$ eigenspace of $n-k$ independent commuting $n$-qubit Pauli operators, known as the \textit{stabilizers}. Stabilizers comprise a considerably large class of quantum codes, including all CSS codes~\cite{steane1996multiple,calderbank1996good}. 

Generally, a stabilizer code is specified by the \textit{stabilizer tableau}, the $n-k$ Pauli operator strings which define the code space. This description is simple and elegant, and has led to remarkable results such as the Gottesman-Knill theorem~\cite{aaronson2004improved}, which efficiently simulates any Clifford operation classically, and the quantum Gilbert-Varshamov bound~\cite{nielsen2002quantum}, which proves by a probabilistic technique that asymptotically good stabilizer codes exist. At the same time, the stabilizer tableau representation is prohibitive in many other ways. Without further structure, it is unclear as to how one might construct desirable codes by simply writing down a clever collection of $n-k$ Pauli strings, or how one might devise and analyze coding algorithms for a given tableau. For this reason, much of the constructive effort in quantum coding theory has relied on developing sophisticated CSS-type product operations that generate CSS codes from nice classical codes~\cite{tillich2013quantum,panteleev2021almost}. Moreover, algorithms to, e.g., decode these constructions typically rely on reducing the problem to a classical code decoding problem, and then relying on classical decoders~\cite{tillich2013quantum,breuckmann2021balanced,panteleev2021degenerate,calderbank1996good}.

In this paper, we argue that a different approach based on graphs to represent stabilizer codes may also prove useful for the development and analysis of quantum codes. The employment of graphs in code construction has a rich history in coding theory. Classical codes are described by a $(n-k) \times n$ binary matrix $H$ whose kernel over $\mathbb{F}_2$ defines the code space. Just as it is not obvious which tables of Paulis constitute good stabilizer codes, so too is it a priori unclear which tables of bits correspond to classical codes. Consequently, graph representations such as Tanner graphs, which encode $H$ into a bipartite graph, were developed to give insight into classical code construction. Tanner graphs built the bridge between expander graphs and classical codes and significantly contributed to modern classical coding through the utilization of graph theory insights~\cite{sipser1996expander}. Even in the quantum realm, graph reasoning has already enabled breakthroughs in constructions of CSS codes. Most famously, the topological structure of Euclidean lattices with periodic boundary conditions motivated the development of the toric and other surface codes~\cite{kitaev2003fault,bravyi2024high}. The toric code was even shown to be efficiently decoded by way of classical graph algorithms, in particular Edmonds' minimum weight perfect matching~\cite{edmonds1965maximum,edmonds1965paths}. Outside of quantum coding, tensor networks have played a key role in quantum simulation and algorithm design.

The critical contributions of graphs in both classical and quantum coding in many specific constructive cases naturally lead to the question of whether one can use graph theory to study additive quantum codes in the most general sense. That is, can we devise graph techniques to reason about quantum stabilizer codes in an inherently general quantum manner, without having to restrict ourselves to CSS codes and thereby reduce the problem to classical codes? This paper takes a step towards answering this question by exploring (a) what the most general graph techniques on stabilizer codes are without relying on classical codes and CSS products and (b) how useful such a representation is for a variety of important coding tasks. 

\subsection{Contributions, scope, and conjectures}
Our first contribution, in the direction of the first question, is the development of a simple graph representation of all stabilizer codes. The general structure of the graph is known as an \textit{encoder-respecting form}, which has the structure of a semi-bipartite graph that maps $k$ input-representing nodes to $n$ output-representing nodes. By semi-bipartite, we mean that inputs may not be connected to each other while outputs generically may. Informally, this result may be stated as follows.
\begin{theorem}[Informal] \label{thm:informal}
    Every stabilizer code is essentially uniquely represented by an encoder-respecting graph satisfying some rules, and conversely every such graph essentially uniquely represents a stabilizer code. There is an efficient compilation algorithm that maps tableaus to graphs in $O(n^3)$ time and vice versa in $O(n^2)$ time.
\end{theorem}
Here, the term ``essentially" masks a more exact statement in which we mod out by local equivalences that do not affect any code properties.
The precise version of Theorem~\ref{thm:informal} is given in Sections~\ref{sec:zxcf} and \ref{sec:inversion}.
This result establishes that our graph representation is sufficiently general to universally represent all stabilizer codes. We devised the compilation algorithm from tableaus to graphs in part for the study of graph properties of well-known codes. As examples, we compile the famous 5-qubit, 7-qubit, and 9-qubit quantum codes and show that each take simple geometric forms respectively of a cone, cube, and tree-star. Conversely, Theorem~\ref{thm:informal} enables us to essentially transform any semi-bipartite graph into a stabilizer code, in such a way that (a) any stabilizer code can be constructed this way and (b) the properties of the code are tied to the properties of the graph. As a consequence, we can leverage the tools of graph theory and graph algorithms to study stabilizer codes. We show that the key properties of a stabilizer code are controlled in a unified way, namely by the corresponding graph's degree.

\begin{theorem}[Informal]
    Let $G$ be a graph and $C(G)$ the corresponding code. The following hold.
    \begin{enumerate}
        \item[(1) ] The distance of $C(G)$ is bounded above by roughly the min-degree of $G$. 
        \item[(2) ] If $G$ is bipartite, then $C(G)$ is a CSS code.
        \item[(3) ] The stabilizer tableau produced by the algorithm in Theorem~\ref{thm:informal} has weights bounded by a quadratic function of the degree of certain nodes in $G$. 
        \item[(4) ] There exists an efficient algorithm which produces an encoding circuit for $C(G)$ whose depth is bounded by (up to a small additive constant) twice the maximum degree of $G$. 
        \item[(5) ] Every constant-depth diagonal gate and the $\sqrt{X}$ can be implemented logically in $C(G)$ with a circuit of depth, respectively, about (up to an additive constant) twice the max-degree and the max-degree.
    \end{enumerate}
\end{theorem}
The formal versions of these statements are given in Section~\ref{sec:inversion}. We emphasize that our universal graph representation is fully distinct from Tanner graph representations, even in the CSS case when the graph is bipartite.
For example, the Tanner graph of a code with a large distance but a low check weight, such as the toric code, would have a very low degree, while condition (1) above requires that a graph using our construction for the same code would have a high degree.

To further strengthen the connection between graphs and codes, we next define a simple, one-player optimization-type game on a graph which we call quantum lights out (QLO), in Section~\ref{sec:inversion}. We prove that QLO unifies many important tasks in quantum coding, which opens a path towards the study of coding algorithms via their correspondence to approximately optimal QLO strategies.

\begin{theorem}[Informal]
    Many coding algorithms such as distance approximation, finding minimum weight stabilizer generators, and decoding are all approximately optimal strategies in the QLO game.
\end{theorem}

The above results demonstrate that graphs in many senses unify various aspects of stabilizer coding in a simple manner. Overall, we give four pieces of evidence that graphs serve as a useful representation. In the first, which we discussed above, we show that graphs enable efficient algorithms which produce encoding circuits and logical-operators whose depths and weights are bounded by the degree. Later on, when we study decoding, we are able to bound the depths directly by the distance of the code for graphs which satisfy a certain property.

In the second aspect, we explore the geometric and topological properties of graphs as tools for building codes. 
Noticing that simple geometric solids realize celebrated quantum codes, we construct codes based on other such solids, including platonic solids and topological transformations of them. For simpler constructions, we are able to calculate the distance by direct computation. As an example, we produce a non-CSS stabilizer code from a dodecahedron, which has parameters $\llbracket 16, 4, 3\rrbracket$. We also construct a $\llbracket 54, 6, 5\rrbracket$ code by lifting the icosahedron to a covering space. For more complicated constructions, we use more indirect analytical techniques based on decoding. The main advantage of graphs as a constructive tool is their flexibility: we show that graphs enable us to more easily produce codes that have a desired numerical distance, by finding graphs which have a similar degree. While the degree alone does not guarantee a distance comparable to the degree, adding additional structural constraints to the graph can achieve a distance comparable to the degree. Graphs also can be chosen to immediately satisfy desired experimental constraints based on locality, geometric structure, etc. Thus, the graph representation may prove helpful tools for code construction at constant-size scales fit for experimental use.

Our third avenue is decoding. In general, we believe that the graph representation serves as an excellent guide for the design of decoding algorithms. Historical evidence of minimum-weight perfect matching on the surface code suggests that graphs are indeed helpful. In this paper, we take a first step in that direction by designing a decoder that employs a simple greedy strategy on the QLO game. We give a general condition on graphs, which we call its \textit{sensitivity}, that sufficiently characterizes a graph's amenity to be decoded well greedily. For concreteness, we also design a simple family of $\llbracket \frac{m 2^m}{m+1}, \frac{2^m}{m+1}\rrbracket$ codes, which we call the hypercube codes. Their distance is provably at most $m$, but we show that the decoder can correct at least $\floor{\frac{m-3}{4}}$ errors on it, so that the distance is also at least $\floor{\frac{m-1}{2}}$. We conjecture, however, that a more careful analysis can prove that the distance is actually $m$. A small case of the hypercube code family is $\llbracket 112, 16, 7\rrbracket$. Moreover, our results imply a lower bound on the distance in terms of the degree, which in turn lead to bounds on encoding circuit depth, logical operator circuit depth, and stabilizer weight entirely in terms of distance.

Although the sensitivity is a natural condition for greedily decodable codes, it is a complicated definition that is difficult to build intuition for the purposes of code construction.
To overcome this issue and build a stronger bridge between our work and the main branches of graph theory, we prove in Lemma~\ref{lemma:girth_sensitivity} that any graph with girth (i.e. shortest cycle length) at least 13 is minimally sensitive and thus optimally decodable by the greedy decoder.
In the same theorem, we prove parallel results for graphs with smaller girths (as small as 5) which satisfy some additional constraints, such as regularity and having certain nodes be sufficiently far from each other.
Consequently, such graphs have a provable distance lower and upper bound.
It is interesting that such a condition on a global property has no dependence at all on $n$, in contrast to classical Tanner graphs which require expansion, a property that explicitly depends on $n$. 
We are further able to show under certain conditions that the distance is exactly the minimum degree. The connection between girth and decodability is particularly interesting because it enables us to leverage results from extremal graph theory to construct codes. In particular, we use a well-known construction of girth-12 high-degree graphs~\cite{benson1966minimal} to build a family of $\llbracket n, \W(n^{4/5}), \Th(n^{1/5}) \rrbracket$ codes in Theorem~\ref{thm:girthy_code}. These constructions in many senses maximally exploit the capabilities of the greedy QLO decoder. 

Because QLO and the greedy decoder are fully general to all stabilizer codes, they allow us to reason about decoding in an inherently quantum manner, as opposed to reducing the decoding problem to classical decoding in the CSS prescription. As such, this approach opens the door to the study of coding algorithms for \textit{non-CSS} codes, including codes which are not even locally equivalent to a CSS code. At the same time, because QLO is general, it is also helpful as a new way to study CSS decoding without finding the underlying classical codes and appealing to classical coding theory. Even with the toric code, which is CSS, decoding is often performed by minimum weight perfect matching, an approach from graph theory, rather than studying the classical constituent codes. In our case, as an example, the $\llbracket n, \W(n^{4/5}), \Th(n^{1/5}) \rrbracket$ code turns out to be CSS but is most easily decoded directly by the greedy QLO strategy rather than classical reductions.

Finally, we exploit the structure of graphs to give a more fine-grained argument about random codes. By randomly connecting edges of our graphs, we give in Theorem~\ref{thm:random_graphs} a large class of codes for which almost all members satisfy a trade-off between rate, distance, and stabilizer weight. More precisely, for a chosen number of physical qubits $n$, rate $R = k/n$ (where $k$ is the number of logical qubits), and parameter $d$, most codes in the family have distance at least $d$. Furthermore, every code in the family has rate $R$ and check weight at most $\sim \frac{R}{1-R} d \log n$, where check weight is the maximum weight of any stabilizer generator. This extends the result of the quantum Gilbert-Varshamov bound, which shows by a coarse-grained probabilistic stabilizer tableau analysis a similar result with a rate-distance trade-off but always near-maximal stabilizer weight. Such a result further exemplifies that graphs enable greater flexibility in code analysis.

Our paper argues at a very general scope, namely at the level of all stabilizer codes. The universality of our work therefore has limitations that do not exist for highly specific quantum low-density parity check (qLDPC) constructions. In particular, even when optimized as much as possible with known techniques, code properties such as check weight, encoding circuit depth, and logical operator circuit depth, will scale with the degree and therefore the code parameters, though not necessarily by a great deal (logarithmic, for example). Thus, our intention is not to use the graph representation as a means for continued study into asymptotically good qLDPC codes with very large constants in the manner of Refs.~~\cite{good1,good2,good3,panteleev2022asymptotically}. Rather, we consider our work to be complementary to qLDPC in the sense that we give general insight into stabilizer codes and new flexible ways to construct codes that may be tailored specifically for a given experimental implementation. As an example, a future quantum long-term memory device may seek a particular large numerical distance threshold but care little for comparably large check weights if the measurement error is sufficiently small. This philosophy underpins our explicit construction of codes which have desirable properties at small scales. We believe also that further applications of graph theoretical techniques may further improve the techniques in this paper to build increasingly practical recipes for code constructions that flexibly adapt to experimental constraints.

More generally, we argue that graph theory provides a promising path forward \textit{generally} across avenues of quantum coding theory. Recent breakthroughs in qLDPC codes, for example, have relied on graphs in their construction to generalize the use of surface code lattices~\cite{bravyi2024high}. Perhaps further generalizations of these specific types of constructive techniques, as well as less universal but more fine-grained (i.e. finding certain graph constructions which map to codes in such a way that some properties scale favorably or not at all with parameters) refinements of the universal graph representation will lead to new constructions of and insights into quantum codes.

\subsection{Related work}
Graph forms have been exploited in many ways in the quantum computation literature, most generally via tensor network representations. One simple case of a graph representation is a graph state. \citet{hu2022improved} has produced a generalized graph state representation of stabilizer \textit{states}, i.e. a stabilizer code with 0 logical qubits. Our reduction from tableau to graph utilizes insights from \citet{hu2022improved}, and in the special case of $n = k$ in our work, there is a transformation that reduces our tableau-to-graph map to their corresponding map. We remark that \citet{mcelvanney2022complete}, building upon \citet{hu2022improved}, derived a different unique canonical form for stabilizer states that are particularly convenient for measurement-based quantum computation. Independently, \citet{yu2007graphical} also utilized graph theory in quantum coding, albeit at a much more general level (including non-stabilizer codes) for constructive and classification purposes.
Additionally, several other works have started exploring the connections between quantum codes and graphs, going in directions ranging from holography~\cite{huberholography}, code concatenation~\cite{beigi-graphconcat}, quadratic forms~\cite{grassl}, and connections with group theory~\cite{schlingemann,schlingemann-werner}.

Parts of our work---particularly the reduction from tableau to graph---utilize rules and notation from the ZX calculus. Recent works have also explored the use of ZX calculus to diagrammatically express quantum codes. A recent work of \citet{kissinger2022phase} developed independently of this work also constructs diagrammatic forms for \textit{CSS codes}.
Their forms are substantially different from ours and serve a different purpose. Primarily, Kissinger establishes a correspondence between CSS codes and ZX diagrams, by non-uniquely mapping CSS codes to such diagrams and mapping a subset of the ZX diagrams (the phase-free ones) to CSS codes. The tableau-to-graph portion of our work extends the conceptual idea of diagrammatic forms inspired by ZX calculus to all stabilizer codes, but takes a significantly different approach and utilizes the result for constructive and algorithmic purposes.

More generally, ZX calculus approaches to quantum error correction have been considered in various ways. Two relatively recent works which utilize the ZX calculus exemplify these different approaches. In the first way, \citet{wu2023zx} represent specifically graph codes with ZX calculus diagrams. This approach, while not general to stabilizer or CSS codes, has shown promise in code analysis. The second, due to~\citet{chancellor2016graphical}, uses the ZX calculus specifically to quantize classical codes in the spirit of CSS code construction. Both of these works give evidence that graphs enable new insights into quantum coding, both in their construction and analysis, and thus motivate our completely general theory of graph representations. Moreover, aside from our tableau-to-graph reduction algorithm which does use elements of the ZX calculus, our work builds everything essentially only from the standard definition graphs. We emphasize that the graphical diagrams of the ZX calculus require additional structure. Therefore, graph representations further simplify related previous works that also devise similar representations based on the ZX calculus. The specific motivations, applications, and analysis in these papers are also distinct from those in this work.

Building upon a preliminary version of this work, \citet{khesin2024equivalence} conducted further analysis of graph representations in the special case of CSS codes, wherein the graphs become bipartite. They specifically analyzed different ways to further simplify the graph representation and potential advantages in doing so.

\section{\label{sec:zxcf}From Tableaus to Graphs}
We begin by constructing the general graph form and compiling any stabilizer tableau into such a graph. More precisely, we will compile an equivalence class of encoding Clifford circuits corresponding to an equivalence class of stabilizer tableaus, and at the end of this section we discuss how to compile starting with a tableau~\cite{gottesman2009introduction}. As a starting point, we appeal to the ZX-calculus, a graph language for linear maps that has become of great interest in quantum information research~\cite{coecke2008interacting,zx1,zx2,zx3,zx4,kissinger2022phase}. The ZX-calculus produces visual diagrams that represent quantum states, circuits, and more. As with any formal logical system, the ZX calculus has a set of rules which may be iteratively applied to transform diagrams into equivalent diagrams. These rules are representations of identities in quantum circuits. The ZX-calculus has become of more interest than ever in fault-tolerant quantum computation and quantum compiler theory because it can explicitly visualize quantum properties such as entanglement in an intuitive manner. It has recently been applied to a host of quantum computation problems, including lattice surgery~\cite{de2020zx} and quantum optimization~\cite{kissinger2020reducing}. Importantly, the ZX-calculus is, for stabilizer tableaus, complete (equalities of tableaus can be derived from corresponding ZX diagrams), sound (vice versa), and universal (every quantum operation can be expressed in the ZX-calculus)~\cite{coecke2008interacting,backens2014zx}. 

We are consequently motivated to leverage the ZX calculus as an intermediate step in the development of a graph presentation. Ultimately, our graphs will be simple graphs and not the more complicated and structured diagrams in the ZX calculus. The diagrams that we map our circuits into are a subset of ZX diagrams that are equivalent by a simple transformation to graphs. To more clearly elucidate the transformation from tableau to graph, we construct the map from a composition of simpler maps. In particular, we first convert the tableau into a circuit. Next we show that up to an equivalence relation, we can efficiently biject the circuits into a class of ZX diagrams of a certain form that satisfies four rules. We thus denote these diagrams \textit{ZX canonical forms} (ZXCFs). Lastly, we remove unnecessary structure from ZXCFs to obtain a graph representation of the code.
The set of transformations that we can efficiently perform is pictorially represented by the following maps. \begin{align} \label{eq:map_sequence}
    \text{Tableau} \longleftrightarrow \text{Encoder} \longleftrightarrow \text{ZXCF} \stackrel{\text{LE}}{\longleftrightarrow} \text{Graph} .
\end{align}
In absence of the last step, everything is completely invertible and has an interpretation as a unique compiler. For the last step, the map between ZXCFs and graphs is invertible modulo \textit{local equivalence} (LE); that is, modulo local (1-qubit) Clifford operations on the outputs.
An immediate consequence for this ZXCF compiler is a diagrammatic method of testing equality between stabilizer codes.

We begin the formalism by noting that properties of a code depend only on the codespace, and thus there is a unitary degree of freedom on the logical space. In particular, encoding circuits for a given stabilizer tableau will produce the same stabilizer code if and only if they have the same image. We therefore make the following definition.
\begin{definition}
    Two circuits are equivalent as \textit{Clifford encoders} if they have the same image over all possible input states. Equivalently, $\CC_1$ and $\CC_2$ are equivalent encoders if there exists a unitary $U$ acting on the inputs of the circuit such that $\CC_1 = \CC_2 U$ or $\CC_2 = \CC_1 U$. 
\end{definition}
Any reference we make to encoding circuits will either implicitly or explicitly refer to them up to this equivalence, and indeed they all map to the same stabilizer code.

\subsection{\label{sec:theory:subsec:construction}ZX canonical form construction}
In the intermediate steps using ZX calculus rules, we follow the standard notation of ZX-calculus graph diagrammatics, specified in~\citet{backens2014zx}; we refer the reader there for the basics of the ZX calculus and transformation rules. 

A ZX diagram is a graph with additional structure. In particular, nodes may be either red or green, nodes may be decorated with local Clifford operators, and edges may be Hadamarded. Additionally, nodes may have a ``free edge'', an edge connected only to one node and dangling on the other side.
Green nodes are associated with $Z$ operators, and red with $X$.
Nodes may have phases, corresponding to an application of $R_Z(\th) = e^{-i \frac{\th}{2} Z}$ for green nodes and $R_X = e^{-i \frac{\th}{2} X}$ for red nodes. We restrict phases to multiples of $\frac\pi2$ to keep our operations Clifford. We colour an edge blue and make it dotted if it has a Hadamard gate on it~\cite{duncan2020graph} (some papers use instead a line with a box, e.g.~\cite{backens2014zx}, and we occasionally also use this convention when convenient).
The circuit takes $k$ input qubits to $n$ output qubits, for $0 \leq k \leq n$.

The specific structure of our graphs will be those in the following form.
\begin{definition}
    An \textit{encoder-respecting form} $\CE$ has only green ($Z$) nodes, and is structured as a \textit{semi-bipartite graph}. A semi-bipartite graph has a left and right cluster, such that left-cluster nodes may have edges only to right-cluster nodes, but right-cluster nodes have no such restrictions. In our case, the left and right clusters are denoted as the input and output clusters, respectively.
    The input cluster $\CI$ has $k$ nodes associated with 
    the $k$ input qubits of the corresponding encoder, and the output cluster $\widetilde{\CO}$ has $n$ nodes, where $n\geq k$. 
    Each input node has a free (not connected to any other nodes) edge---the input edge. Similarly, each output node has a free output edge. The output edges are numbered from $1$ to $n$---in order from left to right on an incomplete stabilizer tableau or top to bottom on Clifford circuit output wires. For convenience, we refer to the node $v$ connected to an output edge numbered $i$ as the output node numbered $i$.
    \label{def:respect}
\end{definition}
The design of the encoder-respecting form graphically illustrates how information propagates from input to output (which edges connect $\CI$ to $\widetilde{\CO}$) as well as the entanglement structure (which edges connect $\widetilde{\CO}$ to $\widetilde{\CO}$) of the underlying encoder.
We emphasize that this idea is only for intuition, as there exist equivalence transformation rules on ZX diagrams that appear to change the connectivity in nontrivial ways but nonetheless represent the same operator.
Note that the structure of $\mathcal{E}$ gives rise to a natural binary ``partial adjacency matrix'' $M_{\mathcal{E}}$ of size $k \times n$, which describes the edges between $\CI$ and $\CO$ akin to the biadjacency matrix of bipartite graphs.

Although this structure appears intuitively appealing, it is insufficient because there exist several equivalence transformations in the ZX calculus that yield an equivalent (but not obviously so) diagram. We therefore constrain an encoder-respecting form into a ZXCF via four additional rules. For notational convenience, we will interchangeably refer to applying an operation on a node's free edge as applying an operation on the node.
\begin{enumerate}
    \item \textit{Edge Rule}: The ZXCF must have exactly one $Z$-node per free edge and every internal edge must have a Hadamard on it. 
    
    \item \textit{Hadamard Rule}: No output node $v$ may both have a Hadamard gate on its output edge and have edges connecting $v$ to lower-numbered nodes or input nodes.

    \item \textit{RREF Rule}: $M_{\mathcal{E}}$ must be in reduced row-echelon form (RREF).

    \item \textit{Clifford Rule}: Let $\CP \seq \widetilde{\CO}$ be the nodes associated with the pivot columns of the RREF matrix $M_{\mathcal{E}}$, so that $|\CP| = k$. The only local Cliffords allowed on nodes are $\set{I, S, Z, SZ, H, HZ}$, but there can be no operations at all on input or pivot nodes. There can be no input-input edges or pivot-pivot edges. 
\end{enumerate}

\begin{figure}
    \centering
    \includegraphics[scale=0.5]{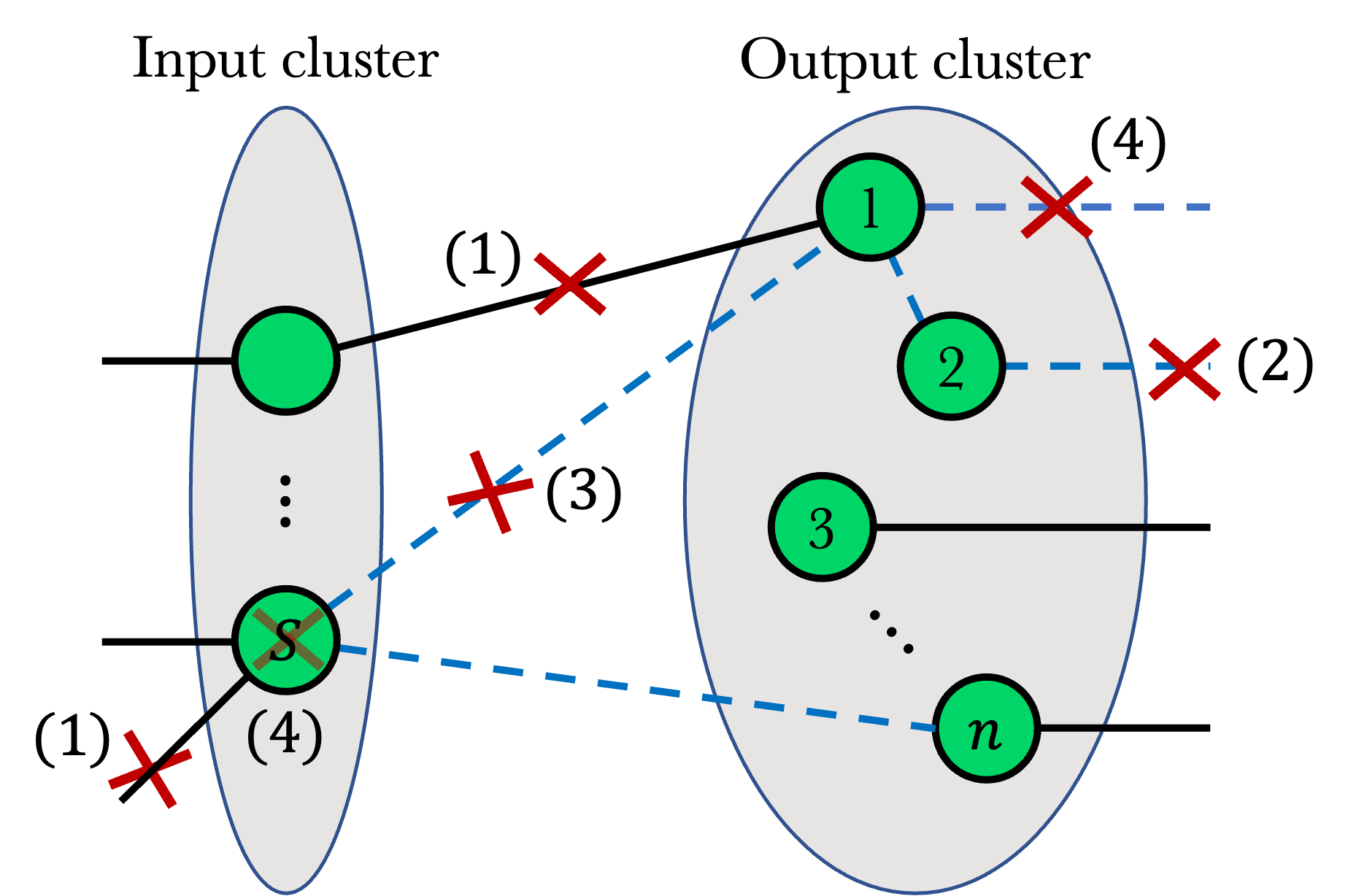}
    \caption{Example of an encoder-respecting form and some ways it might violate the 4 rules. Recall that dotted blue edges indicate the presence of a Hadamard on that edge. Rule 1, the Edge Rule, is violated in two places: one of the internal edges does not contain a Hadamard gate and a $Z$ does not have exactly one free edge. Rule 2, the Hadamard Rule, is violated where the second output node has a Hadamard on its free edge, but since the node is connected to the lower-numbered first output node, this is not acceptable. Rule 3, the RREF Rule, is violated where $M_{\mathcal{E}}$ is not in reduced row-echelon form, as we see that the first input is connected to the first output, meaning no other inputs should be connected to that first output node. Rule 4, the Clifford Rule, is violated twice: one in having a local Clifford operation, $S$, on an input node, and where an $H$ is placed on the first output node, which is a node associated with a pivot column of $M_{\mathcal{E}}$.}
    \label{fog:example_violations}
\end{figure}
We have elected for emphasis to include the $Z$-nodes' only constraint in the edge rule and the no input-input edges constraint in the Clifford rule even though these are redundant with that of the encoder-respecting form.

To provide some visual intuition on these 4 rules, Fig.~\ref{fog:example_violations} depicts a generic example of some possible violations to the rules. We define the four rules such that the ZXCF gives a unique canonical form, is homogeneous in node type, and minimal in terms of decorated operators and number of vertices which do not correspond to a physical or logical qubit. It can be shown~\cite{van2020zx,backens2014zx} that each of these rules corresponds to certain equivalence transformations. For example, row operations on $M_{\mathcal{D}}$ correspond to unitaries on the input. Rather than give all the rules, we will directly devise an encoder to ZXCF map and prove that such a map is efficient and bijective.

\begin{theorem} \label{thm:canonicity}
    Any Clifford encoder has a unique equivalent ZXCF satisfying the Edge, Hadamard, RREF, and Clifford rules. There exists an algorithm, running in worst-case $O(n^3)$ time, that on an input of a Clifford encoder circuit outputs a corresponding ZXCF. This map is bijective up to encoder equivalence.
\end{theorem}
We briefly sketch the proof idea of Theorem~\ref{thm:canonicity} but defer the details to Appendix~\ref{app:sec:compiler}. The proof involves a series of transformations via ZX equivalence rules. In the first step, we apply an isomorphism between circuits and states, and proceed to apply a result of \citet{hu2022improved} to preliminarily represent the circuit as the ZX diagram of a corresponding state. We then transform the diagram into one representing a circuit, and carefully choose a series of equivalence transformations that iteratively force the diagram to satisfy a rule above without at the same time violating a different rule that has previously been satisfied. The resulting product is precisely a ZXCF. In fact, because the steps are equivalence transformations, distinct encoders map to distinct ZXCFs. The only remaining aspect of the proof of Theorem~\ref{thm:canonicity} is whether this one-to-one map is bijective up to encoder equivalence. Since equivalent encoders will have the same stabilizer tableau, we only have to demonstrate the bijection between ZXCFs and stabilizer tableaus. We prove this claim in Section~\ref{sec:compiler}.

\subsubsection{Tableau to encoder}
There are numerous algorithms for mapping a stabilizer tableau into a Clifford encoding circuit~\cite{nielsen2002quantum,gottesman2016surviving}, and it does not matter which algorithm one chooses to produce an encoding circuit. If we quotient out operations which leave the code space invariant, then any such map is also bijective. We describe one such algorithm for completeness.
For a $(n-k) \times n$ tableau, begin by drawing $n$ output wires. At each step, we first simplify the tableau by applying a Clifford operation, and then measure out one of the qubits to remove one row and column from the tableau. Repeating inductively yields a Clifford circuit that takes $k$ qubits to $n$ qubits.

The procedure at each step begins by finding a Clifford operation $U$ that turns the first stabilizer into $Z_1=Z\otimes I\otimes\dots\otimes I$, in the manner dictated by the Gottesman-Knill theorem. $U$ is prepended to the circuit under construction. We then multiply the remaining rows by the first until the entire first column of the tableau has only $I$, with the exception of the first row. This can always be done, since the $n$-qubit Pauli operators in each row of the matrix must commute.
We then post-select on the $+1$ result of a computational basis measurement by applying $\bra0$ in the reverse direction.
When reading the circuit in the forward direction, this is equivalent to initializing an ancillary qubit in the $\ket0$ state.
The effect of $U$ and the measurement is equivalent to applying $\frac{I+P}2$, where $P$ is the first row of the stabilizer tableau.
This is equivalent to post-selecting on the $+1$ measurement result of that stabilizer.
Having measured the qubit, we remove its corresponding row and column from the tableau, and repeat on a tableau of $n-1$ qubits and $n-k-1$ rows, until there are no rows left, at which point the remaining $k$ wires that have not been measured out become the input qubits in the circuit. We conclude by drawing $k$ input wires for those qubits. 


\subsection{\label{sec:compiler}Proof of canonicity}
We next prove that our proposed ZXCF is indeed canonical. In other words, two equivalent stabilizer tableaus---generators of the same subspace of the $n$-qubit Hilbert space---will map to the same ZXCF. 
We proceed by counting the number of stabilizer tableaus and ZXCFs. For notational convenience, let $\Bar{k} := n - k$.


First, the number of stabilizer tableaus with $\Bar{k}$ stabilizers on $n$ qubits is
\begin{align}
\frac{\prod\limits_{i=1}^{\Bar{k}}2\cdot4^n/2^{i-1}-2\cdot2^{i-1}}{\prod\limits_{i=1}^{\Bar{k}}2^{\Bar{k}}-2^{i-1}}=\prod_{i=1}^{\Bar{k}}\frac{2^{2n-i+2}-2^i}{2^{\Bar{k}}-2^{i-1}}.
\label{eq:zx-count}
\end{align}
For each row $i$, there are $2 \cdot 4^n$ possible Pauli strings (including the sign), but the requirement that they commute with previous rows divides the count by $2^{i-1}$. Of these $2 \cdot 4^n/2^{i-1}$ valid strings, $2^{i-1}$ strings are linear combinations of previous strings, with a factor of 2 for strings that differ by a sign from previous strings. 

We have overcounted, however, since there are many ways to find a set of stabilizer generators for a particular tableau. In particular, when choosing a set of generators for a Clifford encoder, we have $2^{\Bar{k}}-1$ choices for the first generator (subtracting the identity). There are then $2^{\Bar{k}}-2$ ways of choosing the next element, as we cannot pick anything in the span of the elements chosen so far. Repeating inductively gives the denominator above.

Next, the number of ZXCF diagrams for the same $n$ and $\Bar{k}$ can be expressed by the following fourfold recursive function $f$ evaluated at $p = o = 0$.
\begin{align}
f(n,\Bar{k},p,o)=\begin{cases}
1 & \text{\hspace{-8.5ex}if } n=\Bar{k}=0 \\
A_{n,\Bar{k},p,o}+B_{n,\Bar{k},p,o}&\text{else}
\end{cases}
\label{eq:zxcf-count}
\end{align}
where
\begin{align}
A_{n,\Bar{k},p,o}=
2^of(n-1,\Bar{k},p+1,o)
\end{align}
if $n \neq \Bar{k}$ and $A_{n,\Bar{k},p,o}=0$ if $n = \Bar{k}$,
and where
\begin{align}
    B_{n,\Bar{k},p,o}=(2^{2p+o+2}+2)f(n-1,\Bar{k}-1,p,o+1)
\end{align}
if $\Bar{k} \neq 0$ and $B_{n,\Bar{k},p,o}=0$ if $\Bar{k} = 0$.

We may intuit the recursion as going through the $n$ output vertices and assigning them to either be pivots or non-pivots. The case $A$ represents the number of ways to assign the remainder of the nodes if we assign the current node to be a pivot. Similarly, the case $B$ counts the cases for when the current node becomes a non-pivot output. This function is computed with a base case of an empty tableau when $n=k=0$. A formal derivation of this function is left to Appendix~\ref{app:recursion}. We solve for an explicit form to obtain
\begin{align}
    f(n,\Bar{k},p,o) = 2^{o (n - \Bar{k})}  \prod_{i=1}^{\Bar{k}} \frac{(2^{n+1} - 2^i) (2^{n - i + 1 + 2 p + o} + 1)}{2^{\Bar{k}} - 2^{i-1}}.
\end{align}
One can verify by standard induction that $f(n,\Bar{k},0,0)$ gives the same expression as Eq.~(\ref{eq:zx-count}), proving that ZXCF is indeed canonical.

\subsection{ZXCF to graph}
In the final step, we reduce a ZXCF directly to a graph. We remark that any local Clifford operation must, by the Clifford rule, be placed only on non-pivot output nodes. However, local Cliffords do not change the weights of the set of correctable errors. Thus, local Cliffords do not change the code distance or any other code parameters. If we also quotient out this equivalence, we may remove all local Cliffords entirely. Since all internal edges have Hadamards, we can remove those as well and simply remember that they exist for inversion. Similarly, we can remove all free edges. The remaining diagram is identically a graph, specified entirely by a collection of nodes and edges. As a consequence, we have completed the map sequence from Eqn.~(\ref{eq:map_sequence}), except for the graph to ZXCF map.
\begin{theorem}
    For any stabilizer code, there exists at least one equivalent code that has a presentation as a semi-bipartite graph with no local Clifford operations.
\end{theorem}

\begin{figure}[ht!]
    \centering
    \includegraphics[width=\linewidth]{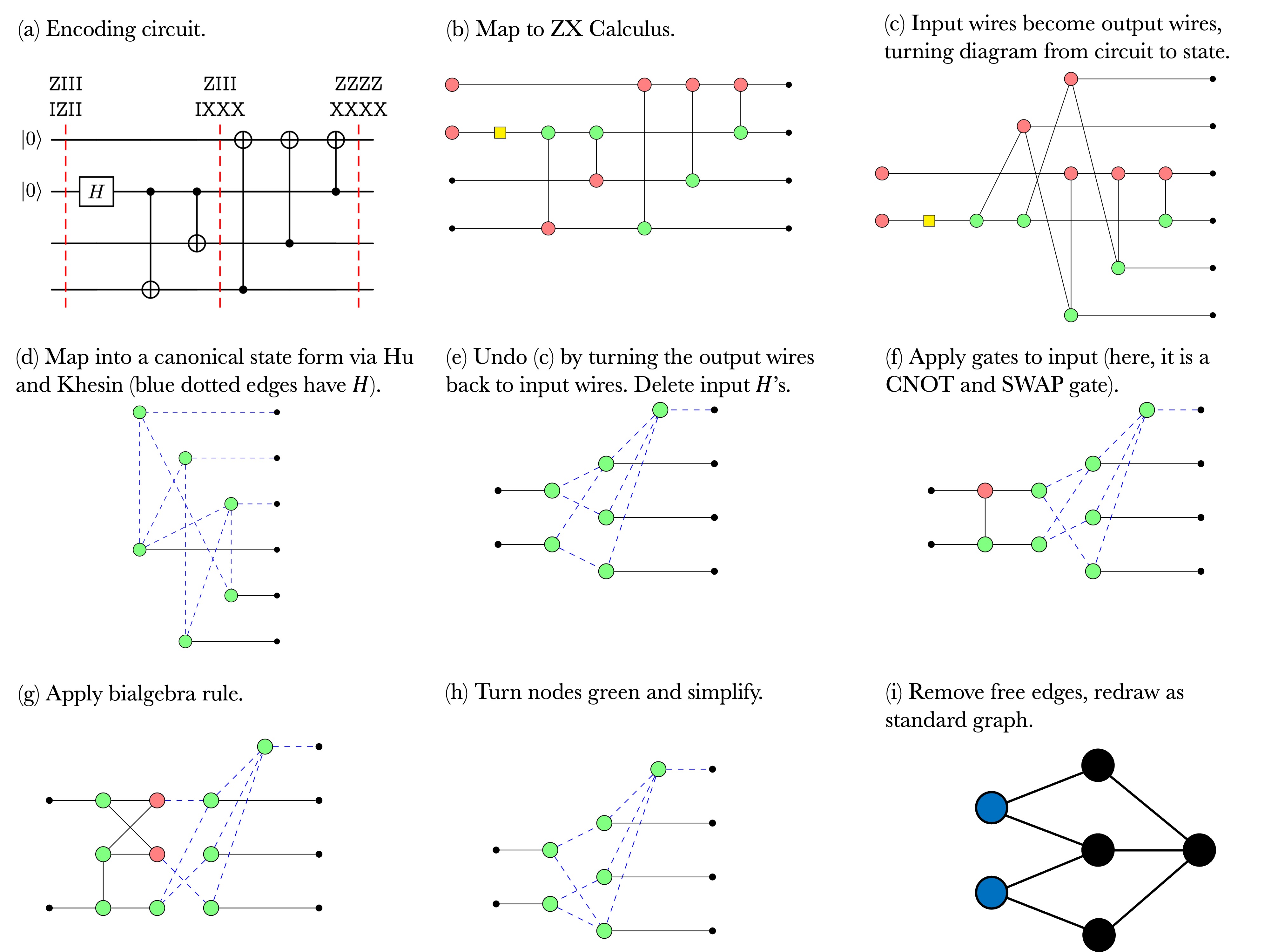}
    \caption{Illustration of the compilation process of a $\llbracket 4, 2, 2 \rrbracket$ code with stabilizers $XXXX$ and $ZZZZ$. In (a), we begin with the encoding circuit of the code. We then map it into the ZX calculus in (b), where yellow boxes (as well as dotted blue edges) represent Hadamards $H$, and the green and red nodes are the standard notation in the ZX calculus. In (c), we turn the circuit into a state by moving input wires to output wires. (d) implements the transformation from \citet{hu2022improved}, turning it into a canonical form which satisfies the Edge and Hadamard rules. We then return the state to a circuit in (e) by moving the output wires from (c) back to input wires. To satisfy the RREF rule, we apply gates on inputs and the bialgebra rule---an equivalence transformation in the ZX calculus---in (f) and (g). The details of these simplification steps are all discussed in Appendix~\ref{app:sec:compiler}. Finally, in (h) and (i) we simplify and remove free edges to turn the diagram into a graph. In the final graph, the inputs are blue (leftmost column) and the outputs are black (middle and right columns).}
    \label{fig:Compilee422}
\end{figure}

For concreteness, Fig.~\ref{fig:Compilee422} illustrates an example run of the compilation process on the $\llbracket 4, 2, 2 \rrbracket$ code, which has stabilizers $XXXX$ and $ZZZZ$. The exact choice of equivalence transformations is dictated by Appendix~\ref{app:sec:compiler}, but the key idea of Fig.~\ref{fig:Compilee422} is that an encoding circuit turns into a ZX diagram, which is simplified by a series of equivalence transformations that ultimately yield a simple graph form. Here, we have coloured the inputs blue and all other nodes black to emphasize that the final graph form can be studied independently of the ZX calculus.

\section{\label{sec:compiler:subsec:applications}
Application to Quantum Codes}
It is illuminating to observe the graph representation of well-known quantum codes. Not only does it inform of geometric structure embedded in the code, but it also provides a guide as to how such codes may be generalized.
We provide three examples for study.
Consider first the nine-qubit code, due to \citet{shor1995scheme}, which uses 9 physical qubits to encode 1 logical qubit. The Shor code may be represented by the stabilizer tableau \begin{align}
\begin{bmatrix}
    Z & Z & I & I & I & I & I & I & I \\
    Z & I & Z & I & I & I & I & I & I \\
    I & I & I & Z & Z & I & I & I & I \\
    I & I & I & Z & I & Z & I & I & I \\
    I & I & I & I & I & I & Z & Z & I \\
    I & I & I & I & I & I & Z & I & Z \\
    X & X & X & X & X & X & I & I & I \\
    X & X & X & I & I & I & X & X & X 
\end{bmatrix} .
\end{align}

\begin{figure}[ht!]
    \centering
    \includegraphics[scale=0.4]{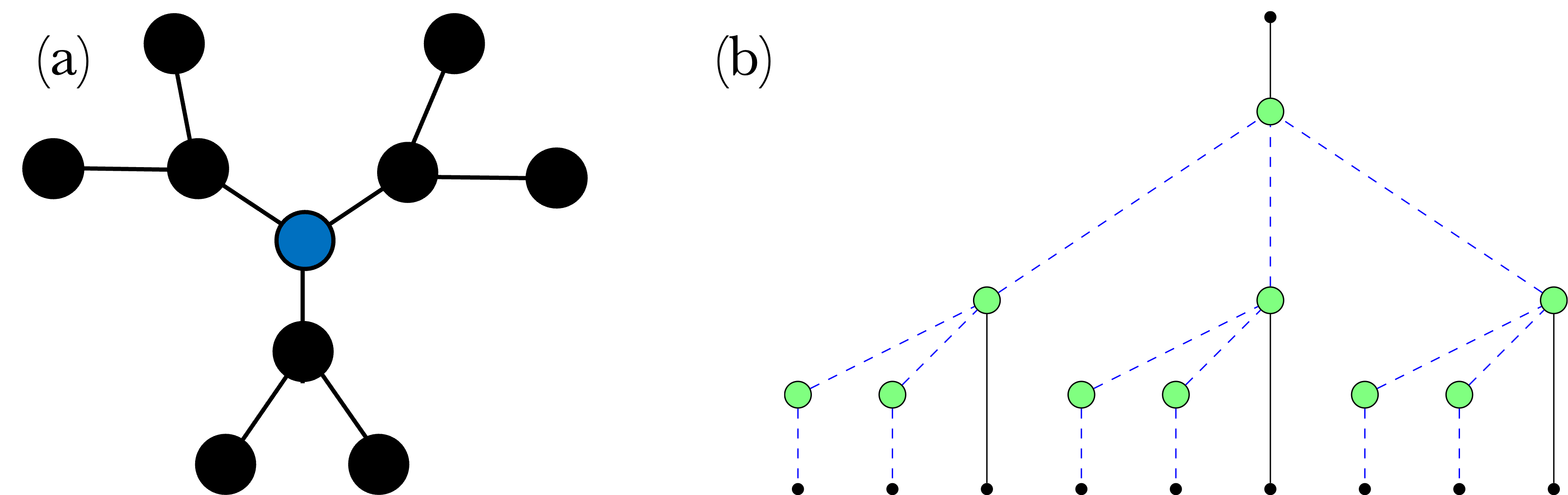}
    \caption{(a) Graph presentation and (b) ZXCF of the 9-qubit code. The input in the graph (a), in the center, is shown in blue. In the ZXCF, the green nodes are those of standard ZX calculus notation, and the dotted blue lines are Hadamarded edges. Note that the graph encoder uses a slightly different basis since its output edges do not contain Hadamard gates.}
    \label{fig:nine-qubit-code}
\end{figure}

In Fig.~\ref{fig:nine-qubit-code}, we give the graph representation in (a) and the ZXCF in (b). We denote in the ZXCF the input node with \textbf{I}, and the free edges are dotted for emphasis. In the graph presentation, the input node is blue and the outputs are black. We first examine the ZXCF. There are three identical sectors of the outputs, two with Hadamarded outputs and one without. This resembles our expectations from examination of the un-normalized qubit representation of the Shor code, $(\ket{000} \pm \ket{111})^{\otimes 3}$. The graph, on the other hand, is a star-shaped tree. It is easy to imagine how one might extend the Shor code into an infinite family of graphs, namely by recursively giving every outer (leaf) node two children.

\begin{figure}[ht!]
    \centering
    \includegraphics[scale=0.475]{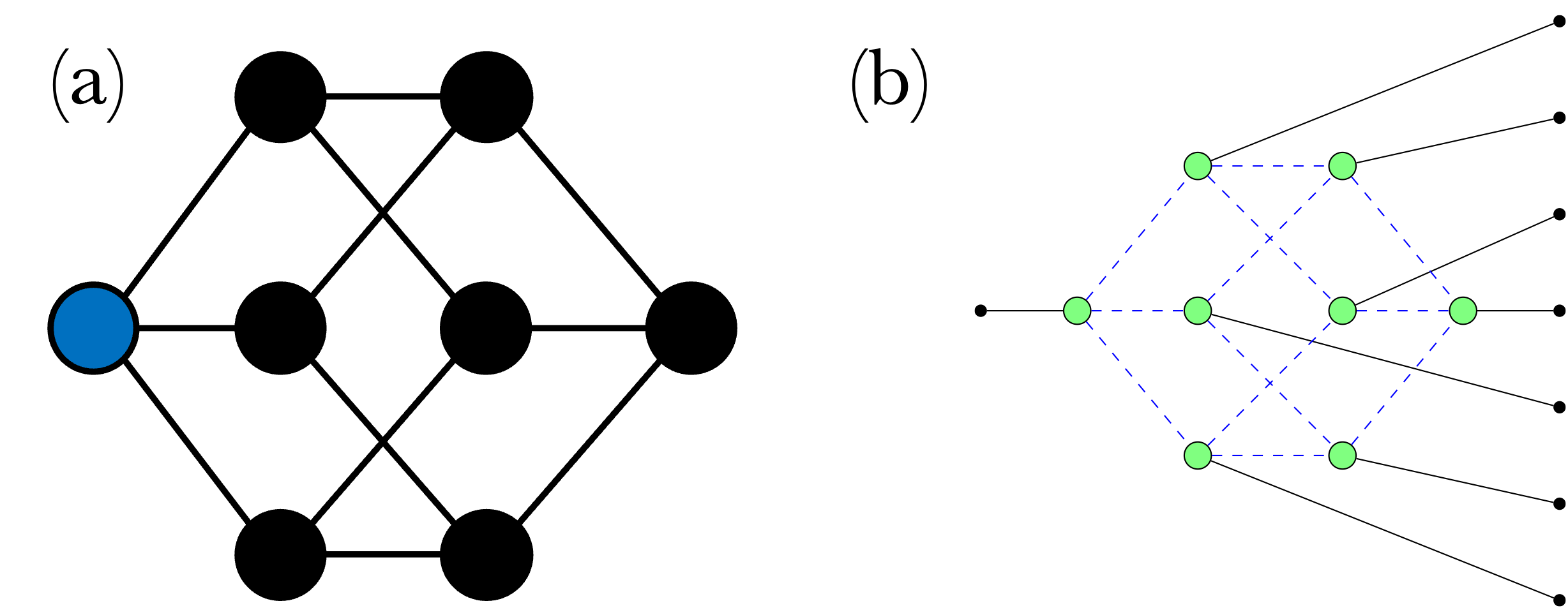}
    \caption{(a) Graph presentation and (b) ZXCF of the 7-qubit code. The input in the graph (a), given by the leftmost node, is shown in blue. The ZXCF in (b) has standard ZX calculus green nodes, and dotted blue lines are Hadamarded edges. Note that the graph encoder uses a slightly different basis as since its output edges do not contain Hadamard gates.}
    \label{fig:7-qubit-code}
\end{figure}

Next, we compile Steane's seven-qubit code, a central construction in many fault-tolerant quantum computation schemes~\cite{steane1996multiple}. The code is represented by the following tableau: \begin{align}
    \begin{bmatrix}
        I & I & I & X & X & X & X & \\ 
        I & X & X & I & I & X & X & \\ 
        X & I & X & I & X & I & X \\
        I & I & I & Z & Z & Z & Z \\
        I & Z & Z & I & I & Z & Z \\
        Z & I & Z & I & Z & I & Z
    \end{bmatrix} .
\end{align}
In our ZXCF, this code takes the form given in Fig.~\ref{fig:7-qubit-code}. In particular, the nodes of the diagram are simply the corners of a cube. A similar picture has been given in a different context in~\citet{duncan2013verifying}. The Steane code's generalization is also evident, via extensions into hypercubes of arbitrary dimension. We explore precisely such a coding scheme in Section~\ref{subsec:decoder}.

We observe that the stabilizers of the 7-qubit code are positioned at exactly the $1$-indices in the binary representation of a node label.
We can see elegant symmetries in Fig.~\ref{fig:7-qubit-code} with the positions of the nodes and their expressions in binary.
The reason why the nodes adjacent to 0 are not 1, 2, and 4, (respectively 001, 010, and 100 in binary) is because the rules of the ZXCF require that the Hadamarded output edges not be connected to lower-numbered nodes.

\begin{figure}[ht!]
    \centering
    \includegraphics[scale=0.475]{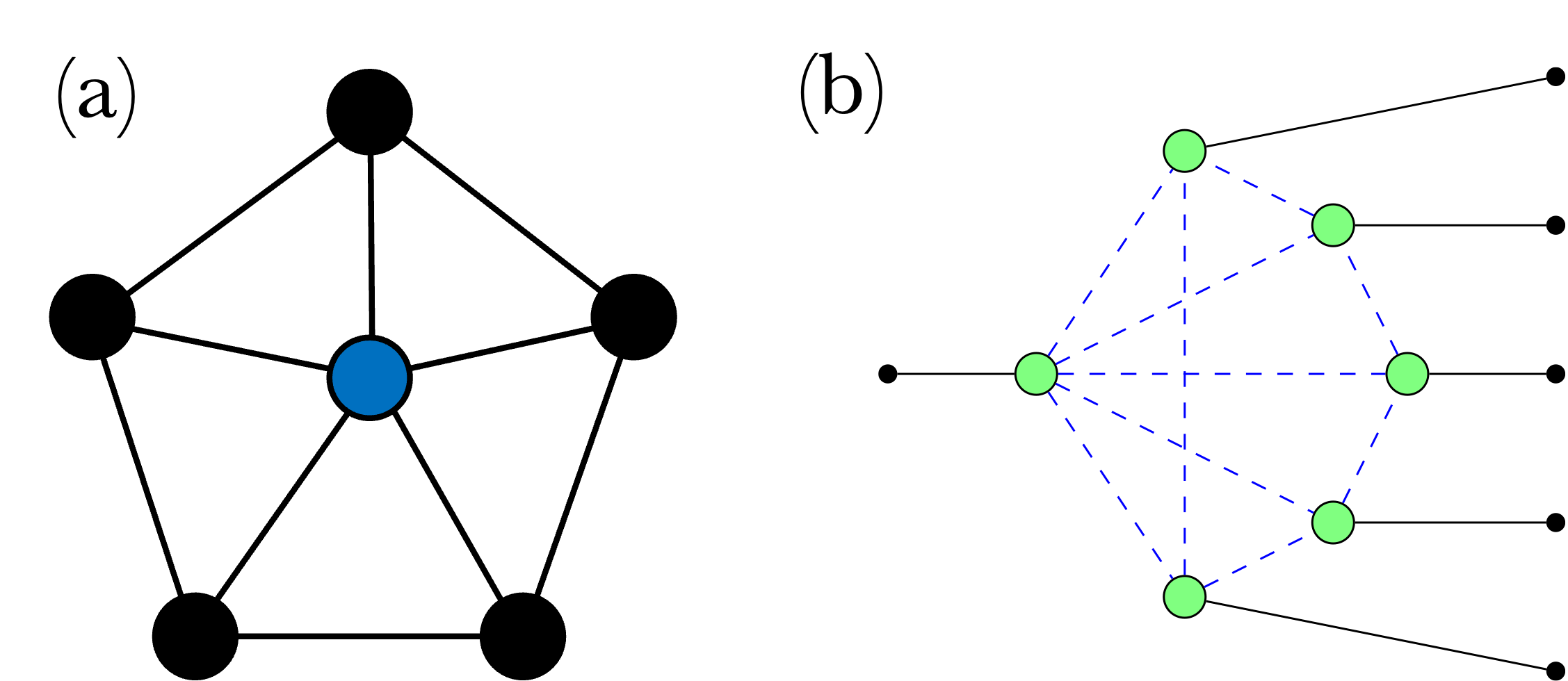}
    \caption{(a) Graph presentation and (b) ZXCF of the 5-qubit code. The input in the graph (a), in the center, is shown in blue. The ZXCF (b) has standard green ZX calculus nodes and dotted blue lines are Hadamarded edges.}
    \label{fig:5-qubit-code}
\end{figure}

As a final example we consider the five-qubit code, which is the smallest code one can achieve for correction of an arbitrary single-qubit error and has been studied experimentally~\cite{gottesman2009introduction,knill2001benchmarking}. Its tableau representation is given by
\begin{align}
    \begin{bmatrix}
        X & Z & Z & X & I \\
        I & X & Z & Z & X \\
        X & I & X & Z & Z \\
        Z & X & I & X & Z 
    \end{bmatrix} .
\end{align}
Our procedure produces the ZXCF shown in Fig.~\ref{fig:5-qubit-code}.

The 5-qubit code takes the shape of either a pentagon with an input in the middle or a cone, depending on how one sets up the graph. Consequently, a simple generalization is to use a $n$-gon, or equivalently add more nodes to the circle in the cone.

Such elegant and simple structure in the 5, 7, and 9-qubit codes suggest that other graphs with similar elegant structure may yield interesting quantum codes, either other constant size codes or larger codes in families in which these are merely small examples. We explore this idea in Section~\ref{subsec:smallcodes}.

\section{\label{sec:inversion}The Inverse Map: from Graphs to Stabilizer Codes}
The bijective relation between stabilizer codes and graphs---particularly those in the ZX canonical form---give justification that graphs are at least as powerful as stabilizer tableaus for the expression of codes. In this section, we lay the foundation for the use of graphs as \textit{constructive tools} for code design. Such a mapping from graphs to stabilizer tableaus can be viewed as an ``inverse map'' to the algorithm given in Section~\ref{sec:compiler}, with some caveats relating to the fact that not all graphs can be equivalently represented by ZXCFs and therefore not all graphs can be mapped into codes.

\begin{figure}[ht!]
    \centering
    \includegraphics[width=0.4\linewidth]{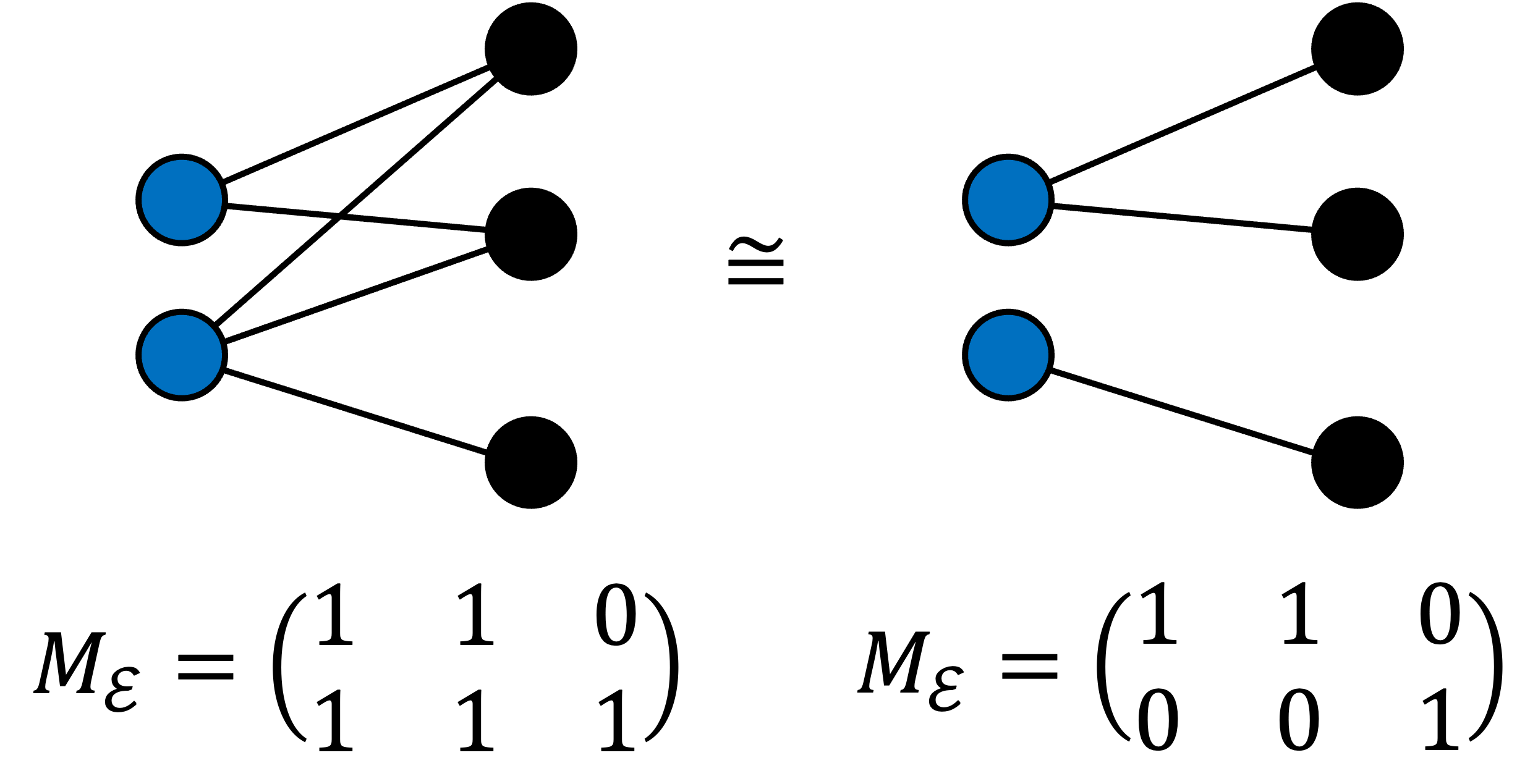}
    \caption{A simple example for which the third output node must be chosen as a pivot. The inputs are on the left side of each graph and colored blue and outputs on the right side are black. The example is given on the left, and an equivalent (via row operations on the partial adjacency matrix between the left and right) graph is given on the right.}
    \label{fig:badpivot}
\end{figure}

Let $G$ be a graph with $n+k$ nodes, separated into $k$ input nodes $\CI$, a choice of $k$ pivot nodes $\CP$, and $n-k$ non-pivot output nodes $\CO$. In the notation of the encoder-respecting form from Definition~\ref{def:respect}, $\CP \cup \CO = \widetilde{\CO}$; we will henceforth directly write $\CP \cup \CO$ to refer to all output nodes. To ensure that our operations are well-defined, there must exist equivalence operations that map $G$ into a ZXCF. The only nontrivial restriction this requirement poses is on the choice of inputs and pivots. The RREF rule ensures that every distinct pivot is connected to exactly one distinct input; in other words, $\CP$ and $\CI$ are perfectly matched. Because row operations on the partial adjacency matrix are equivalences, it suffices to choose pivots such that there exists a row-reduced partial adjacency matrix $M_{\mathcal{E}}$ for which the chosen $k$ pivot nodes correspond to the $k$ pivot columns of $M_{\mathcal{E}}$. A simple example of this subtlety is shown in Fig.~\ref{fig:badpivot}. Here, it is impossible to choose the first two outputs to be the pivots because the third output is forced to be a pivot when we row-reduce. Note, however, that this subtle restriction poses no issue \textit{if every distinct pivot we choose is already connected to exactly one distinct input} (as in the case in Section~\ref{subsec:randcodes}) or if we can observe directly that the pivots we have chosen do correspond to the pivot columns after row reduction. It is also convenient to continue to enforce that inputs are not connected, so as to maintain encoder-respecting form. More generally, we say a choice of inputs and pivots is \textit{valid} if no inputs are connected to each other and each pivot is connected to exactly one input. Henceforth, we will denote $G = (\CI \sqcup \CP \sqcup \CO, E)$ (or equivalently $G = (\CI \cup \CP \cup \CO, E)$) as a graph for which a set of valid inputs and pivots $\CP$ have been specified alongside a set of input nodes $\CI$. The remaining non-pivot output nodes are $\CO$ and the edge set is $E$. Usually, we will implicitly assume inputs are not connected, and just say that a choice of pivots is valid.

Assuming the choice of $\CP$ is valid, $G$ will be equivalent to a ZXCF by a series of equivalence transformations. First, we place a free edge on every node and place a Hadamard on all internal edges, satisfying the edge rule. The Hadamard rule is vacuously satisfied, as we will not place any Hadamard on output nodes. We can row-reduce the partial adjacency matrix to satisfy the RREF rule. Next, we strip input-input edges, which are unitaries on the inputs and thus represent equivalent encoders. Lastly, as discussed in Section~\ref{sec:zxcf}, there is an equivalence relation that strips pivot-pivot edges. This procedure results in a ZXCF with no local Clifford on nodes at all. While there exist encoders that can only be expressed with local Clifford operations that this map cannot produce, the map can always produce a locally equivalent encoder, and is otherwise completely general.

In practice, we never transform our graphs into a ZXCF. In fact, the primary purpose of the ZXCF is to prove that a certain family of graphs is expressive enough to represent every possible stabilizer code. Now that we have established this fact, we can find a more convenient and explicit map from a graph directly to a stabilizer tableau. This map will work so long as every input is connected to exactly one pivot (and vice versa), but otherwise we do not need to strictly abide by the ZXCF rules. The primary way in which we will take advantage of this extra flexibility is to allow pivot-pivot edges. Although no ZXCF has a pivot-pivot edge, our map from graph to tableau remains well-defined. Note that if we removed the pivot-pivot edges of a graph, we may in general change the corresponding code, but if we approach this from the direction of building graphs to make codes, it is worth noting that no additional codes will be expressible by adding pivot-pivot edges that could not have been produced without such edges. Aside from that, knowing that the graph is equivalent to a ZXCF, we can write down the stabilizer generators of its corresponding ZXCF directly. To do so, we first establish notation. Let $\D_{i=1}^n A_i = A_1 \,\D\, \cdots \,\D\, A_n$ for sets $A_1, \hdots, A_n$, where $\D$ is the symmetric difference. The symmetric difference is both commutative and associative, so this iterative procedure is well-defined. 

\begin{definition}[Neighbour sets] \label{def:neighbour_sets}
Let $G = (V, E)$ be a graph with input set $\CI$, pivot set $\CP$, and non-pivot output set $\CO$, so that $V = \CI \sqcup \CO \sqcup \CP$. Let $v \in V$. We define $i(v) = \set{w \in \CI \,|\, w \sim v}$ the inputs connected to $v$ and $p(v) =\set{w \in \CP \,|\, w \sim v}$ the pivots connected to $v$. Similarly, let $o(v) = \set{w \in \CO \,|\, w \sim v}$ be the set of non-pivot outputs connected to $v$. Finally, let $N_o(v) = \set{w \in \CO \cup \CP \,|\, w \sim v}=o(v)\cup p(v)$ be all the output neighbours of $v$. More generally, we also define for a set $A \subseteq V$, $N_o(A) = \Delta_{v \in A} N_o(v)$ and likewise for $i(A),p(A), o(A)$.
\end{definition}
If $v \in \CI$, then $|p(v)| = 1$ since each pivot is connected to a unique input. If $v \in \CP$, $|i(v)| = 1$ for the same reason.
For this reason, we will frequently refer to the input connected to a pivot $v$ as $i(v)$, even though technically the latter is a set containing one node.
The same applies for the pivot $p(v)$ connected to input $v$.
The symmetric difference is useful for correctly counting the parities of Pauli operations applied to certain nodes, i.e. if a Pauli is applied twice in the set, it is effectively not applied at all.

\begin{figure}[ht!]
    \centering
    \includegraphics[width=0.75\linewidth]{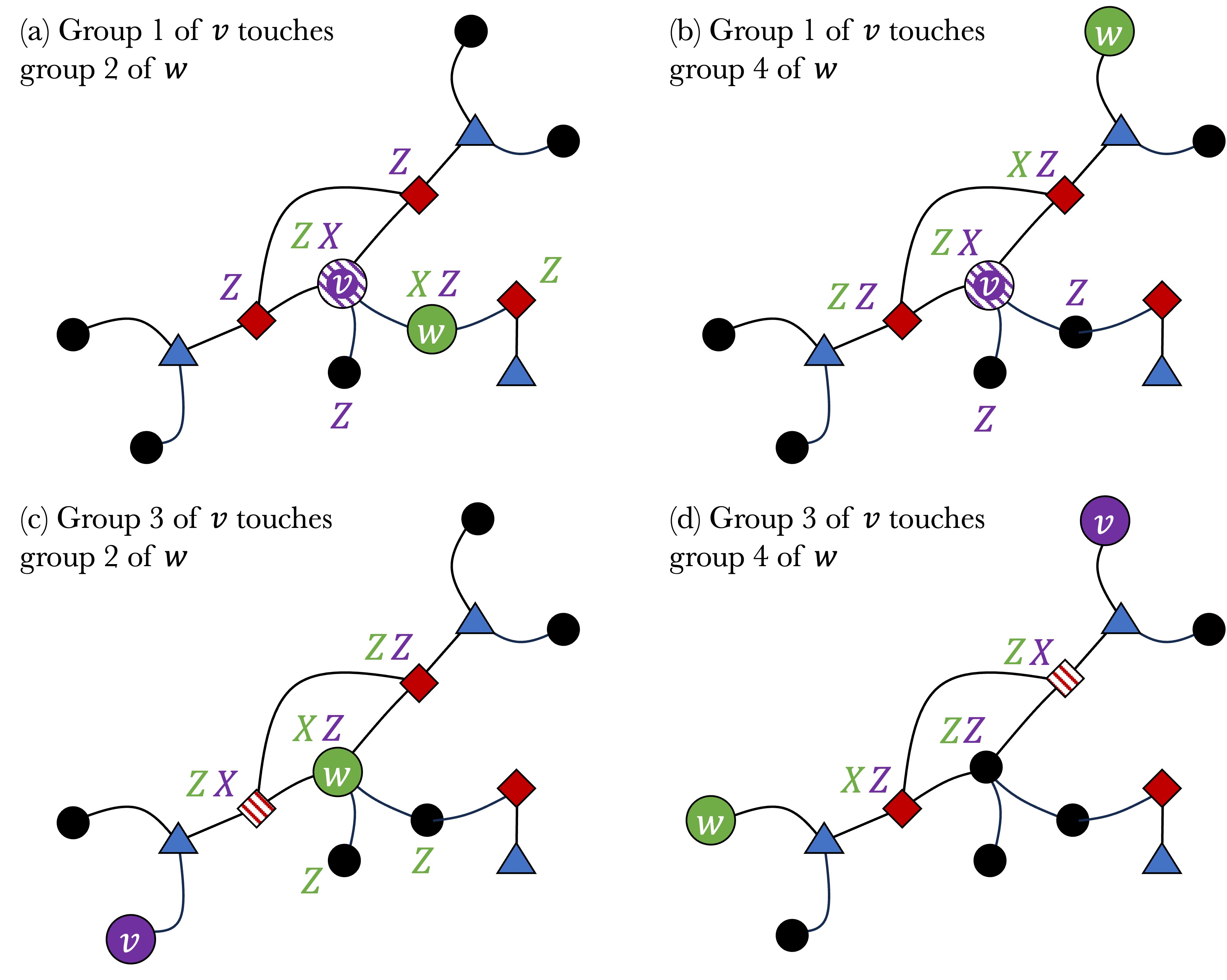}
    \caption{Depiction of the 4 cases that the stabilizers given by Eqn.~(\ref{eq:inverse}) commute. Inputs are blue triangles, pivots are red diamonds, and non-pivot outputs are black circles. For each $v \in \CO$, the support of the stabilizer $S_v$ is partitioned into four groups: $v$, $N_o(v)$, $p(i(v))$, and $N_o(p(i(v)))$, using Definition~\ref{def:neighbour_sets}. $X$'s are placed in the odd groups, and $Z$'s are placed on the even groups. Each case shows that when an odd group of a node $v \in \CO$ (purple) and an even group of a node $w \in \CO$ (green) have an intersection (striped node)---thus giving an anticommutation---there is always a corresponding intersection at another node which cancels out the anticommutation. Therefore, the stabilizers $S_v$ and $S_w$ commute.}
    \label{fig:StabilizersCommute}
\end{figure}

\begin{figure}[ht!]
    \centering
    \includegraphics[width=0.7\linewidth]{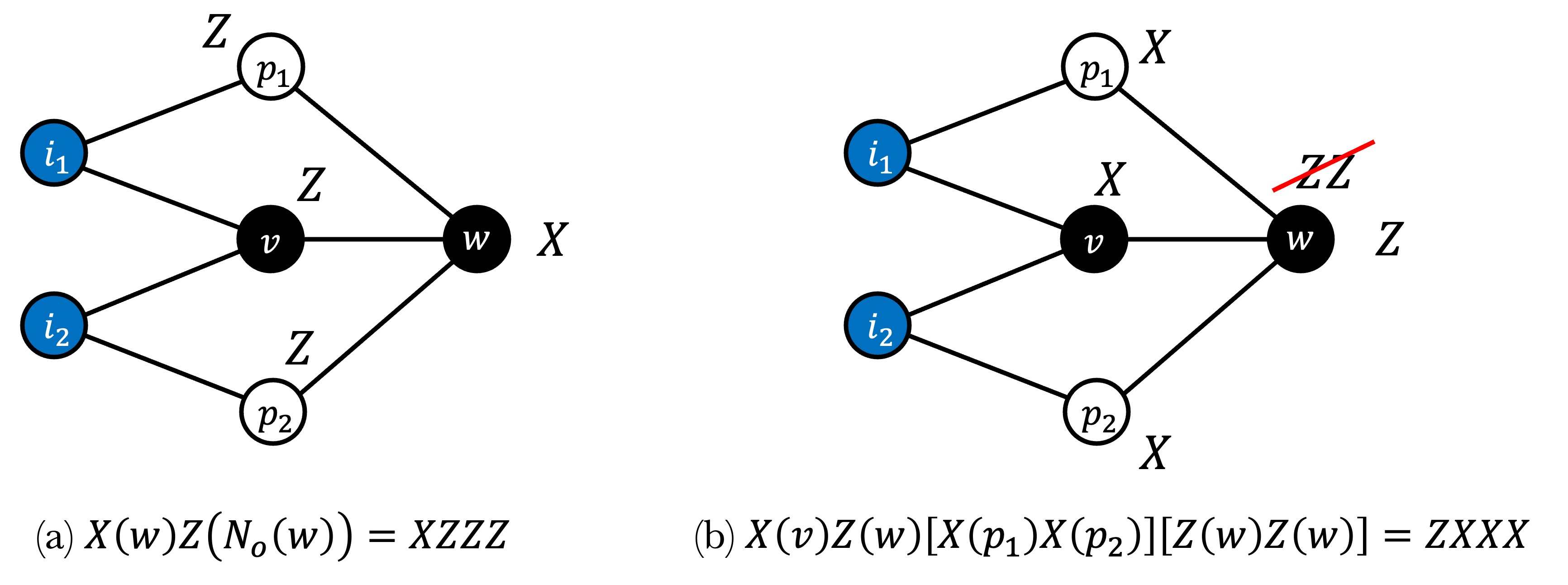}
    \caption{Canonical stabilizers according to Eqn.~(\ref{eq:inverse}) of the $\llbracket 4, 2, 2 \rrbracket$ code. Here $i_1, i_2 \in \CI$ are inputs, $p_1, p_2 \in \CP$ are pivots, and $v, w \in \CO$ are non-pivot outputs. The stabilizer in (a) corresponds to $v$ and the stabilizer in (b) corresponds to $w$, with cancellations of two $Z$'s depicted. The stabilizers differ by a local Clifford, namely a Hadamard on $w$, from the standard presentation of the $\llbracket 4, 2, 2 \rrbracket$ code with stabilizers $XXXX, ZZZZ$.}
    \label{fig:Stabilizers422}
\end{figure}

Every node of the graph represents a qubit---logical if the the node is in $\CI$ and physical otherwise. We can therefore speak of applying operations to the qubits by placing operators to their corresponding nodes.
Let $X(w)$ be a Pauli $X$ on node $w$, and $X(A) = \bigotimes_{v \in A} X(v)$; $Z$ operators are defined analogously. We may also write $X(A)$ and $Z(A)$ for placing a Pauli on every element of a subset $A \subseteq V$ of nodes. Note that we may only directly apply operators to output nodes because these are physical operators. With this notation, we define the following map $\CS : G \mapsto S$ from graph to stabilizer tableau, the output of which we call the \textit{canonical stabilizers} of the code represented by $G$. \begin{align} \label{eq:inverse}
    \CS := \set{X(v) Z(N_o(v)) X(p(i(v))) Z(N_o(p(i(v))))}_{v \in \CO} .
\end{align}
Denote each stabilizer $S_v$ (or, occasionally when more convenient, $S(v)$) for $v \in \CO$.
These stabilizers are readily seen to be a valid generating set. There are exactly $n-k$ generators, and all are independent because only $S_v$ has a $X$ on $v$.
Moreover, we claim that every pair of stabilizers $S_v$, $S_w$ commutes. Partition the support of the Paulis in each stabilizer $S_v$ into four groups: $v$, $N_o(v)$, $p(i(v))$, and $N_o(p(i(v)))$. $S_v$ places $X$'s on the odd groups and $Z$'s on the even groups. We therefore need to check that whenever an odd and an even group intersect, the intersection size has even parity. We show that for two stabilizers $S_v$ and $S_w$, in each of the four cases when an odd group intersects an even group, we can match each element of the intersection uniquely with a distinct element of the intersection. Therefore, the number of anti-commuting overlaps between the stabilizers is even, so the stabilizers commute. We split this argument by the groups. Each case follows directly from the undirected structure of the graph's edges.
The fact that the graph is undirected implies that for $v_1$, $v_2 \in \CO \cup \CP$, $v_1 \in N_o(v_2)$ if and only if $v_2 \in N_o(v_1)$. For example then, if group 1 of $v$ touches group $2$ of $w$, then this means $v \in N_o(w)$. But then $w \in N_o(v)$, so $w$ is in group 2 of $v$ and group 1 of $w$, giving the desired pairing. Analogous arguments prove the remaining 3 cases, which are depicted visually in Fig.~\ref{fig:StabilizersCommute}. Moreover, we show in Fig.~\ref{fig:Stabilizers422} the stabilizers of the graph of the $\llbracket 4, 2, 2 \rrbracket$ code. Note that the ZXCF and graph formalism mod out by local (single-qubit) Cliffords at the end of the encoding process, and thus the stabilizers may differ from the typical presentation by a local Clifford. Here, the typical presentation of the  $\llbracket 4, 2, 2 \rrbracket$ code is given by stabilizers $XXXX, ZZZZ$, while the canonical stabilizers are $XZZZ, ZXXX$. They differ by a Hadamard on the first qubit.

We emphasize that Eqn.~(\ref{eq:inverse}) gives a recipe to construct codes that is essentially independent of ZXCFs (except that inputs and pivots must be matched), even though we will also show that the stabilizers given by Eqn.~(\ref{eq:inverse}) are actually the stabilizers for the ZXCF corresponding to the graph at hand. In principle, one could disregard the entirety of the previous section, \textit{define} Eqn.~(\ref{eq:inverse}) as a map from graphs to stabilizers, and explore their properties. However, it is unclear a priori from this approach how expressive such a class of stabilizers would be and the motivation behind such an assertion would be obscure. The proof of canonicity with the ZXCF guarantees that this recipe is capable of generating all possible stabilizer codes. We next show that such a map generates precisely the stabilizer group of the encoder that the corresponding ZXCF of $G$ represents. Because graphs do not have local Clifford operations associated to them, we can treat $\CS$ as the inverse map from graph to ZXCF, modulo local Cliffords, i.e. up to local equivalence. To do so, we give a fundamental equivalence rule for Paulis on graphs, which arise from equivalence rules in the ZX calculus.

\begin{figure}
    \centering
    \includegraphics[width=0.5\linewidth]{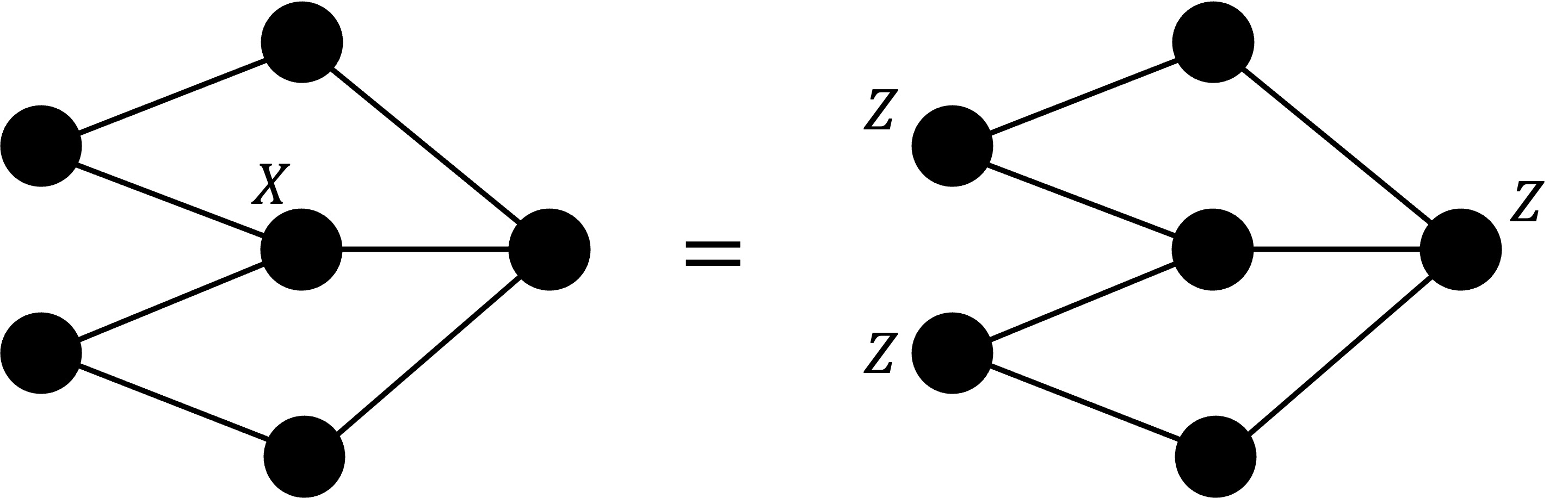}
    \caption{Depiction of the ZX-graph rule on a graph of the $\llbracket 4, 2, 2\rrbracket$ code. A $X$ placed on the central node of the graph is equivalent to a $Z$ on all of its neighbours. The distinctions between inputs, pivots, and non-pivot outputs are intentionally not given, because this rule is agnostic to how we choose to classify the nodes.}
    \label{fig:ZXGraphRules}
\end{figure}

\begin{lemma}[ZX-graph rule] \label{lemma:rules}
Let $G = (V, E)$ be a graph with $V = \CI \sqcup \CO \sqcup \CP$. For any $v \in V$, a $X(v)$ is equivalent to $Z(N(v))$, where $N(v) = \set{w \in V \,|\, (v, w) \in E}$ is the neighbour set of $v$. Note that Paulis placed on input nodes $i \in \CI$ act logically on the corresponding qubit.
\end{lemma}
\begin{proof}
If we endow every node in the graph with a free edge and recover a Hadamard on every internal edge as dictated by the ZXCF rules, then the rules of the ZX calculus show that a $X$ placed on the free edge of a node equals a $Z$ on all of its neighbours, proving the claim.
\end{proof}
As a concrete example, we visualize the ZX-graph rule applied on a graph, which for concreteness is the graph of the $\llbracket 4, 2, 2 \rrbracket$ code with stabilizers $XXXX$ and $ZZZZ$. This illustration is depicted in Fig.~\ref{fig:ZXGraphRules}.

\begin{figure}[ht!]
    \centering
    \includegraphics[width=0.85\linewidth]{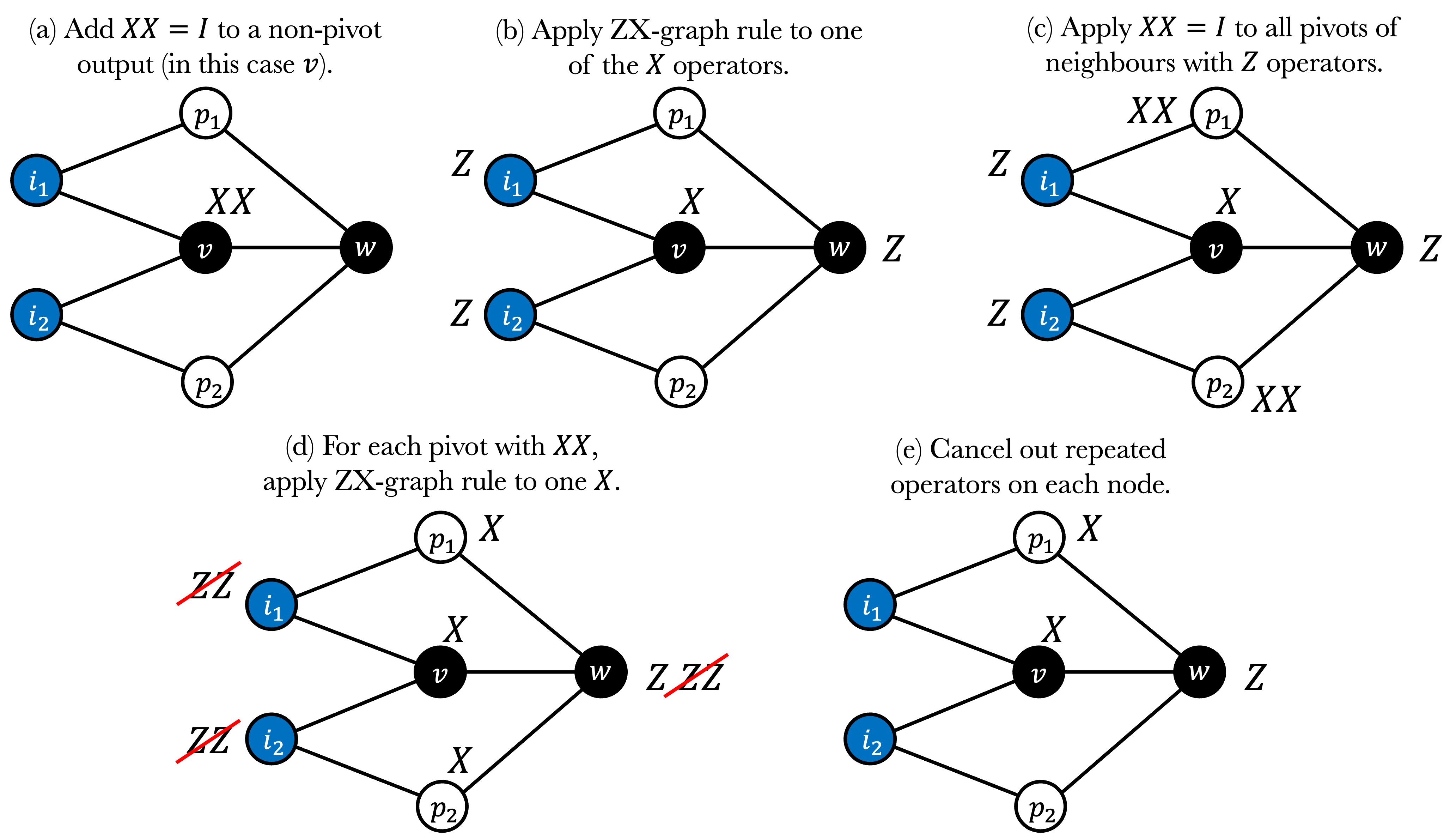}
    \caption{Example derivation of a canonical stabilizer associated with node $v$ in the $\llbracket 4, 2, 2 \rrbracket$ code. Here, $i_1, i_2 \in \CI$ are inputs, $p_1, p_2 \in \CP$ are pivots, and the remaining nodes $v, w$ are non-pivot outputs. We first apply $XX = I$ to $v$. Then, by applying the ZX-graph rule from Lemma~\ref{lemma:rules} and applying $XX = I$ to pivots, we obtain the final stabilizer.}
    \label{fig:GraphRules422}
\end{figure}

The ZX-graph rule in Lemma~\ref{lemma:rules} allow us to \textit{derive} the stabilizers of a graph as given in Eqn.~(\ref{eq:inverse}). For each $v \in \CO$, place $XX = I$ on $v$. Apply the ZX-graph rule on one $X$ to get $Z(N(v))$, where $N(v)$ are the neighbours of $v$. However, some neighbours are inputs, and we can only act directly on physical qubits. We therefore must apply operations on other output nodes to cancel the $Z$'s on $i(v)$ the input neighbours. Observe that for each $i \in i(v)$, placing $XX=I$ on the unique corresponding pivot $p$ for $i$ and applying the ZX-graph rule will cancel out the $Z$ on $i$, but it will also create $Z$'s on $N(p)$. Since a pivot cannot be connected to more than one input, $N(p) = \set{i} \sqcup N_o(p)$, so the extra edges are in $N_o(p)$. Doing so on all pivots yields in total $Z$ operators on $N_o(v)$ and $N_o(p(i(v)))$. Finally, we account for the $X$'s that we did not apply the ZX graph rule to, giving $X(v) Z(N_o(v)) X(p(i(v)) Z(N_o(p(i(v)))))$ which is precisely Eqn.~(\ref{eq:inverse}).
Therefore, we have proven the following theorem.

\begin{theorem}[Inversion]
    Given a graph $G$ with a valid choice of inputs $\CI$ and pivots $\CP$, where a valid choice means that every pivot is connected to exactly one input, the stabilizer tableau associated with the ZXCF equivalent to $G$ is given by Eqn.~(\ref{eq:inverse}). 
\end{theorem}

We remark that if we choose a graph $G$ with no inputs, then the stabilizer tableau of Eqn.~(\ref{eq:inverse}) is exactly that of a \textit{graph state} of $G$, a stabilizer state where each of the $n$ stabilizers are $X$ on a node $v$ and $Z$ on its neighbours.

\begin{figure}
    \centering
    \includegraphics[width=\linewidth]{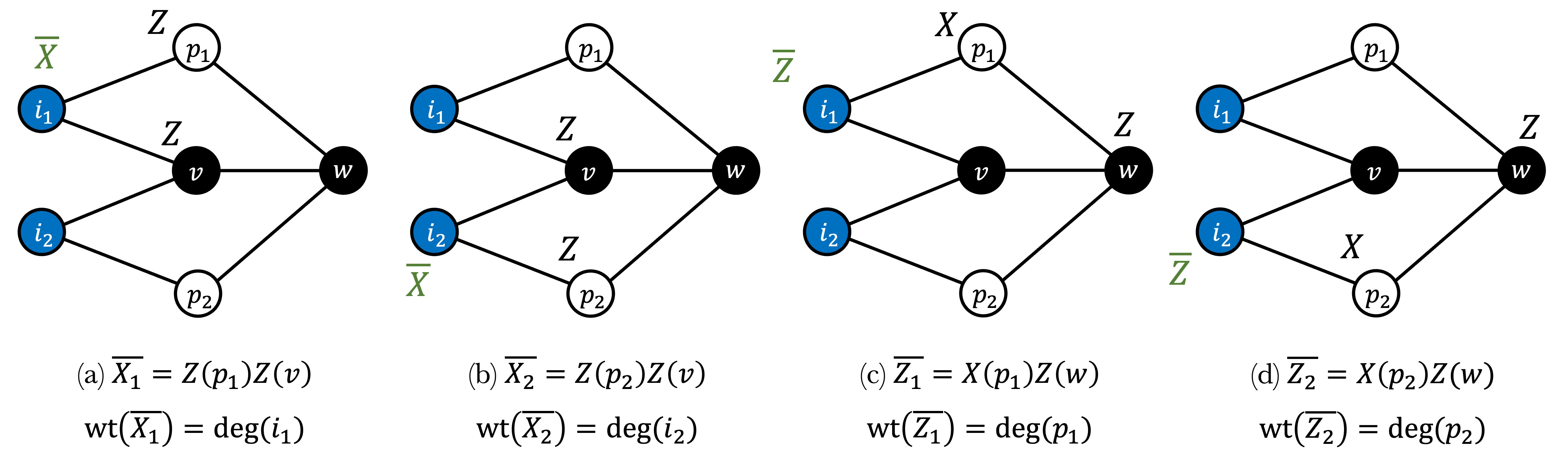}
    \caption{Logical Paulis of the graph of the $\llbracket 4, 2, 2 \rrbracket$ code. $i_1, i_2 \in \CI$ are inputs, $p_1, p_2 \in \CP$ are pivots, and $v, w \in \CO$ are non-pivot outputs. The barred green Pauli is the logical Pauli, and the corresponding physical implementation of the Pauli is shown in unbarred black.}
    \label{fig:LogicalPaulis422}
\end{figure}

Application of the ZX-graph rule on \textit{output} nodes yields the canonical stabilizers; we now show that application of the ZX-graph rule on \textit{input} nodes yields a canonical set of logical Paulis. For each input node $i \in \CI$, construct $\overline{X}_i$ by placing $X(i)$, which is by definition a logical operator on logical qubit $i$. To obtain the physical operator, we apply the ZX-graph rule to get $Z$'s on $N(i) = N_o(i)$. Hence, $\overline{X}_i = Z(N_o(i))$. Similarly, for $i \in \CI$, $\overline{Z}_i = X(p(i)) Z(N_o(p(i)))$, by placing a $X$ on the corresponding pivot $p(i)$ and applying the ZX-graph rule.
A visualization of the logical operators on the graph of the $\llbracket 4, 2, 2 \rrbracket$ code is given in Fig.~\ref{fig:LogicalPaulis422}.

A useful consequence of the structure of the canonical stabilizers is that they reveal when a code is CSS. For our purposes, we say that a code is CSS if there exists a product of single-qubit Cliffords which when conjugated on the stabilizers makes every stabilizer either all $X$ or all $Z$.

\begin{figure}
    \centering
    \includegraphics[width=0.85\linewidth]{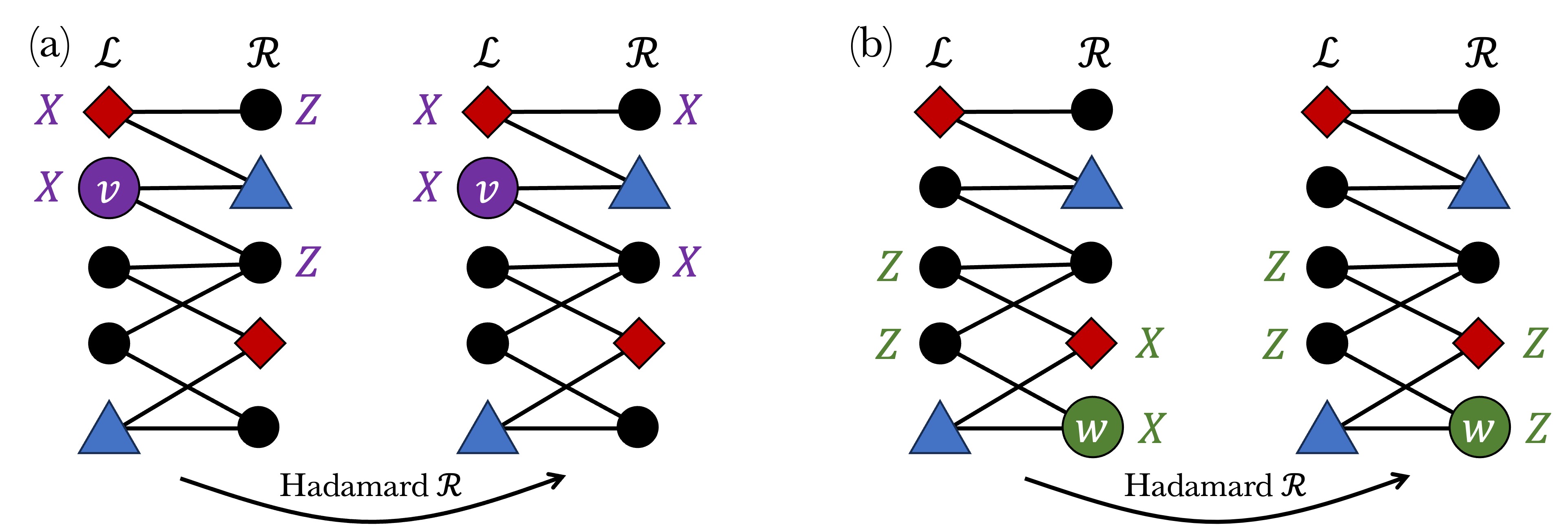}
    \caption{Visualization that bipartite graphs are CSS. Inputs are blue triangles, pivots are red diamonds, and non-pivot outputs are circles. In (a), a left-node $v$ is highlighted. Its stabilizer has only $X$'s in the left side $\CL$ and $Z$'s on the right side $\CR$. In (b), a right-node $w$ is highlighted, and its stabilizer has only $X$'s in $\CR$ and $Z$'s in $\CL$. By Hadamarding all output nodes in $\CR$, stabilizers associated with left nodes such as $v$ become only $X$'s, and stabilizers associated with right nodes such as $w$ become only $Z$'s.}
    \label{fig:BipartiteCSS}
\end{figure}

\begin{theorem}[Bipartite graphs are CSS codes] \label{thm:bipartite_CSS}
Let $G = (V, E)$ be a graph where $V = \CI \sqcup \CO \sqcup \CP$ and $\CI$, $\CP$, and $\CO$ are respectively the input, pivot, and non-pivot output nodes. If $G$ is bipartite, then the corresponding code given by the canonical stabilizers of Eqn.~(\ref{eq:inverse}) is a CSS code.
\end{theorem}
\begin{proof}
If $G$ is bipartite, then we may define $\CL$ and $\CR$ such that $V = \CL \sqcup \CR$ and there are no edges between $\CL$ and $\CR$. Consider any non-pivot output node in the left $v \in \CL \cap \CO$ Then $N_o(v) \subseteq \CR$, $p(i(v)) \subseteq \CL$, and $N_o(p(i(v)) \subseteq \CR$. Likewise if $v \in \CR \cap \CO$ then $N_o(v) \subseteq \CL$, $p(i(v)) \subseteq \CR$, and $N_o(p(i(v)) \subseteq \CL$. Therefore, applying a Hadamard $H$ on all right output nodes $\CR \cap (\CO \cup \CP)$ turns all canonical stabilizers for $v \in \CL \cap \CO$ to have only $X$'s and all canonical stabilizers for $v \in \CR\cap\CO$ to have only $Z$'s. Fig.~\ref{fig:BipartiteCSS} gives a visual proof of the theorem.
\end{proof}

We remark that while we will not need this stronger result in our work, a converse statement to Theorem~\ref{thm:bipartite_CSS} also holds true.
\begin{fact}[\citet{khesin2024equivalence}]
    A code is CSS if and only if its corresponding ZXCF is bipartite.
\end{fact}

\begin{figure}[ht!]
    \centering
    \includegraphics[width=0.33\linewidth]{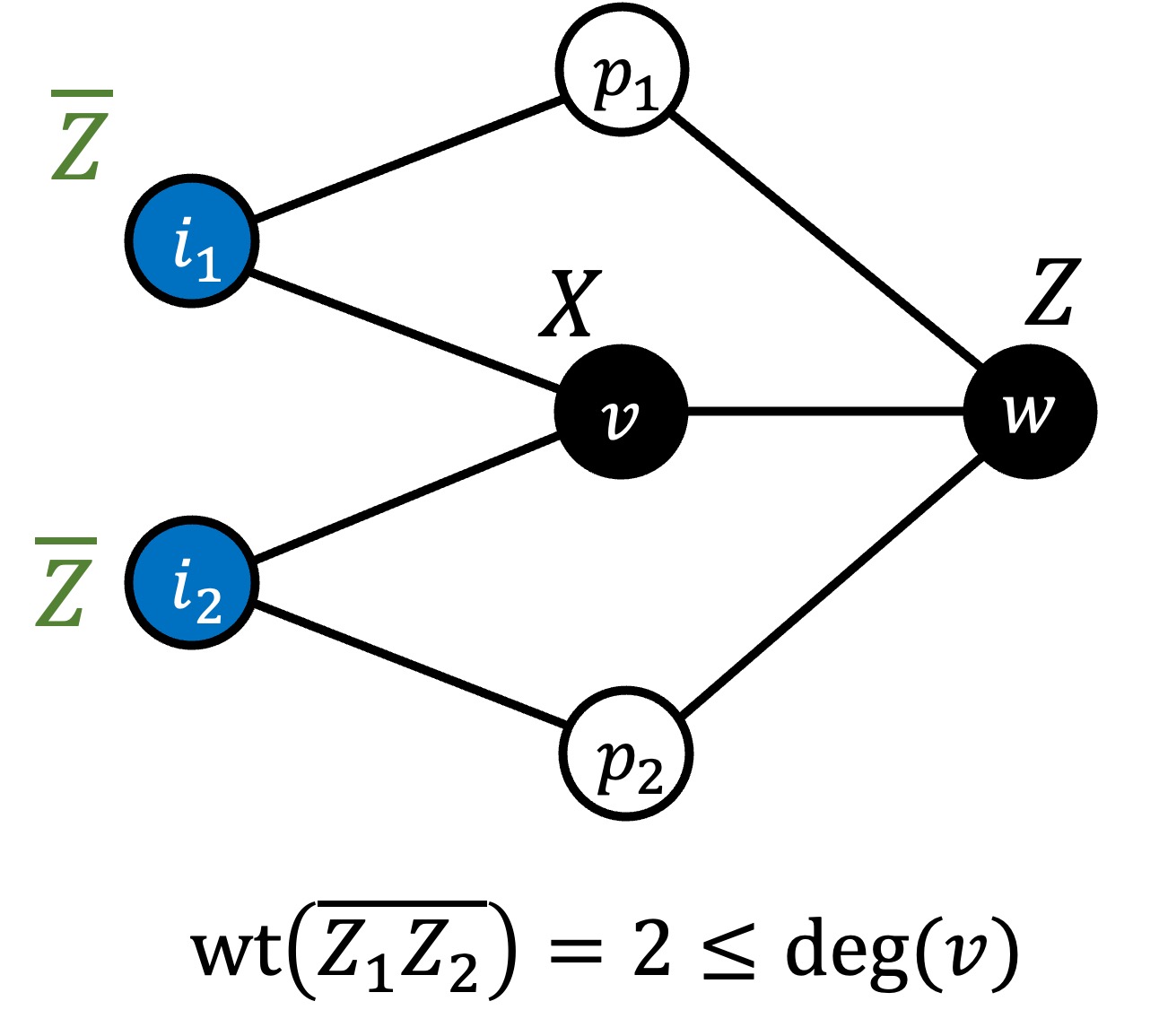}
    \caption{Origin of the distance-degree bound in Proposition~\ref{proposition:degree}, visualized for the graph of the $\llbracket 4, 2, 2 \rrbracket$ code. Here, $i_1, i_2 \in \CI$ are inputs, $p_1, p_2 \in \CP$ are pivots, and $v, w \in \CO$ are non-pivot outputs. By placing a $X$ on an input neighbour $v \in N_o(\CI)$ (where recall $N_o$ is the set of neighbours in $\CO\cup\CP$) and applying the ZX-graph rule from Lemma~\ref{lemma:rules}, we transform it into a mix of logical and physical Paulis on its neighbours. We then remove all the extra physical Paulis on neighbours of $v$ to cancel them out, leaving only the pure logical operator. The whole process requires at most $\deg(v)$ Paulis. This process works over all choices of placing $X$ on $\CI$ or $N_o(\CI)$ and applying the ZX-graph rule, so the degree bounds holds generally across nodes in $\CI \cup N_o(\CI)$.}
    \label{fig:Distance422}
\end{figure}

As a consequence of our canonical logical operator construction, we observe that the degree of $G$ controls the upper bound of the distance as well as the stabilizer weights of $\CS(G)$.
\begin{proposition}[Distance upper bound] \label{proposition:degree}
Let $G$ be a graph with inputs $\CI$, pivots $\CP$, and non-pivot outputs $\CI$. $\CS(G)$ be the tableau of canonical stabilizers as given by Eqn.~(\ref{eq:inverse}).
The distance of the code whose stabilizer tableau is given by $\CS(G)$ is at most the minimum degree of any vertex in $\CI \cup N_o(\CI)$, where $N_o$ is defined in Definition~\ref{def:neighbour_sets}.
\end{proposition}
\begin{proof}
    For each $v\in\CI$, $X(v)$ is a logical operation which is equal to $Z(N_o(v))$.
    The weight of this operation is equal to the degree of $v$.
    Similarly, for each $v\in N_o(\CI)$, the rules of the graph tell us that applying $X(v)$ as well as applying $Z$ to all neighbours of $v$ does not change the circuit effected by the graph.
    Since $v$ neighbours at least one input node, this means that the logical operation formed by the Paulis applied to the inputs is equal to the Paulis applied to the outputs, including $v$.
    This size of this set is equal to $|N_o(v)|+1$, which is at most the degree of $v$, since $v$ neighbours an input.
    The construction of these logical operations proves the desired upper bound on the distance.
    This also shows that the statement of the corollary could be strengthened slightly by replacing the degree of input neighbours by the number of their output neighbours plus one.
\end{proof}
A visualization of the proof of Proposition~\ref{proposition:degree} is given by Fig.~\ref{fig:Distance422}. The below definition is useful for proving structural properties of graphs as codes.

\begin{definition}[$t$-spacing] \label{def:t_spacing}
    Let $G = (V, E)$ be a graph with vertex set $V$ and edge set $E$, and let $V_0 \subseteq V$ be a subset of nodes. We say that $V_0$ is \textit{$t$-spaced} if the distance between every pair of nodes $v_1, v_2 \in V_0$ is at least $t$. By distance between nodes, we mean the minimum path length which begins at $v_1$ and ends at $v_2$.
\end{definition}
In the special case $t = 3$, the subset $V_0$ is known in graph theory as a 2-packing or a strongly independent set.

\begin{proposition} \label{proposition:stab_weight}
    Given a graph $G$, with inputs $\CI$, pivots $\CP$, and non-pivot outputs $\CO$, define \begin{align}
        \d^*_{\CO} = \max_{v \in \CO} \deg(v) ,\quad \d^*_{\CO \CI} = \max_{v \in \CO} |i(v)| ,\quad \d^*_{\CP \CO} = \max_{v \in \CP} |N_o(v)|  
    \end{align}
    where $i$ and $N_o$ are given in Definition~\ref{def:neighbour_sets}. Let $\CS(G)$ be the canonical stabilizers as in Eqn.~(\ref{eq:inverse}).
    Then $\max_{S \in \CS(G)} |S| \leq 1 + \d^*_{\CO} + \d^*_{\CO \CI} \d^*_{\CP \CO}$, where $|S|$ is the weight of a stabilizer $S$. Moreover, if $\CI$ is 3-spaced as in Definition~\ref{def:t_spacing} so that every node $u \in \CO$ is connected to at most one node in $\CI$, then $\max_{S \in \CS(G)} |S| \leq \d^*_{\CO} + \d^*_{\CP \CO}$.
\end{proposition}
\begin{proof}
This result is a direct consequence of Eqn.~(\ref{eq:inverse}).
\end{proof}

In the next section, we will show that another important property, encoding depth, is also bounded above by the degree. We make two remarks regarding the utility of such degree bounds. First, for a general stabilizer code it is unclear as to how one may give a nontrivial upper bound on properties like distance and circuit depth. The degree allows us to algorithmically do so by compiling a stabilizer tableau into a graph and then examining the degree. Second, it is possible, at least in certain cases, to upper bound the degree itself by some function of the code parameters. We show bounds of this type in Sections~\ref{subsec:decoder} and \ref{subsec:randcodes}. When such a bound is possible, it provides nontrivial bounds on code properties in terms of functions of other parameters.

\subsection{\label{subsec:encoding_circuit}Encoding circuit with degree-bounded depth}
A particularly appealing quality of the graph formalism is that it enables us to efficiently and canonically construct encoding circuits which are bounded by a small linear function of the \textit{degree }of the graph. This construction can be taken advantage of in two ways. If we already have a stabilizer code and would like to construct a nice encoding circuit, we can compile the code's tableau into a graph and then apply the procedure below. Note that Theorem~\ref{thm:encoding} assumes a graph presentation, so if the ZXCF has local Cliffords, they are not included in the encoding circuit. They can be appended to the encoding circuit produced by the theorem, at the cost of increasing the depth by 1, since all local Cliffords are single-qubit operations. Alternatively, if we have a graph which by the previously discussed procedure represents a stabilizer code, we can directly write down this encoding circuit and have a guarantee of its depth.

\begin{figure}[ht!]
    \centering
    \includegraphics[width=\linewidth]{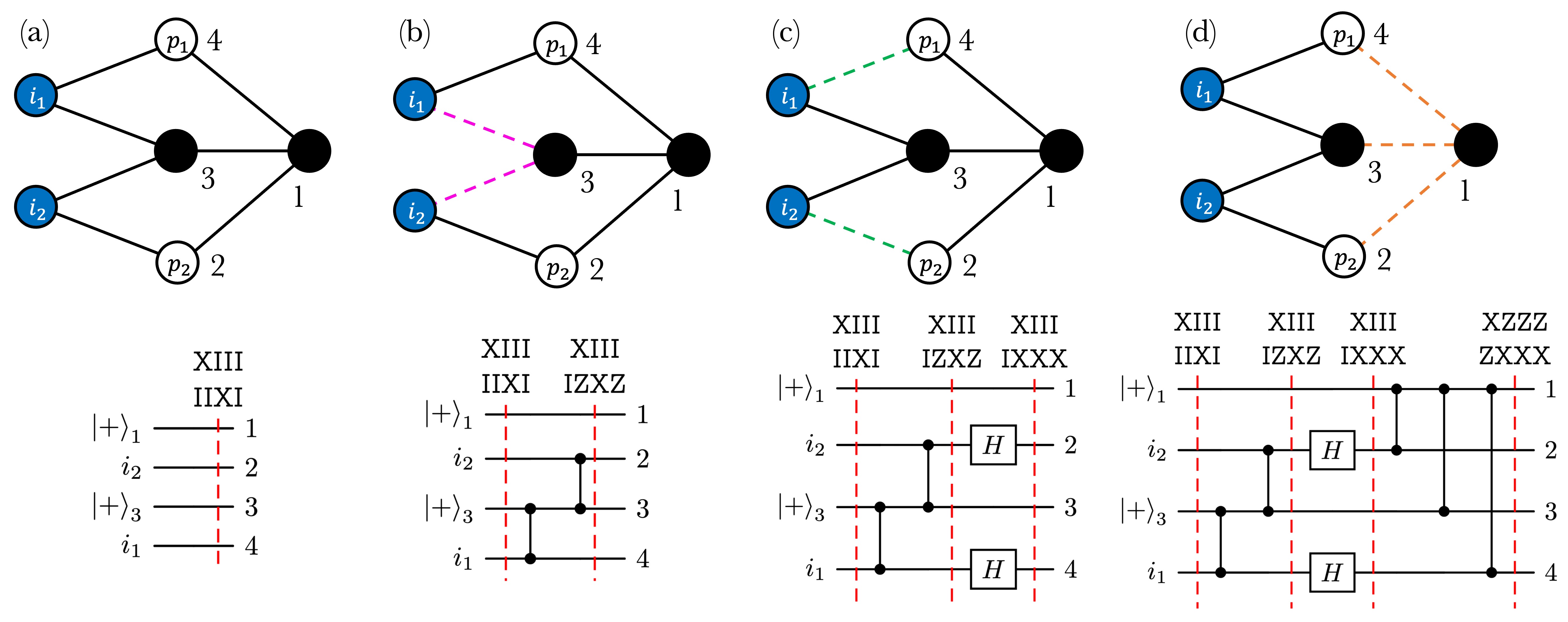}
    \caption{Visualization of the encoding circuit construction from Theorem~\ref{thm:encoding}, depicted for the $\llbracket 4, 2, 2 \rrbracket$ code with stabilizers $XXXX$ and $ZZZZ$. There are four steps in Theorem~\ref{thm:encoding}, which we illustrate (a)-(d) here. At each step, the graph of the code is shown, alongside the current status of the circuit construction. The current stabilizers are also shown at various steps in the circuit. (a) the inputs are labeled $i_1, i_2$ and the outputs are labeled 1, 2, 3, and 4. The two pivots are labeled $p_1$, and $p_2$. We draw 4 wires and label the right-hand side as the outputs. The input side has input wires for the $k = 2$ inputs $i_1$ and $i_2$, and $\ket{+}$ states initialized on the remaining $n-k = 2$ nodes. (b) for every input-output edge (dotted, pink), we add a controlled-$Z$ ($CZ$) gate between the output and the pivot corresponding to the input. (c) for every input (or equivalently, every input-pivot edge---here dotted, green), apply a Hadamard to the corresponding pivot. (d) for every output-output edge (dotted, orange), apply a $CZ$ gate between the corresponding qubits. The final stabilizers are $XZZZ$ and $ZXXX$ which differ from $XXXX$ and $ZZZZ$ by a local Clifford, namely a Hadamard $H_1$. The degree is $3$, so the depth bound stated by Theorem~\ref{thm:encoding} is $9$ which the final circuit visibly satisfies.}
    \label{fig:EncodingCircuit}
\end{figure}

\begin{theorem} \label{thm:encoding}
    Suppose that one may initialize qubits to the $\ket{+}$ state.
    Given a graph $G$ with maximum degree $\mathcal{D}_{\max}$, there exists an efficient algorithm which outputs an encoding circuit $\CC$ for the code with stabilizers $\CS(G)$, with $\CS(G)$ defined by Eqn.~(\ref{eq:inverse}), such that $\operatorname{depth}(\CC) \leq 2 \CD_{\max} + 3$. Here $\operatorname{depth}(\CC)$ is the circuit depth of $\CC$.
\end{theorem}


\begin{proof}
We construct explicitly the transformation from the graph to its encoding circuit\footnote{A related discussion of such a transformation is given in a simultaneous work by one of the authors in \citet{khesin2024equivalence}.}. An informal description is as follows. A graph with $|\CI| = k$ and $|\CO \cup \CP| = n$ requires a circuit taking $k$ input and $n-k$ ancillary qubits to $n$ output qubits. We choose to send the $k$ inputs to the $k$ pivots on the output sides. Edges in $G$ will generally become controlled-$Z$ gates in the encoding circuit, with the exception of edges from $\CI$ to $\CP$, which become a layer of Hadamards. More precisely, construct the circuit $\CC$ as follows.
\begin{enumerate}
    \item[(1) ] \textit{Initialization.} Create $n$ wires, one for each node in $\CO \cup \CP$. Label the output side of the wires by the node. For each wire associated to a pivot $v$, label the input side of the wire by the input corresponding to $v$. The remaining $n-k$ wires unlabeled on the input side are the ancillary qubits. Initialize these unlabeled wires to $\ket{+}$.
    \item[(2) ] \textit{Input-output edges.} For each input $u \in \CI$ and $v \in \CO$, if $v \in o(u)$ then add a $CZ_{u, v}$ to the circuit. The order does not matter since all $CZ$ gates commute.
    \item[(3) ] \textit{Input-pivot edges}. For each input $u \in \CI$ apply $H_u$. These will correspond to the edges between each input and its corresponding pivot.
    \item[(4) ] \textit{Output-output edges.} For each $u \in \CO \cup \CP$ and $v \in N_o(u)$, apply $CZ_{u,v}$.
\end{enumerate}
For concreteness, we give an example of these four steps collectively constructing the encoding circuit for the $\llbracket 4, 2, 2 \rrbracket$ code in Fig.~\ref{fig:EncodingCircuit}.
There are two claims to show: (a) this construction correctly yields an encoding circuit for the code described by $\CS(G)$, and (b) $\operatorname{depth}(\CC) \leq 2 \CD_{\max} + 3$. The first claim is essentially a direct consequence of the rules of the ZX calculus and the form of the ZXCF. Recall that $G$ can be mapped to a ZXCF $\widetilde{G}$ by Hadamarding all edges, endowing each node with a free edge. Initially, work qubits ZX diagrams are given by $Z$ nodes, which are the $\ket{+}$ states used in (1). As we turn the ZX diagram into a circuit, the edges which connect an input node to its corresponding node on the output side of the circuit pick up the gates between them. Since we have chosen the inputs to map to their corresponding pivot on the output side of the circuit, and the only gate between them is $H$, this accounts for (3) above. The remaining edges connect distinct qubits (rather than the same qubit from the input to output side), and such edges are precisely given by $CZ$ gates. This map accounts for (2) and (4). Note that the gates are ordered in this particular way because of the causal structure of the ZX diagram. First, input edges must propagate information to the output edges, so the associated input wires must have their $CZ$ gates to establish their interaction with the ancilla wires. Next, the circuit completes the interactions from input to output via the input-pivot Hadamards. Finally, the output edges interact, so the output-output $CZ$ gates are placed on the ancillary wires. This completes the proof of the first claim, correctness.

To the second claim, we observe that since all $CZ$ gates commute, the depth is given by the minimum number of groupings of $CZ$ gates which act on disjoint sets of qubits. Such a problem reduces to finding the optimal \textit{edge colouring} of $G$, i.e. the minimum number of colours needed to colour every edge, such that no two edges which share a node have the same colour. Although finding the optimal edge colouring is known to be \textbf{NP}-complete, Vizing's edge colouring theorem guarantees that the minimal colouring is at most $\d^* + 1$, and moreover there exists an efficient algorithm which finds such a colouring~\cite{vizing,misra1992constructive}. In sum, (2) and (4) each has depth at most $\CD_{\max} + 1$, and (3) has depth 1, so that the total depth is at most $2\CD_{\max} + 3$ as claimed.
\end{proof}

Although encoding in practice is often performed by a method to prepare the logical zero state, e.g. repeatedly measuring the stabilizers until they are all $+1$, Theorem~\ref{thm:encoding} shows that for codes that have sufficiently small degree, it may be possible to have a much stronger guarantee, namely to efficiently encode any state directly via an encoding circuit. Section~\ref{subsec:smallcodes} gives an example of such an encoding circuit for a code we have constructed, and discusses how the depth may actually be substantially smaller in practice by solving small cases of minimal edge colouring. Moreover, in Section~\ref{subsec:decoder}, we will show that the degree, and thus the encoding circuit depth, can be bounded above by code parameters for certain families of codes. This result will thus directly upper bound the degree of the encoding circuit by code parameters.

\subsection{\label{subsec:gates}Logical non-Pauli gates on graph codes}

Earlier, we showed that the logical Pauli operations can be extracted immediately from the graph representation. Specifically, $\overline{X}_v$ and $\overline{Z}_v$ for $v \in \CI$ are implemented respectively by $Z(N_o(v))$ and $X(p(v)) Z(N_o(p(v)))$. We derived these logical Paulis by placing a Pauli on the free edge of a particular input node, and then applying ZX equivalence rules to push the Pauli through the node into the output.

For a completely general logical operation $\overline{U}$ and a code with graph $G$, the only algorithm to physically implement $\overline{U}$ is to unencode, apply $U$, and then re-encode. Such a technique is typically undesirable, since the encoding circuit is necessarily high-depth to achieve a good distance, which implies that the logical operation's implementation strongly propagates errors and therefore resists fault tolerance.
Nonetheless, if the encoding circuit has a depth which is sufficiently small, it is conceivable that in some cases this generic procedure may be of use. 
Moreover, we already have a generically optimized encoding circuit from Section~\ref{subsec:encoding_circuit}, which in turn implies the following general result.

\begin{theorem}[Generic logical operation] \label{thm:generic_logical_gate}
    Let $G$ be a graph with maximum degree $\CD_{\max}$, and let $U$ be a unitary which can be implemented in depth $\ell_U$. Then there is an efficient algorithm which constructs a circuit $\CC_U$ that logically implements $U$ on the code corresponding to $G$, such that $\operatorname{depth}(\CC_U) \leq 4 \CD_{\max} + 6 + \ell_U$.
\end{theorem}

Conceivably, if a depth of $\CD_{\max}$ is not immediately prohibitive, an additional factor of $4$ significantly reduces the likelihood of practical utility. However, we will show that many important gates, including diagonal Clifford, non-diagonal Clifford, and diagonal non-Clifford gates, can be implemented generically with a reduction on the constant factor. We remark that in the below results we do not claim that the discussed circuits are necessarily the optimal way to implement a particular logical gate. Rather, these results serve as demonstrations that graph techniques enable us to achieve nontrivial improvements in circuit depth relative to the trivial unencode-apply-encode strategy.

\begin{theorem}
    For any graph $G$ with maximum degree $\CD_{\max}$, every diagonal gate $U$ which can be implemented in depth $\ell_U$ can be implemented logically on the code corresponding to $G$ with depth at most $2\CD_{\max} + 4 + \ell_U$.
\end{theorem}
\begin{proof}
    The circuit is still simply the unencode-apply-reencode operation, but with cancellations due to the diagonal structure. Let $E$ be the encoding circuit from Theorem~\ref{thm:encoding}. The circuit is of depth at most $2\CD_{\max} + 3$ because it is of the form $CZ(V) H(V) CZ(V)$, i.e. a layer of $CZ$'s acting on all the nodes $V$ in the graph, a layer of $H$, then another layer of $CZ$. These respectively contribute at most $\CD_{\max} + 1$, $1$, and $\CD_{\max} + 1$ to the encoding circuit depth. The logical operation is given by $CZ(V) H(V) CZ(V)\cdot U\cdot CZ(V) H(V) CZ(V) = CZ(V) H(V) \cdot U\cdot H(V) CZ(V)$ since $U$ commutes with all $CZ$ operations. This circuit has a reduced depth of $(\CD_{\max} + 1) + (1) + (\ell_U) + (1) + (\CD_{\max} + 1) = 2\CD_{\max} + 4+\ell_U$.
\end{proof}
As a corollary, the $CZ$, $S$, and $T$ gates can all be generically applied with a reduction by a factor of $2$ from Theorem~\ref{thm:generic_logical_gate}. 

For a certain non-diagonal Clifford gate, the $\sqrt{X}$ gate, we are able to even further reduce the constant to just 1. This improvement relies on a specific ZX equivalence relation involving a purely graph transformation known as a \textit{local complementation about a vertex} (LCV). The LCV is defined for any graph and has been presented under various guises in the literature~\cite{hu2022improved,backens2014zx,van2020zx}.
We have, in fact, already used the LCV in Appendix~\ref{app:sec:compiler} to enforce the Clifford rule of the ZXCF.
Below, we fix the vertex set $V$ and consider different graphs which can be made with the nodes in $V$.
Given a graph $G = (V, E)$, let $N(u) = \set{v \in V \,:\, (u, v) \in E}$ be the neighbours of $u \in V$ in $G$.
We can assume throughout this paper that $G$ does not permit self-edges; that is, $\forall u \in V,\, (u, u) \notin E$. 

\begin{definition}
Let $V$ be fixed and $G=(V,E)$ be a graph. A \textit{local complementation about a vertex} $v\in V$ is a graph transformation $\text{LCV} \,:\, \mathcal{G} \times V \to \mathcal{G}$ given by $(V, E) \mapsto (V, E \,\D\, K(N(v)))$, where $K(S)$ is the edge set of a complete graph formed by nodes in $S \subseteq V$.
\end{definition}
Intuitively, a LC about a vertex $v$ can be described as follows. Consider the edges of the complete graph built from $N(v)$, and then toggle all these edges in $v$. That is, if $e \in E \cap K(N(v))$, then remove $e$ from $E$, and if $e \in K(N(v)) \setminus E$, add $e$ to $E$.

The LCV builds a bridge between unitary operations on graph codes and inherent graph transformations. Suppose we have a graph $G = (\CI \cup \CP \cup \CO, E)$ and we wish to apply a logical unitary $\overline{U}$; that is, $\overline{U}_v$ for $v \in \CI$. Generically, if we attempt to use the ZX equivalence rules to push $\overline{U}_v$ to the output, the rules will add more local operators and change edges, resulting in $\overline{U}$ applied to the output $\CO$ of a \textit{different graph}, which corresponds to a different code. To make this distinction clearer, let $U_v(G)$ be the operator $U$ applied to $v \in V$ on $G$. If $v \in \CI$, then $U_v(G)$ is a logical operator; otherwise, $U_v(G)$ is a physical operator. For brevity, if $U = U^{(1)} \dots U^{(m)}$, we denote $U_{v}(G) = U^{(1)} \dots U^{(m)}_v(G)$. We also denote $U_{v_1}(G) \otimes \cdots \otimes U_{v_m}(G) = U_{\set{v_1, \dots, v_m}}(G)$.

\begin{figure}
    \centering
    \includegraphics[width=0.25\linewidth]{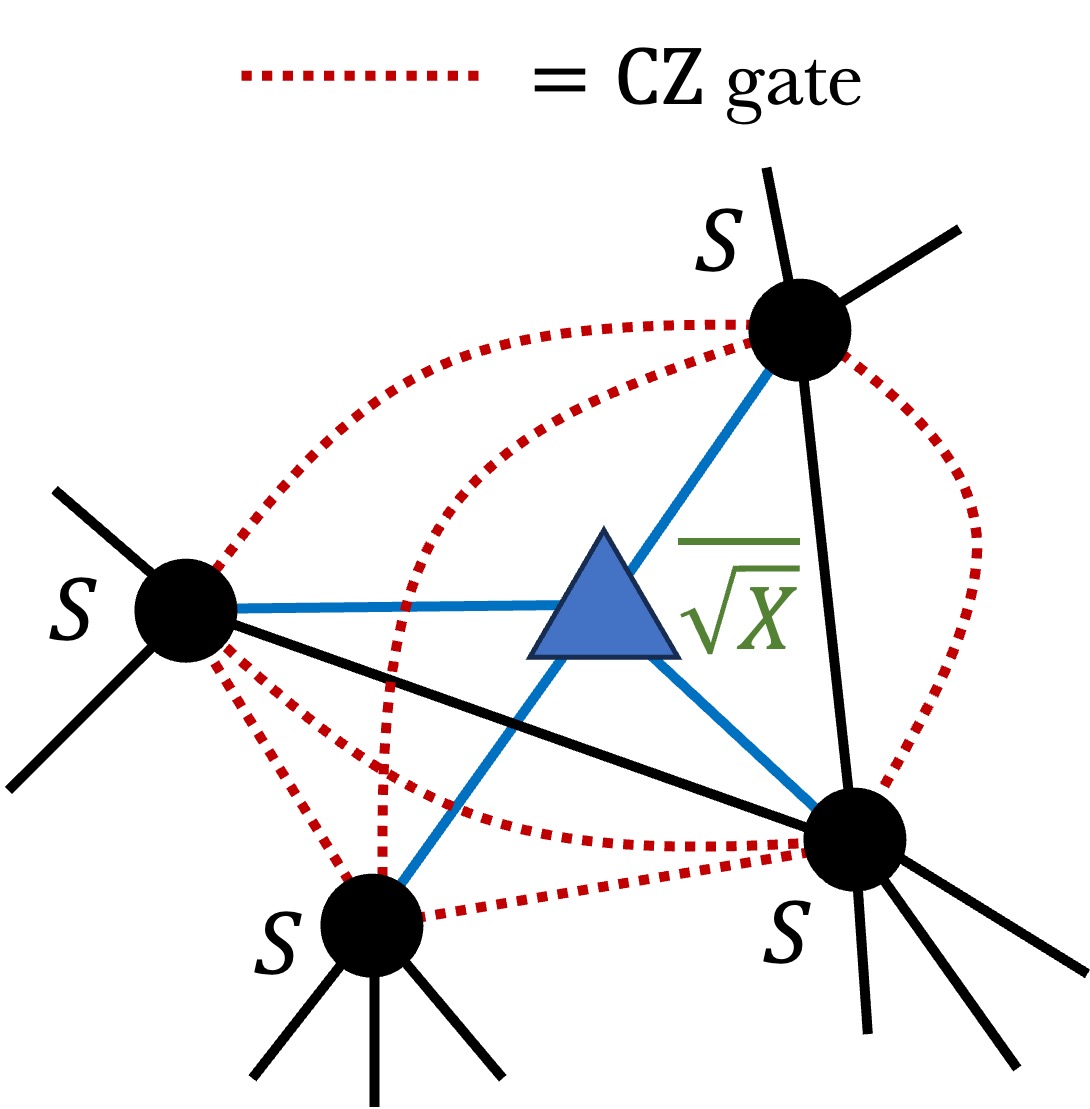}
    \caption{Physical implementation of a logical $\sqrt{X}$ gate in the graph picture. The blue triangle is an input node (corresponding to a logical qubit) on which we wish to apply $\sqrt{X}$. This is accomplished by applying $S$ gates on every neighbour, which are all outputs since input nodes are never connected. Furthermore, controlled-$Z$ gates are applied between every pair of neighbours, depicted as dotted red lines. The implementation of this gate does not depend on how the neighbours of the input are connected to each other. The transformation of edges on the graph is a local complementation about the input vertex.}
    \label{fig:LogicalSqrtX}
\end{figure}

The specific connection between the LCV and the $\sqrt{X} = HSH$ gate is given by the following standard ZX equivalence relation, whose proof can be found in Refs.~\cite{hu2022improved,backens2014zx,van2020zx}. 
\begin{claim}\label{clm:logical}
Let $G = (V, E)$ be a graph. Then $\sqrt{X}_{u}(G) = HSH_{u}(G) =S_{N(u)}(\text{LCV}(G,u))$. Here $H$ is the Hadamard gate and $S = |0\rangle\langle 0| + i |1\rangle\langle 1|$ is the phase gate.
\end{claim}
Therefore, in order to apply a logical $\sqrt{X}$ on some $u \in \CI$, we apply the LCV about $u$, and then apply a $S$ gate on all nodes in $N(u)$. Since $N(u) \subseteq \CO \cup \CP$, all the $S$ gates are physical. Since edges are controlled-$Z$ ($CZ$) gates according to the formalism of the ZX calculus, an equivalent statement of the $\sqrt{X}$ gate implementation on an input $u \in \CI$ is to apply $S$ gates on every neighbour of $u$, as well as $CZ$ gates between every pair of neighbours of $u$. This implementation is depicted in Fig.~\ref{fig:LogicalSqrtX}.

\begin{lemma} \label{lemma:lc-bound}
    Given a graph $G$ representing a code, the graph transformation $\text{LCV}(G, u)$ for $u \in \CI$ can be implemented with a quantum circuit of depth at most $\deg(u)$. There is an efficient algorithm outputting this circuit. 
\end{lemma}
\begin{proof}
    $\text{LCV}(G, u)$ is implemented by $CZ_{vw}$ for all $(v, w) \in K(N(u))$. 
    Since $|N(u)| = \deg(u)$, each node in $K(N(u))$ has degree $\deg(u) - 1$. By Vizing's theorem, there is an efficient algorithm to arrange the $CZ$ gates to have depth at most $(\deg(u) - 1) + 1 = \deg(u)$.
\end{proof}

\begin{theorem}
    Let $G$ be a graph. There is an efficient algorithm which finds a circuit acting on the encoded space of the code given by $G$ that implements the logical $\sqrt{X} = H S H$ gate on an input node $u$ with depth at most $\deg(u) + 1$. Here $H$ is the Hadamard gate and $S = |0\rangle\langle 0| + i |1\rangle\langle 1|$ is the phase gate.
\end{theorem}
The proof is a direct consequence of Claim~\ref{clm:logical} and Lemma~\ref{lemma:lc-bound}.

In general, we do not see a way to completely reduce the implementation depth of an arbitrary Clifford circuit because the Hadamard gate does not appear amenable to further optimization. To have a complete set of Clifford operations which can be implemented in depth strictly better than that of Theorem~\ref{thm:generic_logical_gate}, one solution is to start with a self-dual CSS code, which admits a transversal Hadamard.
In this case, the local complementation simplifications that push a $H$ gate through to the output will cancel out in just the precise way as to produce $H$'s on all output nodes while preserving the graph structure.
$\set{H, \sqrt{X}, CZ}$ is a generating set of the Clifford group, and thus the above results imply that any self-dual CSS code implements logical Cliffords at a depth that is reduced by at least a factor of 2 from the generic procedure of Theorem~\ref{thm:generic_logical_gate}. 

\subsection{\label{subsec:game}Game unification of stabilizer coding algorithms}
Thus far, given a graph, we have described its corresponding stabilizers, given a set of logical operators, and provided a simple upper bound on the distance. To complete our fully graph formalism of stabilizer codes, we will show that important code algorithms---distance calculation or approximation, finding minimum weight stabilizer generators, and decoding---can all be expressed in a unified graph manner, namely by a single class of one-player games on the graph.

\begin{definition}[Quantum lights out] \label{def:QLO}
An instance of a quantum lights out (QLO) game \footnote{We adopt the name quantum lights out because the game appears to be a generalization of a classical game known as lights out, in which a player aims to turn all lights off on a grid with some lights initially on, using moves such that flipping a switch at a given cell toggles all adjacent cell's lights but not their own. QLO generalizes this notion onto a general graph, and adds more flexibility on how lights may be toggled.} on a graph $G$ with inputs $\CI$, pivots $\CP$, and non-pivot outputs $\CO$, is described as follows. Every node is endowed with both a light and a switch, both of which are binary (i.e. on of off), which are initially off. Flipping a switch at node $v$ (a) permanently destroys the light (i.e. the light turns off forever and cannot be turned back on) at $v$ if $v \notin \CI$ and (b) toggles all intact (i.e. non-destroyed) lights on the neighbours of $v$. Let Alice be the player. The graph begins with some initial configuration of lights. The game consists of two rounds: \begin{enumerate}
    \item[(1) ] Depending on the instance, Alice may be allowed to flip switches in the input.
    \item[(2) ] Alice chooses a set of $n$ nodes $v_1, \dots, v_n \in \CP \cup \CO$, subject to some some instance-dependent constraints. For each node $v_i$, in order, Alice chooses whether to toggle the switch at $v_i$ or only to destroy the light on $v_i$. The game ends when the lights reach an instance-specific desired final configuration, usually with all non-input lights off or destroyed.
\end{enumerate}
The specific instance of the QLO game specifies the initial light configuration, whether round (1) exists, the constraints in round (2), and the desired final configuration, e.g. all lights are turned off. Regardless of instance, Alice must make at least one move in round (2). We say that Alice wins in $n$ moves if she makes $n$ choices that lead to the desired final configuration of the game.
\end{definition}
At its core, QLO is a simple way to reformulate the ZX-graph rule from Lemma~\ref{lemma:rules}, alongside some technical details. $X$ operators are equivalent to $Z$ on all neighbours, and in QLO we use a light to indicate the presence of a $Z$ operator (so applying $Z$ toggles a light, and the switch which toggles neighbouring lights is simply a Pauli $X$ operator with the ZX-graph rule. We use this observation repeatedly below.

\begin{figure}[ht!]
    \centering
    \includegraphics[width=0.8\linewidth]{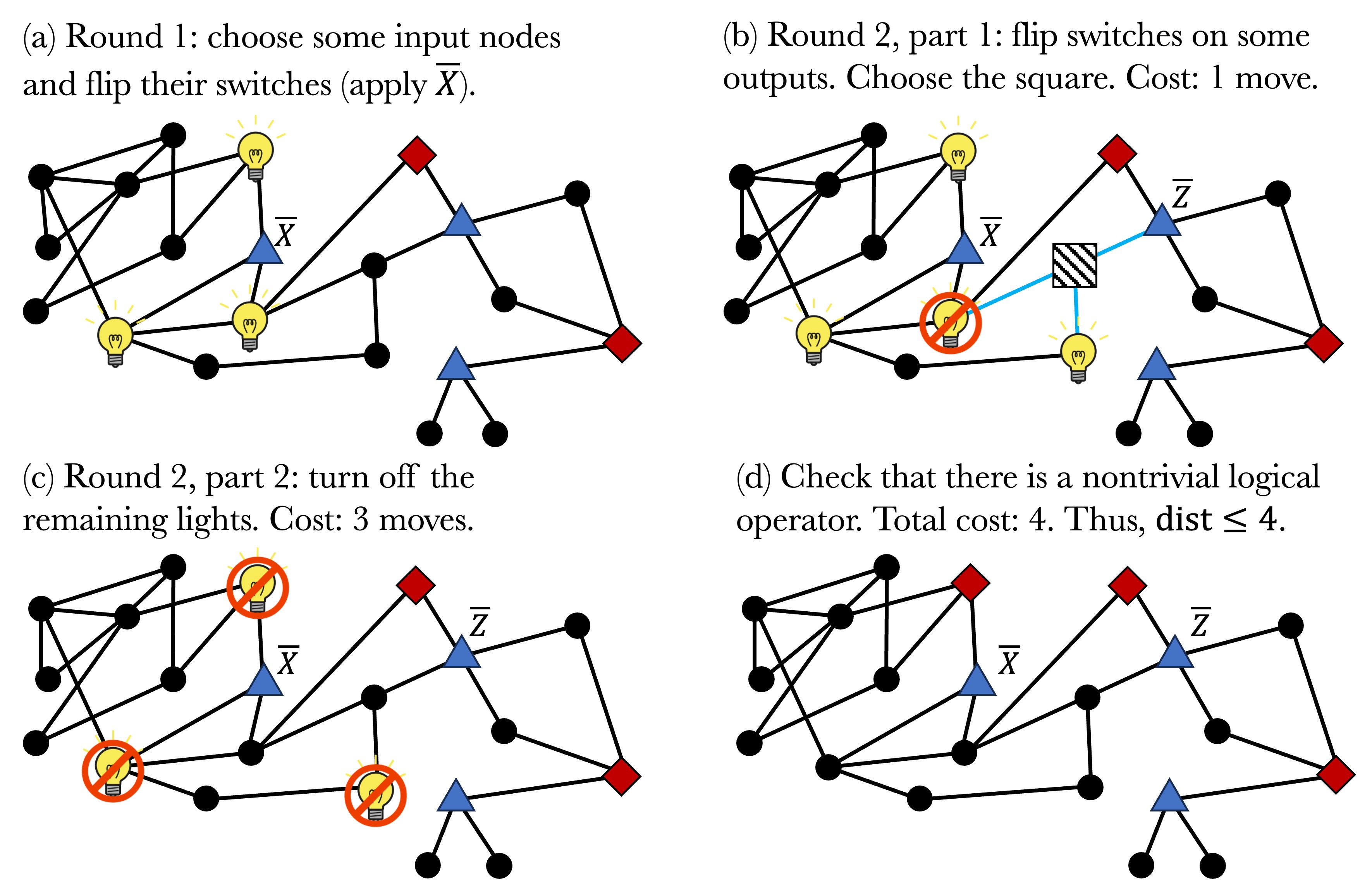}
    \caption{Depiction of distance calculation as a quantum lights out (QLO) game. Here inputs are blue triangles, pivots red diamonds, and non-pivot output nodes black circles. In (a) round 1, Alice may choose to flip switches on inputs, which turn lights on their neighbours. Switch flipping corresponds to applying $X$ operators by the ZX-graph rules of Lemma~\ref{lemma:rules}. In (b) round 2, Alice flips output switches and then turns the remaining lights off directly. The total cost is the number of moves she makes in round (2), and the distance is at most the number of moves. In general, the distance is the minimum cost over all choices of moves, as shown in Theorem~\ref{thm:distanceQLO}.}
    \label{fig:QLO_Distance}
\end{figure}

By correctly choosing the instance of the game, we can represent many important stabilizer coding algorithms as strategies for QLO games. We begin by showing that finding the distance of a code is QLO. Note that the distance is invariant under unitaries on inputs, and hence every equivalent Clifford encoder has the same distance.
\begin{theorem}[Distance is QLO]\label{thm:distanceQLO}
    Given a graph $G$ with inputs $\CI$, pivots $\CP$, and non-pivot outputs $\CO$, denote the distance of the code represented by $d(G)$. $d(G)$ is calculated by the following QLO instance. Initially, all lights are off. In round (1), Alice may flip any number $s$ of switches on input nodes $\CI$. In round (2), Alice has no constraints and may flip any non-input switch or destroy any non-input light. The final configuration consists of all non-input lights off or destroyed, and either at least one input light is on, or $s \geq 1$. Under this instance, \begin{align}
        d(G) = \min_{\CA} m_{\text{D}}(\CA, G) ,
    \end{align}
    where $\CA$ is Alice's strategy and $m_{\text{D}}(\CA, G)$ is the number of moves she made in round 2. As a consequence, every distance approximation algorithm on $G$ to multiplicative error $\e$ is a $\e$-approximately optimal strategy in the distance QLO game instance. Fig.~\ref{fig:QLO_Distance} illustrates distance calculation as a QLO game.
\end{theorem}
\begin{proof}
    As a consequence of the ZX graph rules from Lemma~\ref{lemma:rules}, each of the switches flipped in the $i$th input corresponds to $\overline{X}_i$, i.e. logical $X$ on logical qubit $i$. If the game allowed for flipping a single light at a node, this would correspond to applying a $Z$ on the node. There is no point to applying a $Z$ operator multiple times to the same output node, as they will cancel (and we can order them so that no phase occurs). Hence, instead we simplify by destroying the light altogether as a reminder that re-toggling this light does not increase an operator's weight. Next, toggling lights on all neighbours of a node $v$ correspond to applying $X(v)$. There is no additional Pauli weight if we apply both $X(v)$ and $Z(v)$, so for convenience we also destroy the light on $v$ when we flip the switch. In sum, destroying lights and flipping switches correspond to physical $Z$ and $X$ operators. Lights turned on at inputs after all lights on outputs have been extinguished are logical $Z$ operators. Consequently, the entire distance QLO game consists of applying some logical $X$ operators and translating them into physical operators and logical $Z$ operators. In other words, the physical operator represented by the $n$ moves apply the logical product of the initial logical $X$'s and the final logical $Z$'s, which is a nontrivial logical operator if at least one of the two is not vacuous, i.e. if at least one input light is left on or $s \geq 1$. Since each move increases the physical operator's weight by exactly 1, the minimum number of moves is precisely the distance.
\end{proof}
Since finding the distance of a stabilizer code is \textbf{NP}-complete~\cite{kapshikar2023hardness}, we may immediately conclude that optimally playing QLO with the setup as above is also \textbf{NP}-complete. 

\begin{figure}[ht!]
    \centering
    \includegraphics[width=0.8\linewidth]{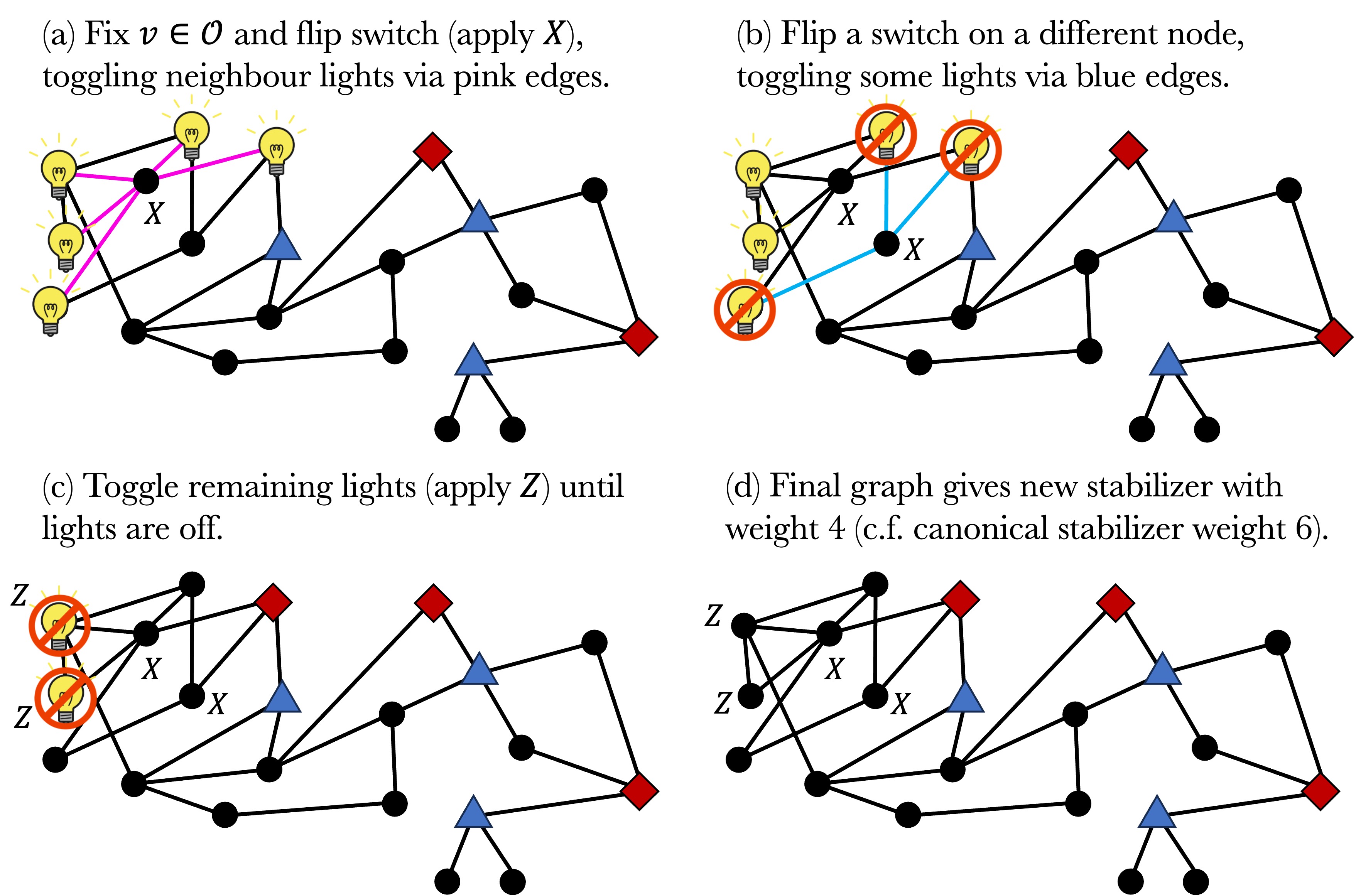}
    \caption{Depiction of minimum weight generator selection as a quantum lights out (QLO) game. Inputs are blue triangles, pivots red diamonds, and non-pivot outputs black circles. Flipping a switch corresponds to a Pauli $X$ and toggling a light corresponds to a Pauli $Z$. In (a) we fix a non-pivot output node $v \in \CO$ whose corresponding stabilizer we wish to find a minimum weight version of. We flip its switch to begin the game, and then in (b) we choose other switches to flip. We then toggle remaining lights in (c) so that all lights are off. The total cost if the number of moves made in the entire process, which in this case is 4. Note that the canonical stabilizers are given by one unique strategy---flip the switch on $v$ then directly turn off all the lights one by one---and in this case the depicted strategy has lower weight than the canonical strategy.}
    \label{fig:QLO_MWGS}
\end{figure}

\begin{theorem}[Minimum weight generator selection is QLO] \label{thm:MWGSQLO}
    Let $\set{S_i}$ be the canonical stabilizer generators of $G$ as given by Eqn.~(\ref{eq:inverse}). Without loss of generality, fix $S_1$ the first stabilizer. Then the minimum-weight stabilizer $S_1'$ which can replace $S_1$ in the tableau, i.e. remains independent of $S_2,\dots, S_{n-k}$, is given by a QLO game instance. Initially, all lights are off. There is no round (1). In round (2), Alice may freely flip switches on any node except $v_1$, the switch in $\CO$ corresponding to $S_1$. She must flip the switch at $v_1$ \textit{exactly once}. She may also destroy any lights in $\CO \cup \CP$. The final configuration consists of all lights destroyed or off. Under this game instance, \begin{align}
        |S_1'| = \min_{\CA} m_{\text{WR}}(\CA, G) ,
    \end{align}
    where $\CA$ is Alice's strategy and $m_{\text{WR}}(\CA, G)$ is the number of moves she made in round 2.
    Thus the minimum weight is equal to the minimum number of moves to end the game.
\end{theorem}

\begin{proof}
    Initially, we apply $X(v_1)$. The canonical stabilizer represents the trivial strategy of flipping pivot switches to remove lights from inputs, and then destroying all remaining lights. However, any sequence of moves that turns all lights off, i.e. effectively applies the identity on the graph, is also a stabilizer. Because we restrict Alice from flipping the switch at $v_1$, the resultant stabilizer is both nontrivial (i.e. not the identity) and independent from all other $S_i$, since $S_1'$ is still the only stabilizer at $X(v_1)$. Hence, the minimum number of moves, plus one for $X(v_1)$, represents the new weight as claimed.
\end{proof}
By examining the proof, we also observe that the act of writing down the canonical stabilizers and the canonical logical operators are also instances of substantially simpler QLO instances. Fig.~\ref{fig:QLO_MWGS} depicts an example strategy to find a stabilizer generator on a node $v$ for a generic graph, which beats the strategy corresponding to the canonical stabilizer.

While the complexity of finding minimum weight generators does not appear to be completely known, this problem is likely about as challenging as distance calculation or decoding since it is precisely the problem of finding short vectors in stabilizer subspaces.

\begin{figure}[ht!]
    \centering
    \includegraphics[width=0.8\linewidth]{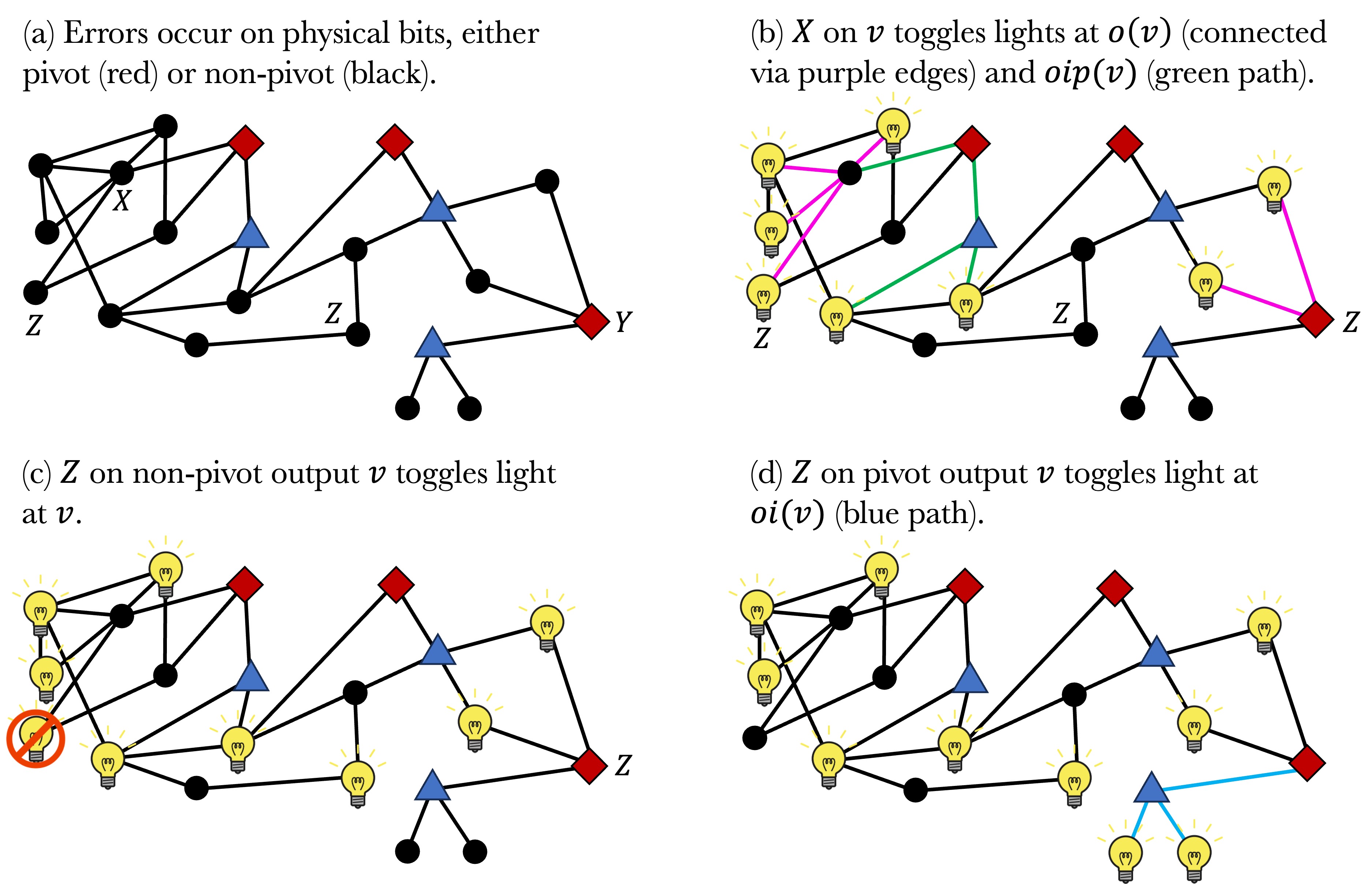}
    \caption{Illustration of decoding a stabilizer code as a quantum lights out (QLO) game. Inputs are blue triangles, pivots are red diamonds, and non-pivot outputs are black circles. For shorthand, we let $oip(v) := o(i(p(v)))$ and $oi(v) := o(i(v))$, where $o(v), i(v)$, and $p(v)$ are as in Definition~\ref{def:neighbour_sets}. In (a), errors occur on physical bits, depicted by some Paulis on output nodes. Here, the lights are toggled not by the ZX-graph rule from Lemma~\ref{lemma:rules}, but by which nodes whose corresponding stabilizer anticommutes with the error (i.e. stabilizers which measure to a violated syndrome). For $X$ on an output node $v \in \CO \cup \CP$, as shown in (b), the affected stabilizers are those corresponding to $o(v)$ the non-pivot output neighbours and $oip(v)$ the non-pivot output neighbours of the inputs corresponding to the pivot neighbours of $v$; their lights thus turn on. Generally, if a node is in $o(v) \cap oip(v)$, it will be toggled twice and not turn on; however, in the depicted graph $o(v)$ and $oip(v)$ are disjoint. A $Z$ on a non-pivot output $v \in \CO$ simply toggles the light on $v$ as shown in (c), while a $Z$ on a pivot $v \in \CP$ toggles the light at $oi(v)$---the non-pivot output neighbours of inputs neighbouring $v$---as shown in (d). The QLO game then consists of finding the smallest number of moves to turn lights off (i.e. fix violated syndromes), and the sequence of moves corresponds to the recovery operation.}
    \label{fig:QLO_Decoding}
\end{figure}

\begin{theorem}[Decoding is QLO]\label{thm:decodingQLO}
    Suppose that after application of a noisy depolarizing (or in general, any mixture of Pauli errors) channel to the codeword, we measure the canonical (or more generally, any known set of) stabilizer generators of a graph $G$ with inputs $\CI$, pivots $\CP$, and non-pivot outputs $\CO$. We thereby obtain syndrome bits $s_1, \dots, s_{n-k} \in \set{\pm 1}$. Then, the recovery map that should be applied is given by a QLO game instance.
    Initially, the light at each vertex $v_i\in\CO$ corresponding to stabilizer $S_i$ is turned on if $s_i = -1$.
    All remaining lights start off.
    In round (1), Alice may flip any number of switches in $\CI$.
    In round (2), Alice has no constraints aside from only being able to act on nodes in $\CO\cup\CP$, as usual.
    The final configuration consists of all non-input lights off or destroyed.
    Under this instance, \begin{align}
        r(\mathbf{s},G) = \min_{\CA} m_{\text{DC}}(\CA,\mathbf{s},G),
    \end{align}
    where $\mathbf{s}$ is the vector of syndromes, $r$ is the number of moves in the optimal recovery map, $\CA$ is Alice's strategy and $m_{\text{DC}}(\CA,\mathbf{s},G)$ is the number of moves she made in round (2). We note that input lights being on does not affect anything in round (2) or in the minimization condition.
    This means that, where convenient, input lights can be completely ignored.
\end{theorem}
\begin{proof}
    Any decoding procedure can be split into steps, where each step involves applying a single-qubit Pauli to an output node. The final recovery map is then given by the product of all the single-qubit Paulis. We begin with a light on at $v_i$ if $s_i = -1$ with no lights on at any pivot. At each step, we will apply a chosen Pauli on a chosen output node. Using casework, we analyze each possible choice of Pauli and node, and which stabilizers for which this operation will flip the syndrome.
\begin{enumerate}
    \item[(1) ] $E = Z(v)$ for $v \in \CO$. $E$ anticommutes with stabilizers which contain $X(v)$. The unique canonical generator which contains $X(v)$ is $S(v)$. Therefore, the syndrome associated with $v$ is flipped.
    \item[(2) ] $E = Z(v)$ for $v \in \CP$. $E$ anticommutes with stabilizers which contain $X(v)$. These are the stabilizers associated with the \textit{non-pivot output neighbours} of the \textit{input} associated with $v$, i.e. $S(w)$ for all $w \in o(i(v))$. Applying a $Z$ to a node $v \in \CP$ negates the syndromes of \textit{all} 
    stabilizers of the vertices $o(i(v))$.
    \item[(3) ] $E = X(v)$ on any $v \in \CO \cup \CP$. $E$ anticommutes with stabilizers which contain $Z(v)$. These are $S(w)$ for $w \in o(v)$ as well as $w \in o(i(p(v)))$ due to the presence of $Z$'s in $S(w)$ on $o(p(i(w)))$ (note that the order of $i$ and $p$ is backwards when we consider stabilizers affected by a node). Applying an $X$ to a node $v \in \CO \cup \CP$ therefore toggles all lights in $o(v) \,\D\, o(i(p(v)))$. The symmetric difference is present because syndromes corresponding to nodes in $o(v) \cap o(i(p(v)))$ are flipped twice and thus identically not flipped at all.
\end{enumerate}
We wish to reframe the action of Paulis on the syndromes in such a way that we can identify all Paulis operators as a choice of either flipping a switch, which toggles lights on neighbours, or toggling a light. (For simplicity, we say toggling instead of destroying a light here, since destroying is merely a simplification so as to avoid double-counting multiple Paulis applied on the same node.) Currently, $Z$ on $v \in \CO$ only affects the syndrome on $v$, so we may think of $Z$ on $v$ as toggling the light on $v$. If we apply $Z$ on $v \in \CP$, we toggle syndromes $oi(v)$. One way to achieve this by using switches and toggles is to flip the switch on $i(v) \in \CI$ and then toggle $v$. By construction, flipping the switch on $i(v)$ toggles lights in the neighbours $N(i(v))$, and $N(i(v)) = \set{v} \cup o(i(v))$ since $i(v)$ is only connected to a single pivot, which is by definition $v$. Hence, if we flip the switch on $i(v)$ and then toggle $v$, we equivalently toggle $oi(v)$ as desired. But flipping switches on nodes in $\CI$ do not cost moves and can be absorbed into the arbitrary allowed action in round (1). Hence, the application of $Z$ on $v \in \CP$ is equivalent to some costless moves in round (1) plus a single toggle of the light in $v$ in round (2).
Similarly, when applying $X$ to $v \in \CO \cup \CP$, we flip the syndromes in $o(v) \,\Delta\, o(i(p(v)))$ and therefore want to toggle exactly those lights using switches and toggles. We may do so by flipping the switch at $v$ as well as all switches in $i(p(v))$; this toggles $o(v) \,\Delta\, o(i(p(v)))$ as desired, and costs only a single move because the switches flipped in $i(p(v))$ are costless and absorbed into round (1). Conversely, any switch flip or light toggle can be associated with the construction of a recovery operator.
Specifically, a switch flip on $\CO \cup \CP$ is associated with a $X$ operator, and a light toggle on $\CO \cup \CP$ is associated with a $Z$ operator. This association differs from the true action only by switch flips on input nodes, which are costless logical operations that do not affect the syndrome and can be absorbed into round (1). Finally, when all lights are off, by our construction, all the syndromes are zero, so the decoding process has finished.
We note that destroying a light at a node after applying the switch at that node represents ambivalence between applying an $X$ or a $Y$, but if Alice is interested in finding the proper recovery map instead of just the map's size, she should leave all lights unbroken.
Then, when she is done toggling switches, she can choose to turn off any lights at vertices whose switches she toggled without having this count as a move in round (2).
We conclude that decoding algorithms are QLO strategies.
\end{proof}
Fig.~\ref{fig:QLO_Decoding} illustrates the origin of decoding as a QLO game. Namely, errors are applied to output nodes, which affect certain stabilizers associated with non-pivot output nodes $v \in \CO$. The affected stabilizers (those giving a violated syndrome) have their lights turned on according to a specific depicted pattern, and the sequence of moves to turn the lights off corresponds to the recovery operation.
We note that construction of the recovery map is very closely related to the calculation of the distance. Indeed, finding the minimum weight recovery operator is also \textbf{NP}-complete. There are, however, more general definitions of optimal recovery that one can consider, which factor in degeneracy classes. In such cases, decoding can become as hard as \textbf{\#P}-complete~\cite{iyer2015hardness}.

It is illustrative to compare the decoding QLO game to a classical analog. For a classical code with a Tanner graph $G = (L, R, E)$, where $L$ and $R$ are respectively the left and right node sets and $E$ is the edge set, we may associate every check node in $R$ with a light and every data node in $L$ with a switch, such that flipping the switch toggles the neighbouring lights. At the start of decoding, a light is on if the check corresponding to that node is violated. We then iteratively choose data bits to flip and update the syndromes. This is precisely equivalent to flipping switches on the data nodes, which toggle the check node lights. When all lights are off, the recovered data must be a codeword. This classical version of lights out on Tanner graphs is thus \textbf{NP}-complete as well. At the same time, classical lights out can be solved efficiently if the graph is a sufficiently good expander, using a simple greedy algorithm~\cite{sipser1996expander}. We later show that if a graph of a stabilizer code satisfies a different but related structural property, then a greedy QLO strategy can also efficiently decode errors.

Had we been considering erasure channels instead of depolarizing channels, the QLO game would look very different.
In the regimes where only a small number of errors has occurred, such as less than half the distance of a given code, then being able to pinpoint the locations of those errors leaves us able to follow a generic recovery algorithm that has no connection to QLO.
For instance, we could replace each erased qubit with the maximally-mixed state, measure all the stabilizers, and search for any recovery operator supported on the erased qubits, all in polynomial time.
It is more interesting to study the erasure channel in the cases where the number of errors approaches the code's threshold and not the distance of the code. We leave this question to future work.

The unified expression of all three algorithms, combined with the many complex techniques to perform optimizations on graphs that have been developed in past decades, suggest that the study of approximating optimal QLO strategies may be a promising pathway for devising coding algorithms.

\section{\label{sec:constructions} Applications of the Graph Formalism}
We proceed to provide constructive evidence in three areas for the utility of a graph representation. First, in Section~\ref{subsec:smallcodes}, we argue that for near-term experimental purposes, the graph formalism provides a simple and flexible prescription for the design of codes with a desired rate $R$ (where $R = k/n$ and $k$ is the number of logical qubits) and distance $d$. Intuitively, this advantage occurs because we can appeal to the geometrical and topological structure of graphs, as well as take inspiration from well-studied graphs, to get close to a distance $d$. We then describe improved results for random codes and construct an efficient graph decoder for a family of stabilizer codes constructed from graphs.

\subsection{\label{subsec:smallcodes}Flexible code design}

Generally, stabilizer tableaus have proven more useful as a description rather than a constructive mechanism. Most well-known stabilizer codes constructed are actually CSS codes~\cite{panteleev2021degenerate,panteleev2022asymptotically,bravyi2024high,kitaev2003fault}, and non-CSS stabilizer codes are usually found by some alternate presentation. The lack of use of stabilizer tableaus is not unusual, as it is simply not obvious as to what collections of Pauli strings construct codes with high distance, easy decoding, gates, geometric properties etc. The universality of the graph representation of stabilizer codes, combined with a rich history of graphs in error correction~\cite{tanner1981recursive,sipser1996expander,bravyi2024high,kissinger2022phase}, suggests that a promising technique to construct certain codes lies in finding sufficiently ``nice'' graphs.

Because of its universal expressive power, the graph representation in this work has parameters that scale with distance. As we showed previously, the distance is bounded above by degree parameters of the graph, and the canonical stabilizers have weights that also scale with the degree. Therefore, the graph representation is best suited for finding codes with a desired approximate numerical values of code parameters, as opposed to asymptotically good LDPC codes. In other words, graphs provide a flexible technique to generate potentially practical small-scale codes.

As an example of this flexibility, if one wanted to create a code with a particular distance, one can set the degree of the graph in accordance with Proposition~\ref{proposition:degree}.
The degree choice gives an upper bound on the distance, so having a sufficiently large degree is a necessary but not sufficient condition to achieve a code with the desired distance. To actually get a good distance, one can either check explicitly for small cases, or use graphs which have a provable distance lower bound. The codes in this section are sufficiently small that we can usually apply the former strategy. However, in a later section, when we investigate decoding algorithms, we will address the second strategy and provide explicit distance lower bounds based on the degree for graphs which satisfy certain structural properties.
Achieving a certain rate translates to having a sufficiently high density of inputs, which in turn can influence the required degree of the input neighbours.
Some experimental devices or algorithms may benefit from codes that have at most a particular number of physical qubits or encode a minimum number of logical qubits, both of which are easy to set in the graph formalism.
If ensuring that the physical qubits behave similarly in a highly uniform manner, a graph could be chosen to be highly regular and symmetric, such as the 5-, 7-, and 9-qubit codes from Section~\ref{sec:compiler:subsec:applications}.
Further still, a graph can be chosen to be planar and local in two dimensions, or perhaps supporting only a fixed number of long-range connections for ease of practical implementation. (To implement a code represented by a graph, we place physical qubits at every output node, so the qubits inherit the geometric properties of the graph. Moreover, the canonical stabilizers are local in the graph in the sense that every qubit supported in the stabilizer associated to $v \in \CO$ is at most 3 edges away from $v$. We note, however, that there certainly exist codes which are, e.g., planar, but whose graph in our representation is not planar.)
On the other hand, a graph embedded into a higher-dimensional space, such as a high-dimensional grid graph for example, could be easier to represent or simulate classically. We will give examples of codes with each of these properties below.

The distance calculations of codes in this section were performed by direct computation. The code is freely available for use~\cite{Lu_graphcodes}.

\begin{figure}[ht]
\centering
\begin{tikzpicture}
    \begin{pgfonlayer}{nodelayer}
        \node [style=input,label={\large $I_1$}] (1) at (0.00, 2.00) {};
        \node [style=input,label={\large $I_2$}] (2) at (1.25, 2.00) {};
        \node [style=output,label={\large 16}] (3) at (-1.00, 1.25) {};
        \node [style=output,label={\large 9}] (4) at (2.25, 1.25) {};
        \node [style=output,label={\large 10}] (5) at (2.75, -0.00) {};
        \node [style=pivot,label={\large 4 ($P_3$)}] (6) at (3.75, -0.00) {};
        \node [style=output,label={\large 3}] (8) at (3.00, 2.00) {};
        \node [style=pivot,label={\large 2 ($P_2$)}] (9) at (1.50, 3.00) {};
        \node [style=pivot,label={\large 1 ($P_1$)}] (10) at (-0.25, 3.00) {};
        \node [style=output,label={\large 8}] (11) at (-1.75, 2.00) {};
        \node [style=output,label={\large 15}] (12) at (-1.50, -0.00) {};
        \node [style=output,label={\large 7}] (13) at (-2.50, -0.00) {};
        \node [style=output,label={\large 14}] (14) at (-1.00, -1.25) {};
        \node [style=output,label={\large 13}] (15) at (0.00, -2.00) {};
        \node [style=output,label={\large 12}] (16) at (1.25, -2.00) {};
        \node [style=output,label={\large 11}] (17) at (2.25, -1.25) {};
        \node [style=pivot,label={\large 6 ($P_4$)}] (20) at (-0.25, -3.00) {};
        \node [style=output,label={\large 5}] (21) at (1.50, -3.00) {};
        \node [style=input,label={\large $I_3$}] (22) at (3.00, -2.00) {};
        \node [style=input,label={\large $I_4$}] (23) at (-1.75, -2.00) {};
    \end{pgfonlayer}
    \begin{pgfonlayer}{edgelayer}
        \draw (1) to (12);
        \draw (1) to (4);
        \draw (1) to (10);
        \draw (2) to (3);
        \draw (2) to (5);
        \draw (2) to (9);
        \draw (3) to (14);
        \draw (3) to (11);
        \draw (4) to (17);
        \draw (4) to (8);
        \draw (5) to (16);
        \draw (5) to (6);
        \draw (6) to (8);
        \draw (6) to (22);
        \draw (8) to (9);
        \draw (9) to (10);
        \draw (10) to (11);
        \draw (11) to (13);
        \draw (12) to (15);
        \draw (12) to (13);
        \draw (13) to (23);
        \draw (14) to (16);
        \draw (14) to (23);
        \draw (15) to (17);
        \draw (15) to (20);
        \draw (16) to (21);
        \draw (17) to (22);
        \draw (20) to (23);
        \draw (20) to (21);
        \draw (21) to (22);
    \end{pgfonlayer}
\end{tikzpicture}
\caption{The $\llbracket 16, 4, 3\rrbracket$ dodecahedral code. Blue nodes are inputs, black nodes are outputs, and orange nodes are pivots. This code is the optimal packing of input nodes into the dodecahedral graph so that no inputs have a path between them of length less than 3. As the dodecahedron has odd cycles, it is not a bipartite graph and therefore this code is a non-CSS stabilizer code. The labels indicate an ordering of the input and output nodes as well as the pivots (lowest numbered nodes next to each input). There are no extra input-pivot edges but we allow pivot-pivot edges where this does not change our constructions. An ordering can be chosen on the nodes to avoid all pivot-pivot edges.}
\label{fig:dodeca}
\end{figure}

\subsubsection{The dodecahedral code}
We now showcase a few examples of novel small codes that we constructed from the graph formalism.
As a first example, we examine the 7-qubit Steane code in Fig.~\ref{fig:7-qubit-code}, which takes the shape of a cube.
Such a geometric structure motivates the exploration of other highly regular geometric solids as graph codes, such as the platonic solids.
While the graphs of a tetrahedron and octahedron perform rather poorly as codes due to their particular symmetries, the graph of a dodecahedron represents a relatively good code. Fig.~\ref{fig:dodeca} exhibits the dodecahedral code, which has parameters $\llbracket 16, 4, 3\rrbracket$.
Since the degree of every node is 3, the dodecahedral code saturates its distance-degree bound.
To maintain a distance equal to the degree, no two inputs may share a neighbour (as otherwise there exists a logical operator of lower weight), so the maximum number of inputs that can be selected is 4. We remark that this graph is not a CSS code, being non-bipartite. Furthermore, having distance 3, the dodecahedral code does not improve upon Hamming-type quantum codes in parameters, e.g. the $\llbracket 15, 7, 3 \rrbracket$ quantum Hamming code~\cite{chao2018fault}. Nonetheless, it is interesting that simply inserting another platonic solid into the graph formalism has provided a new, nontrivial, and non-CSS code.

\begin{figure}[ht]
    \centering
    \resizebox{0.42\linewidth}{15em}{
    \begin{tikzpicture}[scale=0.3]
    \begin{pgfonlayer}{nodelayer}
		\node [style=none,] (0) at (0, 0) {\scriptsize $I_1$\hspace{3ex}\phantom.};
		\node [style=none] (1) at (0, -1) {\scriptsize $I_2$\hspace{3ex}\phantom.};
		\node [style=none] (2) at (0, -3) {\scriptsize $I_3$\hspace{3ex}\phantom.};
		\node [style=none] (3) at (0, -5) {\scriptsize $I_4$\hspace{3ex}\phantom.};
		\node [style=z-node] (4) at (0, -2) {};
		\node [style=z-node] (5) at (0, -4) {};
		\node [style=z-node] (6) at (0, -6) {};
		\node [style=z-node] (7) at (0, -7) {};
		\node [style=z-node] (8) at (0, -8) {};
		\node [style=z-node] (9) at (0, -9) {};
		\node [style=z-node] (10) at (0, -10) {};
		\node [style=z-node] (11) at (0, -11) {};
		\node [style=z-node] (12) at (0, -12) {};
		\node [style=z-node] (13) at (0, -13) {};
		\node [style=z-node] (14) at (0, -14) {};
		\node [style=z-node] (15) at (0, -15) {};
		\node [style=z-node] (20) at (1, 0) {};
		\node [style=z-node] (21) at (1, -8) {};
		\node [style=z-node] (22) at (2, -1) {};
		\node [style=z-node] (23) at (2, -9) {};
		\node [style=z-node] (25) at (3, -10) {};
		\node [style=z-node] (26) at (3, -3) {};
		\node [style=z-node] (27) at (4, -5) {};
		\node [style=z-node] (28) at (4, -13) {};
		\node [style=z-node] (29) at (5, 0) {};
		\node [style=z-node] (30) at (5, -14) {};
		\node [style=z-node] (31) at (6, -1) {};
		\node [style=z-node] (32) at (6, -15) {};
		\node [style=z-node] (33) at (7, -3) {};
		\node [style=z-node] (34) at (7, -4) {};
		\node [style=z-node] (35) at (7, -5) {};
		\node [style=z-node] (36) at (7, -6) {};
		\node [style=z-node] (37) at (7, -6) {};
		\node [style=h-box] (38) at (8, 0) {};
		\node [style=h-box] (39) at (8, -1) {};
		\node [style=h-box] (40) at (8, -3) {};
		\node [style=h-box] (41) at (8, -5) {};
		\node [style=z-node] (42) at (9, 0) {};
		\node [style=z-node] (43) at (9, -1) {};
		\node [style=z-node] (44) at (9, -2) {};
		\node [style=z-node] (45) at (9, -3) {};
		\node [style=z-node] (46) at (9, -4) {};
		\node [style=z-node] (47) at (9, -5) {};
		\node [style=z-node] (48) at (9, -6) {};
		\node [style=z-node] (49) at (9, -7) {};
		\node [style=z-node] (50) at (9, -8) {};
		\node [style=z-node] (51) at (9, -10) {};
		\node [style=z-node] (52) at (10, -9) {};
		\node [style=z-node] (53) at (10, -11) {};
		\node [style=z-node] (54) at (9, -12) {};
		\node [style=z-node] (55) at (9, -14) {};
		\node [style=z-node] (56) at (10, -13) {};
		\node [style=z-node] (57) at (10, -15) {};
		\node [style=z-node] (58) at (11, 0) {};
		\node [style=z-node] (59) at (11, -7) {};
		\node [style=z-node] (60) at (12, -1) {};
		\node [style=z-node] (61) at (12, -2) {};
		\node [style=z-node] (62) at (12, -3) {};
		\node [style=z-node] (63) at (12, -9) {};
		\node [style=z-node] (64) at (14, -4) {};
		\node [style=z-node] (65) at (14, -11) {};
		\node [style=z-node] (66) at (15, -5) {};
		\node [style=z-node] (67) at (15, -12) {};
		\node [style=z-node] (68) at (16, -6) {};
		\node [style=z-node] (69) at (16, -14) {};
		\node [style=z-node] (70) at (13, -2) {};
		\node [style=z-node] (71) at (13, -8) {};
		\node [style=z-node] (72) at (17, -7) {};
		\node [style=z-node] (73) at (17, -15) {};
		\node [style=z-node] (74) at (12, -13) {};
		\node [style=z-node] (75) at (12, -11) {};
		\node [style=z-node] (76) at (11, -12) {};
		\node [style=z-node] (77) at (11, -10) {};
		\node [style=none] (78) at (18, -15) {\scriptsize\phantom.\hspace{3ex}16};
		\node [style=none] (79) at (18, -14) {\scriptsize\phantom.\hspace{3ex}15};
		\node [style=none] (80) at (18, -13) {\scriptsize\phantom.\hspace{3ex}14};
		\node [style=none] (81) at (18, -12) {\scriptsize\phantom.\hspace{3ex}13};
		\node [style=none] (82) at (18, -11) {\scriptsize\phantom.\hspace{3ex}12};
		\node [style=none] (83) at (18, -10) {\scriptsize\phantom.\hspace{3ex}11};
		\node [style=none] (84) at (18, -9) {\scriptsize\phantom.\hspace{3ex}10};
		\node [style=none] (85) at (18, -8) {\scriptsize\phantom.\hspace{3ex}9};
		\node [style=none] (86) at (18, -7) {\scriptsize\phantom.\hspace{3ex}8};
		\node [style=none] (87) at (18, -6) {\scriptsize\phantom.\hspace{3ex}7};
		\node [style=none] (88) at (18, -5) {\scriptsize\phantom.\hspace{7ex}6 ($P_4$)};
		\node [style=none] (89) at (18, -4) {\scriptsize\phantom.\hspace{3ex}5};
		\node [style=none] (90) at (18, -3) {\scriptsize\phantom.\hspace{7ex}4 ($P_3$)};
		\node [style=none] (91) at (18, -2) {\scriptsize\phantom.\hspace{3ex}3};
		\node [style=none] (92) at (18, -1) {\scriptsize\phantom.\hspace{7ex}2 ($P_2$)};
		\node [style=none] (93) at (18, 0) {\scriptsize\phantom.\hspace{7ex}1 ($P_1$)};
	\end{pgfonlayer}
	\begin{pgfonlayer}{edgelayer}
		\draw [style=h] (37) to (35);
		\draw [style=h] (34) to (33);
		\draw [style=h] (31) to (32);
		\draw [style=h] (30) to (29);
		\draw [style=h] (27) to (28);
		\draw [style=h] (25) to (26);
		\draw [style=h] (22) to (23);
		\draw [style=h] (21) to (20);
		\draw (0.center) to (20);
		\draw (20) to (29);
		\draw (31) to (22);
		\draw (22) to (1.center);
		\draw (2.center) to (26);
		\draw (26) to (33);
		\draw (34) to (5);
		\draw (27) to (35);
		\draw (37) to (6);
		\draw (8) to (21);
		\draw (23) to (9);
		\draw (10) to (25);
		\draw (13) to (28);
		\draw (14) to (30);
		\draw (32) to (15);
		\draw (27) to (3.center);
		\draw (35) to (41);
		\draw (40) to (33);
		\draw (39) to (31);
		\draw (29) to (38);
		\draw [style=h] (56) to (57);
		\draw [style=h] (55) to (54);
		\draw [style=h] (53) to (52);
		\draw [style=h] (51) to (50);
		\draw [style=h] (48) to (49);
		\draw [style=h] (46) to (47);
		\draw [style=h] (45) to (44);
		\draw [style=h] (43) to (42);
		\draw (32) to (57);
		\draw (55) to (30);
		\draw (28) to (56);
		\draw (54) to (12);
		\draw (11) to (53);
		\draw (51) to (25);
		\draw (23) to (52);
		\draw (50) to (21);
		\draw (7) to (49);
		\draw (48) to (37);
		\draw (41) to (47);
		\draw (46) to (34);
		\draw (40) to (45);
		\draw (44) to (4);
		\draw (39) to (43);
		\draw (42) to (38);
		\draw [style=h] (69) to (68);
		\draw [style=h] (66) to (67);
		\draw [style=h] (65) to (64);
		\draw [style=h] (63) to (62);
		\draw [style=h] (61) to (60);
		\draw [style=h] (59) to (58);
		\draw (42) to (58);
		\draw (60) to (43);
		\draw (44) to (61);
		\draw (62) to (45);
		\draw (46) to (64);
		\draw (66) to (47);
		\draw (48) to (68);
		\draw (59) to (49);
		\draw (52) to (63);
		\draw (69) to (55);
		\draw [style=h] (73) to (72);
		\draw [style=h] (71) to (70);
		\draw [style=h] (75) to (74);
		\draw [style=h] (76) to (77);
		\draw (93.center) to (58);
		\draw (60) to (92.center);
		\draw (91.center) to (70);
		\draw (70) to (61);
		\draw (62) to (90.center);
		\draw (89.center) to (64);
		\draw (66) to (88.center);
		\draw (87.center) to (68);
		\draw (72) to (86.center);
		\draw (72) to (59);
		\draw (71) to (85.center);
		\draw (71) to (50);
		\draw (63) to (84.center);
		\draw (83.center) to (77);
		\draw (77) to (51);
		\draw (79.center) to (69);
		\draw (78.center) to (73);
		\draw (73) to (57);
		\draw (54) to (76);
		\draw (76) to (67);
		\draw (67) to (81.center);
		\draw (74) to (56);
		\draw (53) to (75);
		\draw (75) to (65);
		\draw (65) to (82.center);
		\draw (80.center) to (74);
	\end{pgfonlayer}
\end{tikzpicture}
}\hspace{3ex}
\raisebox{-0.9em}{
\resizebox{0.43\linewidth}{16.2em}{
\begin{tikzpicture}
\node[scale=10]{
\begin{quantikz}
    \lstick{\LARGE $I_1$}   & \ctrl{8}\gategroup[16,steps=4,style={dashed,inner sep=0pt},label style={label
position=below,anchor=north,yshift=-0.2cm}]{\LARGE $CZ$} & \qw & \qw & \qw & \ctrl{14}\gategroup[16,steps=3,style={dashed,inner sep=0pt},label style={label
position=below,anchor=north,yshift=-0.2cm}]{\LARGE $CZ$} & \qw & \qw & \gate{H}\gategroup[16,steps=1,style={dashed,inner sep=0pt},label style={label
position=below,anchor=north,yshift=-0.2cm}]{\LARGE $H$} & \ctrl{1}\gategroup[16,steps=2,style={dashed,inner sep=0pt},label style={label
position=below,anchor=north,yshift=-0.2cm}]{\LARGE $CZ$} & \qw & \ctrl{7}\gategroup[16,steps=2,style={dashed,inner sep=0pt},label style={label
position=below,anchor=north,yshift=-0.2cm}]{\LARGE $CZ$} & \qw & \qw\gategroup[16,steps=5,style={dashed,inner sep=0pt},label style={label
position=below,anchor=north,yshift=-0.2cm}]{\LARGE $CZ$} & \qw & \qw & \qw & \qw & \qw & \text{\LARGE 1 ($P_1$)}\\
    \lstick{\LARGE $I_2$}   & \qw & \ctrl{8} & \qw & \qw & \qw & \ctrl{14} & \qw & \gate{H} & \control{} & \qw & \qw & \ctrl{1} & \qw & \qw & \qw & \qw & \qw & \qw & \text{\LARGE 2 ($P_2$)}\\
    \lstick{\LARGE $\ket+$} & \qw & \qw & \qw & \qw & \qw & \qw & \qw & \qw & \ctrl{1} & \qw & \qw & \control{} & \ctrl{6} & \qw & \qw & \qw & \qw & \qw & \lstick{\LARGE 3}\\
    \lstick{\LARGE $I_3$}   & \qw & \qw & \ctrl{7} & \qw & \qw & \qw & \ctrl{1} & \gate{H} & \control{} & \qw & \qw & \ctrl{6} & \qw & \qw & \qw & \qw & \qw & \qw & \text{\LARGE 4 ($P_3$)}\\
    \lstick{\LARGE $\ket+$} & \qw & \qw & \qw & \qw & \qw & \qw & \control{} & \qw & \ctrl{1} & \qw & \qw & \qw & \qw & \ctrl{7} & \qw & \qw & \qw & \qw & \lstick{\LARGE 5}\\
    \lstick{\LARGE $I_4$}   & \qw & \qw & \qw & \ctrl{8} & \qw & \qw & \ctrl{1} & \gate{H} & \control{} & \qw & \qw & \qw & \qw & \qw & \ctrl{7} & \qw & \qw & \qw & \text{\LARGE 6 ($P_4$)}\\
    \lstick{\LARGE $\ket+$} & \qw & \qw & \qw & \qw & \qw & \qw & \control{} & \qw & \ctrl{1} & \qw & \qw & \qw & \qw & \qw & \qw & \ctrl{8} & \qw & \qw & \lstick{\LARGE 7}\\
    \lstick{\LARGE $\ket+$} & \qw & \qw & \qw & \qw & \qw & \qw & \qw & \qw & \control{} & \qw & \control{} & \qw & \qw & \qw & \qw & \qw & \ctrl{8} & \qw & \lstick{\LARGE 8}\\
    \lstick{\LARGE $\ket+$} & \control{} & \qw & \qw & \qw & \qw & \qw & \qw & \qw & \ctrl{2} & \qw & \qw & \qw & \control{} & \qw & \qw & \qw & \qw & \qw & \lstick{\LARGE 9} \\
    \lstick{\LARGE $\ket+$} & \qw & \control{} & \qw & \qw & \qw & \qw & \qw & \qw & \qw & \ctrl{2} & \qw & \control{} & \qw & \qw & \qw & \qw & \qw & \qw & \lstick{\LARGE 10}\\
    \lstick{\LARGE $\ket+$} & \qw & \qw & \control{} & \qw & \qw & \qw & \qw & \qw & \control{} & \qw & \ctrl{2} & \qw & \qw & \qw & \qw & \qw & \qw & \qw & \lstick{\LARGE 11}\\
    \lstick{\LARGE $\ket+$} & \qw & \qw & \qw & \qw & \qw & \qw & \qw & \qw & \qw & \control{} & \qw & \ctrl{2} & \qw & \control{} & \qw & \qw & \qw & \qw & \lstick{\LARGE 12}\\
    \lstick{\LARGE $\ket+$} & \qw & \qw & \qw & \qw & \qw & \qw & \qw & \qw & \ctrl{2} & \qw & \control{} & \qw & \qw & \qw & \control{} & \qw & \qw & \qw & \lstick{\LARGE 13}\\
    \lstick{\LARGE $\ket+$} & \qw & \qw & \qw & \control{} & \qw & \qw & \qw & \qw & \qw & \ctrl{2} & \qw & \control{} & \qw & \qw & \qw & \qw & \qw & \qw & \lstick{\LARGE 14}\\
    \lstick{\LARGE $\ket+$} & \qw & \qw & \qw & \qw & \control{} & \qw & \qw & \qw & \control{} & \qw & \qw & \qw & \qw & \qw & \qw & \control{} & \qw & \qw & \lstick{\LARGE 15}\\
    \lstick{\LARGE $\ket+$} & \qw & \qw & \qw & \qw & \qw & \control{} & \qw & \qw & \qw & \control{} & \qw & \qw & \qw & \qw & \qw & \qw & \control{} & \qw & \lstick{\LARGE 16}
\end{quantikz}
};
\end{tikzpicture}
}}
\caption{Two presentations of the encoding circuit of the dodecahedral code, labeled as in Fig.~\ref{fig:dodeca}. On the left, the presentation is of a ZX diagram. Yellow boxes indicate $H$ gates and blue dashed edges are, as usual, Hadamarded.
On the right, the presentation is of a standard quantum circuit.
Each dashed block represents a set of transversal (depth-$1$) operations, so that the circuit depth is the number of blocks, in this case 6. The presentation here is partially optimized, by inspection, to improve upon the upper bound given in Theorem~\ref{thm:encoding}. It is possible with further heuristics to produce a depth-5 (but more visually complicated) circuit, which is provably optimal.}
\label{fig:dodeca-encoder}
\end{figure}

To complete our discussion of the dodecahedral code, we apply Theorem~\ref{thm:encoding} to give the code a low-depth encoding circuit. Fig.~\ref{fig:dodeca-encoder} illustrates two representations of the encoding circuit. On the left, we draw a ZX diagram which appears as an intermediate step in the algorithmic construction in Theorem~\ref{thm:encoding}. On the right, we translate the ZX diagram into standard quantum circuit notation.
The translation is an immediate consequence of the ZX calculus rules: a $Z$ node with a single output is a $\ket+$ state and a Hadamarded edge between two $Z$ nodes implements a $CZ$ gate. Corresponding to the theorem, the $CZ$ gates before the layer of Hadamards form the input-output edges; the Hadamards effectively form input-pivot edges; and the $CZ$ gates after the layer of Hadamards form the edges from non-inputs to non-inputs.
We observe also that the application of 
the ZX calculus simplification rules to the diagram in Fig.~\ref{fig:dodeca-encoder}, such as merging $Z$ nodes together, recovers the graph representation of the dodecahedral code itself. In this sense the encoding circuit appears almost immediately from the graph representation.

While our results about the worst-case encoder depth suggest that it may have a depth of up to 9, we have applied some analysis by inspection to further optimize the depth to 6.
Further compilation of this circuit would allow us to reduce the depth further.
This is because output-output edges commute with every gate in the circuit as they are not on the input-pivot wires containing the only non-diagonal gates.
In particular, the 5 gates in the last block can be moved to the second, first, fourth, first, and first blocks, respectively, resulting in a circuit of depth 5.
This is the minimum possible depth as that is the number of gates acting on some qubits, such as the first input.

\begin{figure}[ht!]
    \centering
    \includegraphics[width=0.5\linewidth]{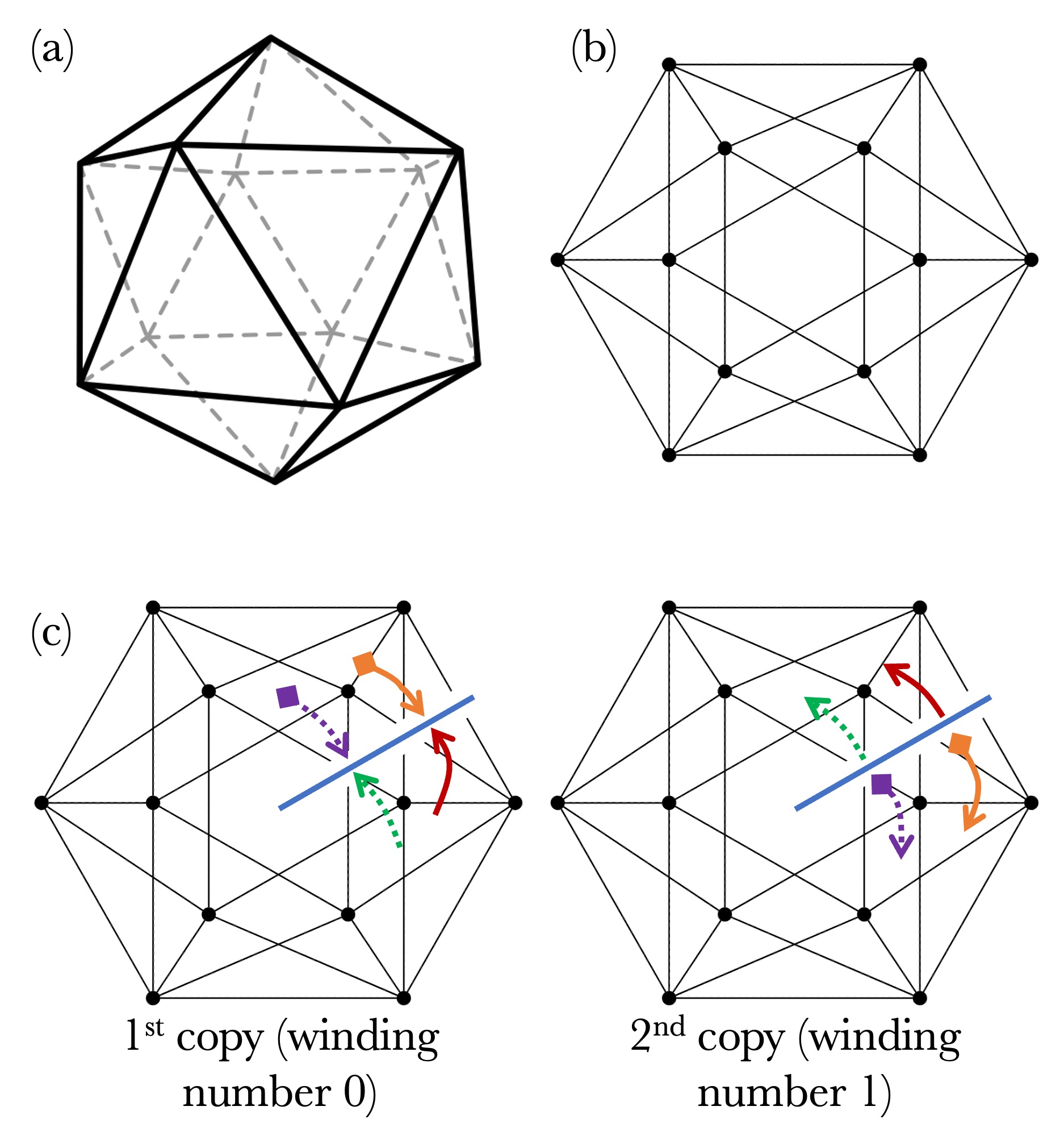}
    \caption{Double-covering the icosahedron. We begin with an icosahedron, shown in (a). The graph version of this solid is given in (b). To double-cover it, we make two copies of the icosahedron and ``spear" it through the center, illustrated with a solid blue line. In the first copy, as we move (clockwise) counterclockwise and pass through the spear, we ``teleport" to the second copy and continue as shown by the (solid orange arrows with the square tails) solid red arrows. Similarly, in the second copy, as we move (clockwise) counterclockwise and pass through the spear, we ``teleport" to the first copy and continue as shown by the (dotted purple arrows with the square tails) dotted green arrows. This idea can be extended to a general $t$-covered icosahedron by associating each copy with a winding number. Going (clockwise) counterclockwise through the spear teleports to the copy with winding number one (lower) higher mod $t$.}
    \label{fig:icosa}
\end{figure}

\subsubsection{Covering spaces and the 5-covered icosahedral code}
We next consider a code based on the graph of an icosahedron, depicted in Fig.~\ref{fig:icosa}(a)-(b).
Since the vertices of an icosahedron have degree 5, we can hope that a code based on this graph would have distance 5. Unfortunately, we find that the code underperforms the degree bound because a $Y$ gate on two antipodal vertices is equivalent to applying a $Z$ to each vertex.
Hence, a logical $Y$ on any input is a logical operation of weight 3, with one $Y$ on the vertex opposite the input and two more on any two antipodal vertices.
This observation also implies that if we wish to add more than one input to the icosahedron, our distance will be reduced further, as a logical operation would just consist of $Y$'s on two vertices opposite the inputs.
To get around this issue, we lift the graph into a finite covering space.
Specifically, we can imagine taking a double cover of the icosahedron by ``spearing'' the icosahedron through the centers of two opposite faces and attaching a winding number to each vertex, which is computed modulo 2. Such a construction yields a double-covered icosahedron. Fig.~\ref{fig:icosa} illustrates the double-covered icosahedron in an intuitive manner, namely by gluing two icosahedrons together across the spear.
We find that the double-covered graph with a single input has distance 5.
In fact, we can improve the rate slightly by taking a $t$-cover of the icosahedron. Specifically, when $t = 5$ we can fit 6 input vertices, where the inputs are not vertices of the punctured faces, and each of which is wound $300^\circ$ around from the previous input.
In Fig.~\ref{fig:icosa}, if we puncture the two central faces, the inputs in the 5 copies of the icosahedron will be on copies of the 6 vertices along the perimeter, each rotated $300^\circ$ from the last.
This construction creates a $\llbracket 54, 6, 5\rrbracket$ code.

The ideas used to create the five-covered icosahedron can be extended more generally to extend otherwise finite graphs.
In fact, arbitrarily complicated covers can be proposed, leading all the way up to the universal cover of the graph.
If regularity and a finite set of vertices are desired, some periodic boundary conditions can be established to keep the graph finite.
In general, such covering space graphs may provide a promising path towards code design recipes in this framework. In the case of Cayley graphs, similar ideas have already been utilized to construct quantum LDPC codes from Cayley graphs of free groups with generators comprised of columns of a classical parity check matrix~\cite{couvreur2013construction}.

\subsubsection{Planar and higher-dimensional lattice-type codes}
Even more generally, lattice-type codes enable the construction of families of codes with increasing degree (and thus, potentially, increasing distance) while also preserving some notions of regularity that are often desirable in code construction.
If we wish to remain in two dimensions and also have a Euclidean lattice, we are restricted in the sense that the triangular lattice, which has degree 6, is the highest-degree such lattice.
We numerically confirmed that such lattices do yield distance-6 codes.
Note that while surface codes also have Euclidean planar connections, they can allow for much higher distances.
This comes at the trade-off of having only a constant number of logical qubits.
If we were to construct a large two-dimensional triangular lattice, we can achieve a different trade-off: capping our distance at 6, but being able to place numerous inputs in the lattice, achieving a constant rate.

Beyond degree 6, we can either raise the dimension or embed graphs in non-Euclidean spaces. For example, an embedding of triangular surfaces into hyperbolic spaces yields the Poincar\'e's disk map. However, because the Poincar\'e disc packs infinitely many points into a finite space, we would have to construct finite approximations and then study the effect of artificially establishing a finite boundary. Analysis of such effects appears difficult and is out of the scope of this work. Nonetheless, such classes of codes in non-Euclidean spaces may prove useful, and even connected to HaPPY codes and other hyperbolic-type codes~\cite{pastawski2015holographic,higgott2024constructions,breuckmann2021quantum} which use graphs in hyperbolic space.

The other path---increasing the dimension---is more amenable to analysis. A simple, general recipe to construct a code from a lattice is to cut off the lattice in each dimension at a finite number of intervals, endow the finite version with periodic boundary conditions in each direction, and then embed some inputs and pivots within them such that the choices of pivots are valid and inputs are not connected.

One immediate advantage of this approach is a robustness to deformation that does not hold for CSS codes. In particular, 
we showed in Theorem~\ref{thm:bipartite_CSS} that bipartite graphs correspond to CSS codes. For lattices with periodic boundary conditions, the bipartite condition is equivalent to the length of every single dimension of the torus being even. Such a requirement is very non-local and does not change the central architecture of the code.
Even for CSS torii, the addition of one vertex far from any inputs such that it forms an odd-length cycle will make the code no longer CSS but will not meaningfully affect any of the code's parameters.

When constructing families of codes in the graph formalism, it is important to consider the geometry of the graph as well as the placement of the inputs.
To make use of the graph formalism for computing stabilizers and logical operations, the identification of pivot nodes is paramount.
While the ZXCF does not allow pivot-pivot edges, these do not affect the expressions of the graph formalism.
Thus, for easy pivot selection, it suffices for all of the inputs to be separated by paths of length at least 3 from each other in the graph. Although such a restriction is not necessary for code construction, it does allow for any input's neighbour to be chosen as its pivot as that node would necessarily not border any other inputs by the distance condition. The downside to such graphs, however is that their rate is bounded as $R \leq \frac{1}{q}$, where $q$ is the degree of the graph (assuming it is regular). In other words, they have a strong rate-distance trade-off. Nonetheless, such graphs may be interesting at small scales, as we have already seen. 

In higher dimensions, perhaps the simplest lattice we can utilize to build a code is the boolean hypercube in an arbitrary $m$ dimensions. We represent the hypercube by labeling nodes as length-$m$ bitstrings and connecting two nodes if their Hamming distance is exactly 1, i.e. they differ by one bit flip. We observe immediately that the hypercube code is CSS for all $m$, since the hypercube is bipartite. 
To select inputs, we observe that if we wish to place inputs such that they are separated by paths of length at least 3, we can leverage classical coding theory to provide this guarantee automatically. Specifically, we consider the \textit{classical} $[2^r - 1, 2^r - r - 1, 3]$ Hamming codes~\cite{hamming1950error}. Letting $m = 2^r - 1$, we can equivalently express the parameters as $[m, m - \log(m+1), 3]$, where the code is perfect when $m$ is one less than a power of 2. Note that all logarithms in this paper are taken to be base 2. Since this code has $m - \log(m+1)$ logical bits, there are a total of $\frac{2^m}{m + 1}$ codewords represented as length-$m$ bitstrings. Because the Hamming code's distance is 3, all such codewords are guaranteed to be 3-spaced in the hypercube graph.
Therefore, by choosing the input nodes in the $m$-dimensional hypercube to be those whose bitstring is a codeword in the $[m, m - \log(m+1), 3]$ Hamming code, we obtain a code with $\frac{2^m}{m+1}$ logical qubits, $2^m -\frac{2^m}{m+1} = \frac{m 2^m}{m+1}$ physical qubits, and a distance $d(m)$ bounded above by the degree $m$. That is, a $\llbracket \frac{m 2^m}{m+1}, \frac{2^m}{m+1}, d(m)\rrbracket$ family of codes. Due to the 3-spacing of inputs, any choice of pivots uniquely connected to a corresponding input will do; later, we will show that it is both possible and particularly convenient to choose the pivots such that they are also separated by paths of length at least 3. We will also later prove in Section~\ref{subsec:decoder} that $d(m) \gtrsim m/2$, up to an additive error of 2, and give evidence that $d(m) = m$. For example, for $m = 7$, the code has parameters $\llbracket 112, 16, 7 \rrbracket$. (This distance was computed numerically; we later discuss this in more detail alongside the lower bound.) Asymptotically, the code scales as $\llbracket n, \Th(\frac{n}{\log n}), \Th(\log n)\rrbracket$.

\begin{definition}
\label{def:hypercube}
    Let the family of \textit{hypercube codes} be the set of $\llbracket \frac{m 2^m}{m+1}, \frac{2^m}{m+1}, D_m\rrbracket$ codes defined above, parametrized by $m\geq3$.
\end{definition}

To improve the rate further without the strong distance-rate trade-off barrier $R \leq \frac{1}{d}$, we will need to use the most general form of graph codes, namely those with inputs separated by paths of length 2. We conclude this section with some remarks on how such codes may be constructed.
In a grid lattice on a $m$-dimensional torus, we wish to pack inputs more closely while being mindful of the fact that our distance will be bounded above by the number of output nodes next to each output node.
This would mean packing inputs with a graph distance of fewer than 3 between them.
Such a packing will reduce the distance below the degree of the graph, but we can hope to try to not let the distance be reduced ``too much'' while (hopefully) considerably improving the packing density of inputs and thus the rate of the code.

One such construction is as follows.
First, consider a foliation of the $m$-torus into sequentially numbered ``layers'' of $(m-1)$-dimensional toric hypersurfaces that together stack into a $m$-dimensional torus.
In each layer, we will consider sets of vertices grouped by their \textit{index}, the sum of their integer coordinates taken mod 3.
We can assume that the size of each torus dimension is a multiple of 3 to make sure these sets are well-defined.
Note that the index does not consider the $m$th coordinate, as that is determined by the number of the layer that a node ends up in.
\begin{enumerate}
    \item[(1) ]
In layer 1, we set all nodes of index 0 to be inputs, with pivots in layer 2.
    \item[(2) ]
In layer 2, we set all nodes of index 0 or 1 to be pivots.
    \item[(3) ] 
In layer 3, we set all nodes of index 1 to be inputs, with pivots in layer 2.
    \item[(4) ] 
We repeat this 3-layer structure as many times as desired.
\end{enumerate}
Such a pattern is essentially a 3-layered, high-dimensional analog of a checkerboard pattern.
Again, we see that our distance upper bound is $d$, as vertices such as those of index 1 in layer 1 have $d$ inputs among their $2d$ neighbours. More sophisticated techniques are necessary, however, to lower-bound the distance.
However, the rate of this code is a constant $\frac27$, which is achieved at any scale of this code, without the need for extreme numbers of qubits for the asymptotic behaviour of the code parameters to be expressed.
Additionally, we note that if an odd number of 3-layer structures are used or if any of the dimensions of each layer is odd, the code graph will have a cycle of odd length, meaning that the resulting code will not be a CSS code.

These examples showcase a wide variety of families and constructions, both small and large, that can be achieved with the graph formalism. While some are sufficiently small to admit brute-force computational distance calculations, others are more difficult to analyze and will require additional distance-bounding tools to understand. We take a first step in this direction in the next section.

\subsection{\label{subsec:decoder}Decoding algorithms and analysis}
We continue our discussion by constructing a simple greedy decoding algorithm on graphs. We give a sufficient-condition characterization of graphs for which the greedy decoder successfully recovers errors up to half of the theoretical maximum. That is, graphs which satisfy this certain property have a provable distance \textit{lower bound} based on its degree. Finally, we will show that the  $\llbracket \frac{m 2^m}{m+1}, \frac{2^m}{m+1}\rrbracket$ boolean hypercube code, defined in Section~\ref{subsec:smallcodes}, satisfies this characteristic and thus is efficiently decodable. 

Denote for simplicity $oip(v) := o(i(p(v)))$ and $oi(v) = o(i(v))$, where $o(\cdot), i(\cdot)$, and $p(\cdot)$ are as in Definition~\ref{def:neighbour_sets}. In Theorem~\ref{thm:decodingQLO}, we showed that decoding amounts to strategies in QLO. We there showed that Pauli operations on nodes are equivalent to a combination of switches and toggles on outputs and switches on inputs, which could cleanly reframed to fit the round (1) and round (2) structure of the QLO game framework. Here, we will be more direct, and directly apply Paulis to toggle corresponding lights. Recall from the proof of Theorem~\ref{thm:decodingQLO} that $Z$ on $v \in \CO$ toggles the light on $v$, $Z$ on $v \in \CP$ toggles all lights on $oi(v)$, and $X$ on $v \in \CO \cup \CP$ toggles all lights in $o(v) \,\D\, oip(v)$. ($Y$ errors are treated as both an $X$ and a $Z$ error.) In this manner we may equivalently ignore the existence of lights on pivots entirely, as they are never toggled.

In analyzing strategies for decoding, we will find it helpful to introduce a metric, which we call the illumination. Informally, the \textit{illumination} of a node $v$ is a weighted count of how many lights are on, out of all those that are toggled by flipping $v$'s switch. The set of lights toggleable by $v$ are either in $o(v)$ or $oip(v)$ (we ignore the lights which are in both, as flipping $v$'s switch toggles them twice, effectively doing nothing). We count each light in $o(v)$ as 1 light, but we count each light in $oip(v)$ only as a fraction of a light. This is because flipping a pivot switch $p \in p(v)$ can toggle many lights in $oip(v)$ at once. Therefore, we will see that the illumination quantifies how many moves we can save by flipping the switch at $v$.

\begin{figure}[ht!]
    \centering
    \includegraphics[width=0.28\linewidth]{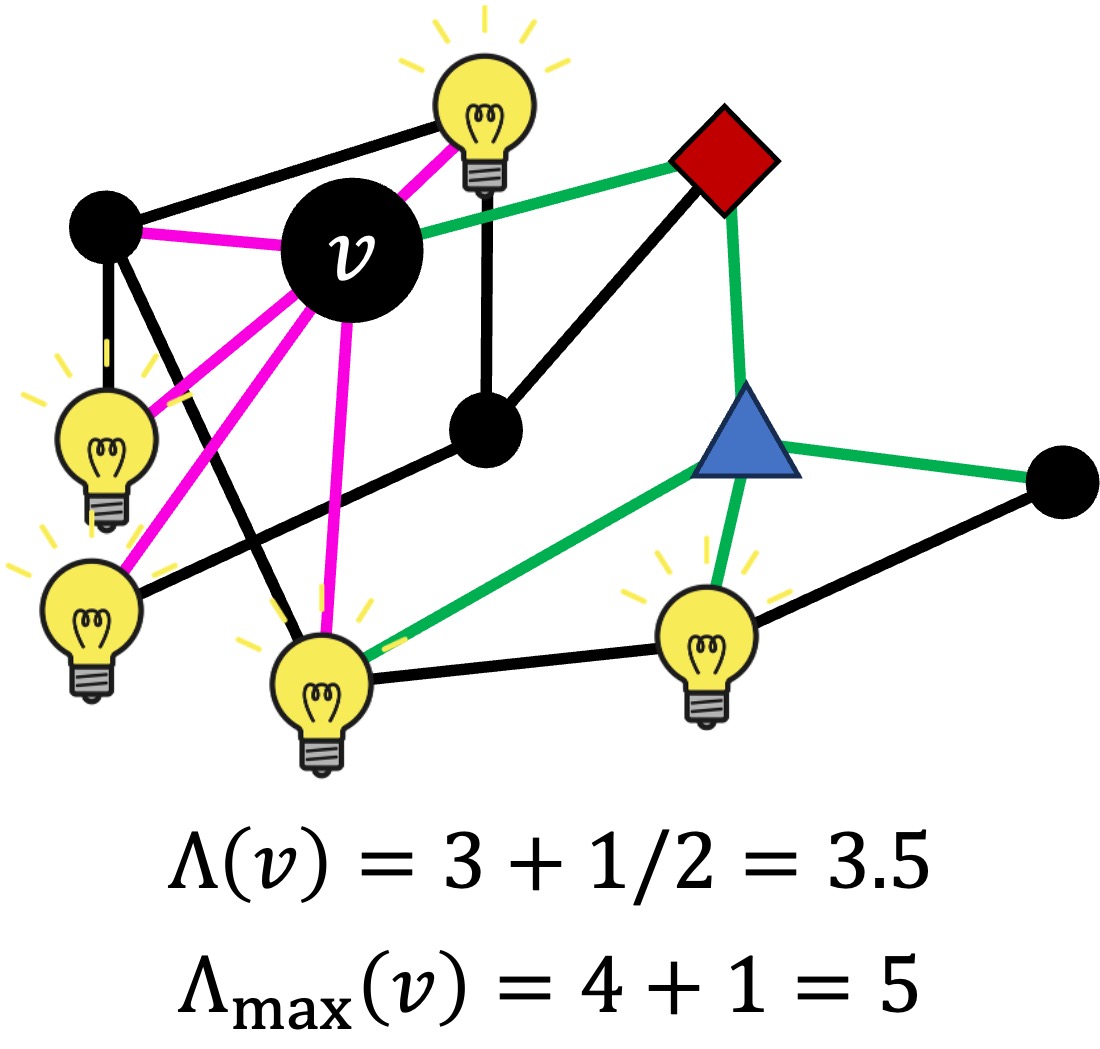}
    \caption{A depiction of the illumination of a node $v$ for a given configuration of lights. The set $L=o(v)\,\D\,oip(v)$ does not include the node directly below $v$, as it is connected by both a pink edge (denoting the paths from $v$ to $o(v)$) and a green edge (denoting the paths from $v$ to $oip(v)$). As there are 3 lights on in $o(v)\cap L$, this contributes 3 to the illumination of $v$.
    One of two lights is on in $oip(v)\cap L$, which contributes $\frac12$ to the illumination.
    The max illumination is 5, which can be computed by turning on all the lights in the figure, or by adding $|o(v)\cap L|=4$ and all pivot neighbours $p$ of $v$ such that $oi(p)$ has at least one light in $L$.}
    \label{fig:illumination}
\end{figure}

\begin{definition}\label{def:illumination}
    Let $G = (V, E, \set{o, i, p})$ be a graph with input, output, and valid pivot nodes. For any node $v\in\CO\cup\CP$, let $L=o(v) \,\D\, oip(v)$. Then the \textit{illumination} of $v$, denoted $\L(v)$, is given by
    \begin{equation}
        \L(v)=|\set{\text{lights on in $o(v)\cap L$}}|+\sum_{\substack{p\in p(v)\\|oi(p)\cap L|>0}}\frac{|\set{\text{lights on in $oi(p)\cap L$}}|}{|oi(p)\cap L|}.
    \end{equation}
Note that each pivot $p$ around $v$ has an associated set of nodes $oi(p)$ that can be toggled with one operation---applying a $Z$ to $p$. Thus, the number of lights on is normalized to a weight of 1, as if all of the lights in $oi(p)$ were concentrated in a light at $p$.
We also define the \textit{max-illumination} analogously, as \begin{align}
    \L_{\text{max}}(v)\coloneqq|o(v)\cap L|+|\set{p\in p(v)\mid |oi(p)\cap L|>0},
\end{align}
which represents the maximum possible value of $\L(v)$ across all light configurations.
    
For $v\in\CI$, we define $\L(v)=|\set{w\in o(v)\mid \text{light at $w$ is on}}|$ and $\L_\text{max}(v)=|o(v)|$.
Note that both for $v\in\CO\cup\CP$ and for $v\in\CI$, $\L_\text{max}(v)\leq \deg(v)$.
\end{definition}

Applying $X$ to a node $v\in\CO\cup\CP$ or $Z$ to $v\in\CI$ will decrease the illumination of $v$ by $\L(v)-\left(\L_\text{max}(v)-\L(v)\right)=2\L(v)-\L_\text{max}(v)$ lights (if this value is negative, then the illumination has actually \textit{increased}).
Additionally, note that while the illumination $\L$ depends on a contextual ``current'' configuration of lights, the max-illumination $\L_\text{max}$ depends only on the graph and its assignment of inputs and pivots.
It is useful to compare the notion of illumination to a classical analog. Classically, when considering a Tanner graph of a code, we may label each data node by the number of check nodes it neighbours which are violated. Decoders such as the flip algorithm from \citet{sipser1996expander} use this metric to correctly deduce the error, by flipping data bits whose node has more violated syndromes than not. The notion of illumination is quite similar in that it also captures violated syndromes around a node. In our graph representation however, the presence of pivots complicates how we measure the amount of violation, giving rise to the more technical definition of illumination.

\begin{algorithm}\caption{Greedy graph decoder} \label{alg:greedy}
\DontPrintSemicolon
  \SetKwFunction{FMain}{GreedyDecoder}
  \SetKwProg{Fn}{Function}{:}{}
  \Fn{\FMain{$G$}}{
        $\text{explored} \leftarrow [],\, (v_0, n_0) \leftarrow (\text{nil}, 0)$\;
        \While(\texttt{// STAGE 1: X error recovery loop}){True}{
            \For{$v \in \CO \cup \CP$}{
                $\text{gap} \leftarrow 2\L(v)-\L_\text{max}(v)$\;
                \tcp{If the illumination of $v$ would decrease by applying $X(v)$, `gap' is positive}
                \If{$\text{gap} > n_0$}{
                    $v_0 = v,\, n_0 = \text{gap}$\;
                }
            }
            \If{$v_0 \in \textrm{explored}$ \textbf{or} $n_0 < 1$}{
                \textbf{break}\;
            }
            \texttt{Recover}$_X(v_0)$ \tcp{toggle lights at $o(v_0) \,\D\, oip(v_0)$, destroy light at $v_0$}
            \texttt{Append} $v_0$ \textbf{to} explored\;
        }
        $\text{explored} \leftarrow [],\, (v_0, n_0) \leftarrow (\text{nil}, 0)$ \tcp{Reset tracking variables}
        \While(\texttt{// STAGE 2: Z error \textit{on pivots} recovery loop}){True}{
            \For{$v \in \CP$}{
                $\text{gap} \leftarrow 2\L(i(v))-\L_\text{max}(i(v))$\;
                \If{$\text{gap} > n_0$}{
                    $v_0 = v,\, n_0 = \text{gap}$\;
                }
            }
            \If{$v_0 \in \textrm{explored}$ \textbf{or} $n_0<1$}{
                \textbf{break}\;
            }
            \texttt{Recover}$_Z(v_0)$ \tcp{toggle lights at $oi(v_0)$}
            \texttt{Append} $v_0$ \textbf{to} explored\;
        }
        \For(\texttt{// STAGE 3: Z error \textit{on non-pivot outputs} recovery loop}){$v \in \CO$}{
            \If{\texttt{LightIsOn}$(v)$}{
                \texttt{Recover}$_Z(v)$ \tcp{Destroy light at $v$}
            }
        }
        \KwRet
  }
\end{algorithm}

\begin{figure}[ht!]
    \centering
    \includegraphics[width=0.9\linewidth]{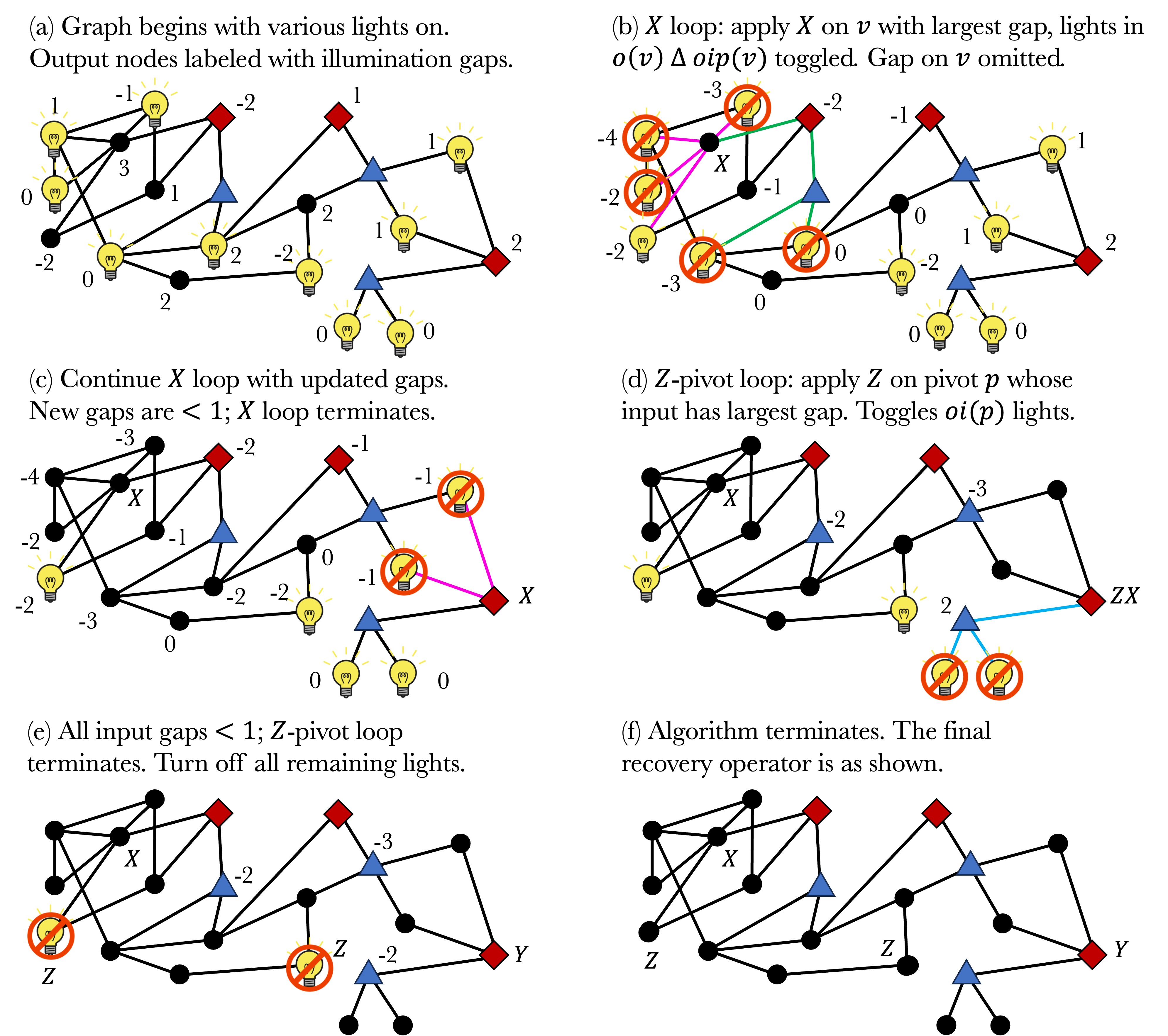}
    \caption{Visualization of Algorithm~\ref{alg:greedy} run on the final graph in Fig.~\ref{fig:QLO_Decoding}. Inputs are blue triangles, pivots are red diamonds, and non-pivot outputs are black circles. In (a)-(c), non-pivot output nodes are labeled with their illumination gaps, given by $2 \L(v) - \L_{\max}(v)$, where $\L$ and $\L_{\max}$ are given in Defintion~\ref{def:illumination}. These are the nodes checked by the $X$ loop. In (d)-(e), the illumination gaps of inputs are labeled; these are the nodes checked by the $Z$-pivot loop. The final recovery map shown is precisely the error we applied originally to the graph in Fig.~\ref{fig:QLO_Decoding}.}
    \label{fig:GreedyDecoder}
\end{figure}

The greedy decoder, Algorithm~\ref{alg:greedy}, is a strategy in the decoding instance of QLO. In particular, the operations permitted are \texttt{Recover}$_X$ and \texttt{Recover}$_Z$, which apply the corresponding Paulis that effectively flip switches or destroy lights depending on the node $v$ chosen. We showed earlier in Theorem~\ref{thm:decodingQLO} how a QLO strategy maps to a physical recovery operation.
The construction of the algorithm is based on the following intuition. If $v \in \CO$ suffers a $X$ error, the stabilizers at $o(v) \,\D\, oip(v)$ will measure $-1$, so in the QLO game the lights in $o(v) \,\D\, oip(v)$ will turn on. Perhaps there will be some $Z$ errors in $o(v) \subseteq \CO$, each of which correspond to turning on a single light in $o(v)$.
Therefore, assuming it is difficult to turn on more than half of the lights in $o(v)$, a good guess is that if more than half of the lights are on in $o(v)$ then $v$ has suffered an $X$ error, at which point we can apply $X$ on $v$ to reverse the error, via \texttt{Recover}$_X$.
However, applying $X$ to $v$ toggles both $o(v)$ and $oip(v)$.
Toggling $oip(v)$ is the same as toggling $oi(p)$ for each pivot $p$ in $p(v)$.
Since this can be toggled by applying a single $Z$ to $p$ (effectively toggling ``the light at $p$'', or equivalently, $oi(p)$, we treat the status of the lights in $oi(p)$ as having the same collective importance as one light in $o(v)$, and we therefore normalize $oi(p)$'s contribution to the gap by dividing the respective term by $|oi(p)|$ for each $p$.
This is all summed up by the aggregate quantity $\L(v)$, the illumination of $v$ as defined in Definition~\ref{def:illumination}.
At each step, we apply the single move that will most decrease the illumination of any given node, starting with nodes in $\CO\cup\CP$ and continuing with nodes in $\CI$.
We only perform moves that decrease the illumination by at least 1, as that is how many moves it would take to decrease it by other means.
We depict the action of Algorithm~\ref{alg:greedy} in Fig.~\ref{fig:GreedyDecoder}. We begin with the QLO decoding game from Fig.~\ref{fig:QLO_Decoding} built from the application of some Pauli error. When the algorithm terminates, we observe that the recovery operation is precisely the original Pauli error applied, indicating that the greedy algorithm decoded correctly in this example.

Applying \texttt{Recover}$_X$ may itself turn on more lights in other places, so to avoid issues related to parallelism, we work in sequence and apply $X$ recoveries one at a time, in order of what most decreases the node's illumination, a process we will refer to as the first stage of the algorithm. A similar story holds for $Z$ errors on $v \in \CP$. They are detected by stabilizers in $oi(v)$, so we apply the analogous greedy recovery operation there in the second stage. Finally, for remaining lights on in $v \in \CO$, we directly destroy the lights with $Z$ operations in the third and final stage of Algorithm~\ref{alg:greedy}.

The key property within this intuition that seems to enable greedy-type decoding is the idea that it must be difficult to trick Algorithm~\ref{alg:greedy} into believing that an $X$ error on some $v$ occurred by using a small amount of moves to increase the illumination of $x$. We now formalize this property to give a simple performance guarantee of the greedy decoder.

In this definition, we are specifically interested in how much can an operation at node $v$ change the illumination of another node.

\begin{definition} \label{def:sensitive}
    Let $G = (V, E, \set{o, i, p})$ be a graph with with vertices $V = \CI \sqcup \CP \sqcup \CO$, inputs $\CI$, pivots $\CP$, non-pivot outputs $\CO$, edges $E$, and $o, i, p$ as in Definition~\ref{def:neighbour_sets}. We say that $G$ is \textit{$B$-sensitive} or that the \textit{sensitivity} of $G$ is $B$, where $B\in\mathbb{R}$ is the maximum amount by which the illumination $\L(v)$ of any node can change from the application of certain single-qubit Pauli operations, when starting with all lights off. Specifically, if $v\in\CO\cup\CP$, we allow any Pauli operation and if $v\in\CI$, we only consider Pauli $Z$ operations.
    Quantitatively, denoting the change in the illumination of $v$ due to applying the Pauli $P$ on vertex $w$ by $\D_{P(w)}(\L(v))$, we have
    \begin{equation}
    B\coloneqq\max\left(\max_{\substack{v,w\in\CO\cup\CP\\w\neq v\\P\in\set{X,Y,Z}}}\D_{P(w)}(\L(v)),\max_{\substack{v\in\CI\\w\in\CO\cup\CP\\w\neq p(v)}}\D_{Z(w)}(\L(v))\right).
    \end{equation}
\end{definition}
To give some intuition about what the notion of sensitivity represents, imagine a graph representing a social network in which nodes are people and edges connect people who are friends.
If it is common for two vertices (people) to have many shared neighbours (mutual friends), then it will be possible to toggle many lights on around the first vertex by flipping the switch on the second.
If mutual friends are scarce and two nodes never share more than one mutual friend, then it will be difficult to turn on many lights around a given vertex by flipping switches at other vertices.
The precise description of illumination used in calculating sensitivity also includes terms such as $oip(v)$, which care about more distant vertices than merely the immediate neighbours, but counting mutual friends can serve as a good heuristic for the sensitivity of a graph before inputs and pivots have been assigned.
Altogether, the sensitivity definition forms a sufficient condition to place a relatively tight guarantee on the distance of the code and the greedy decoder's performance.

\begin{definition}\label{def:brightness}
    Let $G = (V, E, \set{o, i, p})$ be  a graph with vertices $V = \CI \sqcup \CP \sqcup \CO$, inputs $\CI$, pivots $\CP$, non-pivot outputs $\CO$, edges $E$, and $o, i, p$ as in Definition~\ref{def:neighbour_sets}.
    We define the \textit{brightness} $\ell\in\mathbb{N}$ of graph $G$ as the minimum max-illumination across all vertices in $V$, as defined in Definition~\ref{def:illumination}. Specifically,
    \begin{align} \label{eq:delta_*}
         \ell\coloneqq \min_{v\in V}\L_\text{max}(v),
     \end{align}
     where $\L_{\max}$ is given in Definition~\ref{def:illumination}. Note that by definition, $\ell \leq \min_{v \in V} \deg(v)$ since $\L_{\max}(v) \leq \deg(v)$. By Proposition~\ref{proposition:degree}, $\ell \leq \operatorname{dist}(C(G))$, where $C(G)$ is the code corresponding to $G$ and $\operatorname{dist}$ is the distance.
\end{definition}
Below, we will see that $\L_{\max}(v)$ quantifies the difficulty of tricking Algorithm~\ref{alg:greedy} at node $v$. The brightness $\ell$ uniformly lower bounds $\L_{\max}(v)$ across all $v$, which will allow us to state guarantees and distance lower bounds in terms of $\ell$.

\begin{theorem}[Distance lower bound]\label{thm:distanceBound}
    Let $G$ be a $B$-sensitive graph with brightness $\ell$ as defined in Definition~\ref{def:sensitive} and Definition~\ref{def:brightness}. Let $C(G)$ be the code that $G$ represents.
    Then, $\text{dist}(C(G))\geq\ceil{\frac{\ell}B}+1$.
\end{theorem}
\begin{proof}
    The intuition behind this proof is that a small sensitivity $B$ makes it difficult to turn off the lights around a node in few moves, which forces the weight of any logical operator to be large.
    We continue to adopt the picture in which $Z$ on $v \in \CO$ toggles the light on $v$, $Z$ on $v \in \CP$ toggles lights in $oi(v)$, and $X$ on $v \in \CO$ toggles lights in $o(v) \,\D\, oip(v)$.
    
    Every stabilizer contains an $X$ on a node in $\CO$, so any logical operator $U$ must either include at least one $X$ or have at least one $Z$ on a pivot, where $Y$ operations are treated as both $X$ and $Z$.
    This is because any operator consisting of no $X$'s and no $Z$'s on pivots must contain only $Z$'s on vertices in $\CO$, in which case it would either be the identity operator or have an anti-commuting stabilizer and thus not be a logical operator.
    We may split the operation of $U$ into a sequence of single qubit Paulis applied onto different qubits. Thus, without loss of generality, we may choose the first Pauli in this sequence to be either a $Z$ operator on $v \in \CP$ or an $X$ operator on $v \in \CO \cup \CP$. This first operator hence 
   increases the illumination of a vertex $v$ from 0 to $\L_\text{max}(v) \geq \ell$.
    
    At the same time, the final configuration has all lights off, and thus every node ends with illumination 0. Consequently, the remaining Paulis in the sequence must reduce the illumination at $v$ from at least $\ell$ to 0.
Suppose that the initial Pauli was $X$ on $v \in \CO \cup \CP$. Then the illumination can only decrease by at most $B$ with each move, as each move is equivalent to applying a Pauli recovery operation. Therefore, at least $\ceil{\frac{\ell}B}$ Paulis after the first are necessary to reduce the illumination at $v$ to 0.
Suppose instead that $U$ has no $X$ but has a $Z$ on some pivot $p \in \CP$. Then after applying $Z(p)$, the illumination $\L(i(p))$ will be at least $\ell$ and will also require $\ceil{\frac{\ell}B}$ Pauli $Z$ operations to reduce to 0.
    Combined with the first move to increase the illumination, this results in a minimum of $\ceil{\frac{\ell}B}+1$ moves, which by Theorem~\ref{thm:distanceQLO} is a lower bound for the distance.
\end{proof}

\begin{theorem}[Greedy guarantee] \label{thm:greedy_guarantee}
     Let $G$ be a $B$-sensitive graph with brightness $\ell$ as defined in Definition~\ref{def:sensitive} and Definition~\ref{def:brightness}. Let $C(G)$ be the code that $G$ represents.
     Then the greedy graph decoder given by Algorithm~\ref{alg:greedy} corrects at least $\ceil{\frac{\ell+1}{2B}}-1$ errors. In other words, the less sensitive a graph is, the more effective the greedy decoder. The runtime of Algorithm~\ref{alg:greedy} is $O(n \CD_{\max}^2 + \min(n,\CD_{\max}^4) \mathcal{W})$, where $\CD_{\max}$ is the maximum degree of $G$ and $\mathcal{W}$ is the weight of the error.
\end{theorem}

\begin{proof}
    The decoder will fail if it ever corrects an error that did not occur, or if it ever fails to correct one that did. Suppose that the decoder has applied the correct recovery operators in both the $X$ loop (stage 1) and the $Z$ on pivots loop (stage 2). Then the recovery operation in stage 3 ($Z$ on non-pivot outputs) must also be correct since at that point the only remaining possible Paulis that can be applied are $Z$'s on non-pivot outputs, which each just turn the light of its own node off with no interaction with other lights.
    It suffices, therefore, to prove the correctness of stage 1 and stage 2.
    Additionally, if we focus on the first mistake the decoder makes, either applying an unnecessary recovery or moving on from a stage without recovering all errors.
    We can use the fact that at the time of the decoder's first error, all preceding stages have been decoded correctly, and therefore we only have to be concerned about error types that we are meant to recover at the current or later stages.
    Specifically, when recovering $Z$ errors on pivot nodes, we do not have to be concerned about remaining $X$ operations to recover.
    This observation allows us to use our definition of sensitivity from Definition~\ref{def:sensitive}, which we carefully formulated to not consider how $X$ operations may influence the illumination of input nodes.
    
    Suppose that at some point the decoder applied a $Z$ on a pivot or an $X$ recovery operation on node $w$ where none was needed. Then the \texttt{gap} variable on a corresponding node $v$ must be at least 1, meaning that we have $1\leq2\L(v)-\L_\text{max}(v)\leq2\L(v)-\ell$. Here, either $v = w$ if the error is $X$ or $v = i(w)$ if $w \in \CP$.
    This implies $\L(v)\geq\frac{\ell+1}2$. But $\L(v)$ can change by at most $B$ per operation, and thus per error, so such a scenario can only occur if there are at least $\ceil{\frac{\ell+1}{2B}}$ errors. Hence Algorithm~\ref{alg:greedy} can still correct $\ceil{\frac{\ell+1}{2B}}-1$ errors.

    Now suppose that at some point the decoder fails to recover a $Z$ error on $w \in \CP$ or an $X$ error. Since everything was correctly recovered up to this point, the only way for the decoder to miss this recovery is if the error on $w$ occurs alongside many other errors which collectively bring down the \texttt{gap} of some corresponding node $v$ to $< 1$. Again, $v = w$ if an $X$ error occurred and otherwise $v = i(w)$. If so, then $1>2\L(v)-\L_\text{max}(v)\geq\ell-2\left(\L_\text{max}(v)-\L(v)\right)$, implying $\L_\text{max}(v)-\L(v)>\frac{\ell-1}2$.
    Thus, it would take more than $\frac{\ell-1}{2B}$ or at least $\floor{\frac{\ell-1}{2B}}+1$ additional errors to reduce the illumination enough for the decoder to make a mistake. Combined with the original error on $w$, this would require a total of $\floor{\frac{\ell-1}{2B}}+2$ errors for a mistake, so Algorithm~\ref{alg:greedy} can still correct $\floor{\frac{\ell-1}{2B}}+1\geq\ceil{\frac{\ell+1}{2B}}-1$ errors. This completes the proof of the correctness guarantee.
    
    The runtime of the algorithm can be computed as follows. The very first act of the algorithm is to calculate \texttt{gap} on every node. This has the same asymptotic complexity as calculating $\L(v)$, which takes $O(n\CD_{\max{}}^2)$ time since $|o(v)|+|oip(v)|=O(\CD_{\max{}}^2)$ for all $v$. Next, we iterate through the $X$ loop. Whenever we make a recovery, we must update the \texttt{gap}.
    Any time we apply an $X$ recovery operation, we affect $O(\CD_{\max{}}^2)$ lights, each of which can be next to $O(\CD_{\max{}})$ inputs, which in turn affect the illuminations of $O(\CD_{\max{}})$ other nodes each.
    The total number of nodes whose illumination is affected is thus $O(\CD_{\max{}}^4)$, which we would need to update $\CW$ times, where $\CW$ is the weight of the error.
    However, this total number of affected nodes cannot exceed $n$, so we get that the total runtime is $O(n\CD_{\max{}}^2+\min(n,\CD_{\max{}}^4)\CW)$. The $X$ recovery loop subsumes the other two loops in complexity, so the asymptotics are the same when we account for the remaining loops.
\end{proof}

The guarantees we have given are phrased in terms of brightness, assuming a sufficiently small sensitivity. Both of these definitions are technical, so it is useful to be able to lower bound these quantities generically in terms of simpler quantities, such as degree-type functions. We will first give a relatively general lower bound for the brightness, so that we can use the results of Theorem~\ref{thm:distanceBound} and Theorem~\ref{thm:greedy_guarantee} to directly bound the distance and greedy guarantee from below by a simpler function of the graph.
To do so, we will need to recall a standard definition from graph theory.

\begin{definition}[Girth] \label{def:girth}
Let $G$ be a graph. We say that the \textit{girth} of $G$ is the length of the shortest cycle in $G$, where a cycle is a sequence of vertices that are adjacent in $G$ such that last vertex is equal to the first. The length of the cycle does not double-count the last vertex.
We denote this $\operatorname{girth}(G)$.
\end{definition}

Recall from Definition~\ref{def:t_spacing} that a subset of nodes in a graph is $t$-spaced if every path between two nodes in the subset has length at least $t$.

\begin{lemma}[Girth to brightness] \label{lemma:3away_is_good}
Let $G  = (\CI \sqcup \CP\sqcup \CO, E)$ be a graph with inputs $\CI$, pivots $\CP$, outputs $\CO$, and edge set $E$.
Suppose that $G$ satisfies at least one of the following conditions:
\begin{enumerate}
    \item[(1)] $\text{girth}(G)>6$.
    \item[(2)] $\text{girth}(G)>4$ and $\CP$ is 3-spaced.
    \item[(3)] $\text{girth}(G)>4$ and $\CI$ is 3-spaced.
\end{enumerate}
Then the brightness of $\ell$ of $G$, as given by Definition~\ref{def:brightness}, satisfies
\begin{equation}\label{eq:max-brightness}
    \ell=\min\left(\min_{v\in \CO\cup\CP}|N_o(v)|,\min_{v\in \CI}|o(v)|\right)=\min\left(\min_{v\in \CO\cup\CP}\deg(v)-|i(v)|,\min_{v\in\CI}\deg(v)-1\right),
\end{equation}
where $N_o(v)$, $i(v)$, and $o(v)$ are as in Definition~\ref{def:neighbour_sets}.
\end{lemma}
\begin{proof}
We note that the definition of brightness enforces that the quantity in the right-hand side of Eqn.~(\ref{eq:max-brightness}) is an upper bound for $\ell$.
This way, we just need to show that the brightness is no smaller than the quantity above.
In the case of $v\in\CI$, the max-illumination gives us that $\ell\geq\min_{v\in\CI}|o(v)|$, and since $v$ neighbours 0 inputs and 1 pivot, $|o(v)|=\deg(v)-1$.
We now consider the case where $v\in\CO\cup\CP$.

One reason for why the max-illumination $\L_\text{max}(v)$ could be less than $|N_o(v)|$ is if the some element $w \in o(v)$ was not in $o(v) \,\D\, oip(v)$, meaning that $w \in oip(v)$ as well.
If $\text{girth} > 4$, this cannot occur, as the $w$ would have to be both adjacent to $v$ as well as connected to $v$ with a sequence of 3 edges, passing through a pivot and an input, which would form a cycle of length 4.

In all cases of the claim we have $\operatorname{girth}(G) > 4$, so by the above $o(v) \cap oip(v) = \emptyset$ and hence $o(v) \subseteq o(v) \,\D\, oip(v)$.
The only other reason for why we night have $\L_\text{max}(v)<|N_o(v)|$ is if there are two pivots $p_1, p_2$ in $p(v)$ with nontrivial intersection: $oi(p_1) \cap oi(p_2) \neq \emptyset$. Then, since $oip(v) = \D_{p \in p(v)} oi(p)$, $oi(p_1) \cap oi(p_2)$ could be disjoint from $oip(v)$.
However, if the pivots are 3-spaced, $|p(v)| \leq 1$ so the above case cannot occur. Alternatively, if the inputs are 3-spaced, no node in $oi(p_1)$ can intersect with a node in $oi(p_2)$. Otherwise, this node in $\CO$ would neighbour 2 inputs, giving a path of length 2 between inputs.
Lastly, if no 3-spacing is guaranteed, then we must have a cycle of the form $v\to p_1\to i_1\to o\to i_2\to p_2\to v$, which is impossible if $\text{girth}(G)>6$.
These 3 sufficient conditions prove the lemma.
\end{proof}

\begin{corollary}\label{corollary:max-bright}
     If $G$ is a $q$-regular graph such that $\text{girth}(G)>4$ and the inputs of $G$ are 3-spaced (as in Definition~\ref{def:t_spacing}), then the brightness $\ell$ (as in Definition~\ref{def:brightness}) of $G$ satisfies $\ell=q-1$.
\end{corollary}
\begin{proof}
    By 3-spacing, $|i(v)|\leq1 \,\forall v$, so the claim follows from condition (3) of Lemma~\ref{lemma:3away_is_good}.
\end{proof}

Before we proceed to give a similarly general bound on sensitivity, we construct an explicit family of codes which has both low sensitivity and large brightness, so that the greedy decoder can correct errors almost up the maximum weight allowed by the distance-degree upper bound of Proposition~\ref{proposition:degree}.
Previously, we defined in Section~\ref{subsec:smallcodes} the hypercube code with parameters $\llbracket \frac{m 2^m}{m+1}, \frac{2^m}{m+1}\rrbracket$. We label the nodes of the hypercube by their length-$m$ binary string representation. Denote $e_i$ the $i$th standard basis vector in $\Z_2^m$ where $1\leq i\leq m$, i.e. $(e_i)_i = 1$ and $(e_i)_{j \neq i} = 0$. Moreover, we showed by injection of the classical Hamming code that the input nodes of the hypercube code are 3-spaced. Before we calculate the sensitivity of the hypercube, we give a useful lemma.


\begin{lemma}[Hypercube 3-spacing] \label{lemma:hypercube_separation}
    It is possible to choose pivots on the hypercube code, as defined in Definition~\ref{def:hypercube}, such that the pivots of the code are 3-spaced, as defined in Definition~\ref{def:t_spacing}.
\end{lemma}
\begin{proof}
    The minimum path length between inputs is already $3$. By homogeneity of the hypercube, it is enough to choose the pivots in a uniformly consistent way for every input. One example which works is to set $p(v) = \set{v + e_1}$ for all $v \in \CI$. This assignment will never choose $p(v) \in \CI$ since the inputs are separated by paths of length at least 3. Now suppose there exists a path between two pivots $p_1$ and $p_2$ with length less than $3$. In other words, with at most 2 bit flips, we can go from $p_1$ to $p_2$, so $p_2 = p_1 + e_a + e_b$ for some $a, b \in [m]$. Since $p_1 = i_1 + e_1$ and $p_2 = i_2 + e_1$, where $i_j$ is the input corresponding to $p_j$, $i_2 = i_1 + e_a + e_b$, which contradicts the fact that inputs are separated by paths of length at least 3.
\end{proof}

We note that there are several other transformations that can be applied to the hypercube to choose a set of pivots.
More generally, to assign a set of pivots on the hypercube, it suffices to choose any map from the set of vertices of the hypercube to itself, such that the graph distance between two nodes is preserved and every vertex is mapped to an adjacent one.
In the proof of Lemma~\ref{lemma:hypercube_separation}, we used a reflection about the first axis by adding $e_1$, although we could have just as easily chosen any $e_i$.
Another acceptable map would have been to choose any two axes, and rotate the hypercube $90^\circ$ (or $270^\circ$) in the plane formed by those axes.
This would leave most of the components of each vector intact, and modify the two bits corresponding to the chosen axes forwards (or backwards) in a cycle $00\to10\to11\to01\to00$.
This would ensure every vertex moves to a neighbouring spot, and as rotations are isometries, the distance condition would also be satisfied.

\begin{figure}[ht!]
    \centering
    \includegraphics[width=0.65\linewidth]{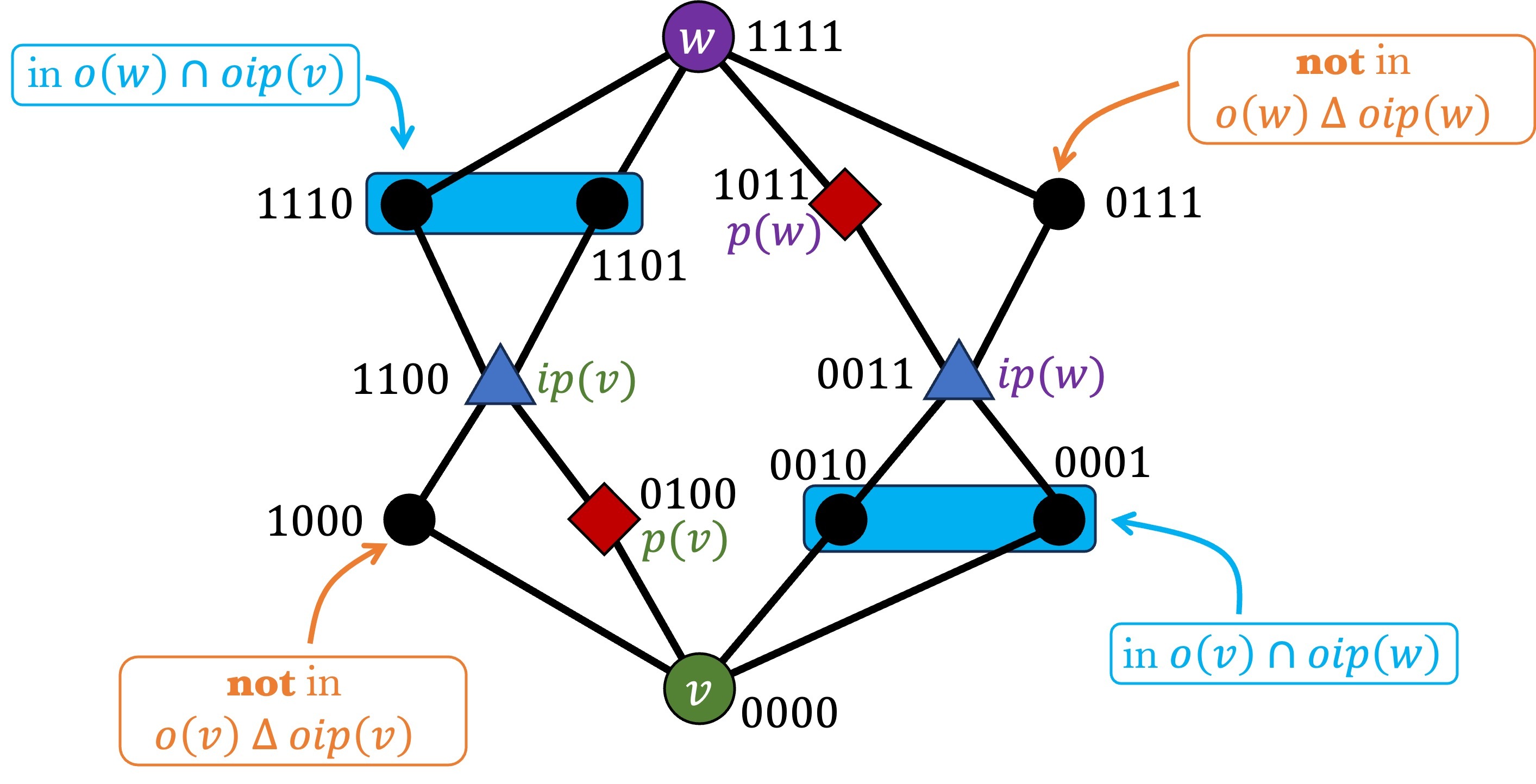}
    \caption{A depiction of how a switch flipped at $w$ can affect the illumination of $v$ for the $m$-dimensional hypercube code.
    Nodes are labeled with the first four bits of a $m$-bit bitstring identifying their vertex in the hypercube graph. Pivots are shown as red diamonds and inputs as blue triangles.
    Flipping the switch at $w$ toggles the lights in $o(w)\,\D\,oip(w)$. This does not include the top right vertex, 0111, due to the symmetric difference.
    This can toggle lights in $o(v)$ in the bottom right, affecting the illumination of $v$ by 2, as well as in $oip(v)$ (top left), affecting the illumination of $v$ by up to $\frac2{m-2}$, since the input node shown at 1100 has $m$ neighbours, of which one is a pivot and another, shown as 1000 in the bottom left, might not be in $o(v)\,\D\,oip(v)$.}
    \label{fig:hypercube-sensitivity}
\end{figure}

\begin{theorem}[Hypercube sensitivity] \label{thm:hypercube_sensitivity}
    The $m$-dimensional hypercube code, as defined in Definition~\ref{def:hypercube}, has a sensitivity of $B \leq 2+\frac{2}{m-2}$.
\end{theorem}
\begin{proof}
The nodes of the hypercube are labeled by their length-$m$ bitstring. The sensitivity is a maximum over a certain quantity over all nodes. By symmetry of the hypercube, the sensitivity is exactly this certain quantity for any fixed node, and without loss we choose the node to be $v = 0^m$. The general idea is that when we take neighbour-type operations $p(\cdot), o(\cdot), i(\cdot)$ on a bitstring with weight $M$, the resultant bitstring's weight can only change by 1 due to the connectivity of the hypercube.

    We need to show that no operation allowed by Definition~\ref{def:sensitive} on a vertex $w$ can change the illumination of $v$ by more than $2 + \frac{2}{m-2}$.
    We also recall Lemma~\ref{lemma:hypercube_separation}, so our inputs and pivots will both be 3-spaced.
    First note that if $v\in\CI$, no $Z$ operation can toggle more than one light in $o(v)$ since the inputs are 3-spaced.
    
    Suppose we apply $X$ on some $w \in \CO \cup \CP$, which toggles lights in $o(w)$ and $oip(w)$. In order for this operation to affect the illumination of $v$, $o(w) \,\D\, oip(w)$ must intersect $o(v)$ or $\bigcup_{p \in p(v)} oi(p)$. These are all an odd distance from $v$; likewise $o(w)$ and $oip(w)$ are an odd distance from $w$. But the hypercube is bipartite, so if the $X$ on $w$ is to affect the illumination of $v$, the bitstrings corresponding to $v$ and $w$ must have the same parity.
    Thus, we only need to consider $w$ whose bitstring has even weight.
    We have that $v=0^m$, so if the bitstring of $w$ has too large a weight, $X$ on $w$ cannot affect $o(v)$ or $oip(v)$ at all.
    If $w$ has weight greater than 6, it cannot affect the illumination of $v$, as $oip(v)$ and $oip(w)$ cannot intersect. If $w$ has weight exactly 6, then $X$ on $w$ affecting the illumination of $v$ would require a distance of 2 between a weight 2 input $ip(v)$ and a weight 4 input $ip(w)$, which is impossible since the inputs are 3-spaced.

    Suppose $w$ has weight 2.
    Without loss of generality, the vertex $w$ has bitstring $110\dots0$ and $o(w)$ intersects $o(v)$ at $10\dots0$ and $010\dots0$.
    However, for an operation at $w$ to affect $v$ any further requires either $oip(w)$ intersecting $o(v)$ or $oip(v)$ intersecting $o(w)$.
    In either case, we need to draw a chain of four steps from $v$ to $w$, which cannot go through $10\dots0$ and $010\dots0$.
    Without the trailing zeros, this chain will look like $000\to001\to(101\text{ or }011)\to111\to110$ without loss of generality.
    However, the central node in the chain would then have to be the input $ip(w)$ or $ip(v)$, in which case either $100$ or $010$ would be a neighbour of this input. Let us call this weight-1 node $u$. We note that $u$ is a neighbour of all three of $v$, $w$, and the input above, which is either $ip(w)$ or $ip(v)$.
    The node $u$ would therefore \textit{not} be in at least one of the symmetric differences $o(v) \,\D\, oip(v)$ or $o(w) \,\D\, oip(w)$, meaning that either the light at $u$ was not turned off by $w$ or it was not one of the lights illuminating $v$ to begin with.

    Meanwhile, if $w$ has weight 4, then we only need to consider the interactions of $o(v)$ and $oip(w)$ as well as $o(w)$ and $oip(v)$.
    This case is illustrated in Fig.~\ref{fig:hypercube-sensitivity}.
    Any two nodes in the hypercube share at most two neighbours. Applying this fact to $v$ and $ip(w)$ as well as $w$ and $ip(v)$ shows that $oip(w)$ decreases the illumination of $v$ by at most 2, while $o(w)$ turns off at most 2 lights in $oip(v)$.
    Since $ip(v)$ and $v$ share the neighbour $p(v)$, they will also share one other weight-1 neighbour, which will therefore not be in the set $o(v) \,\D\, oip(v)$. Hence $\left|oip(v)\cap\left(o(v) \,\D\, oip(v)\right)\right|=m-2$, by subtracting out from the $m$ neighbours the above node as well as one additional node which is a pivot.
    Thus, any operation on $w$ toggles at most 2 lights in $o(v)$ and at most 2 lights out of $m-2$ relevant ones in $oip(v)$, changing the illumination of $v$ by up to $2+\frac{2}{m-2}$.
    \end{proof}
    
The greedy guarantee of Theorem~\ref{thm:greedy_guarantee} states that a graph with low sensitivity is greedy-correctable up to about $\ell$ errors, where $\ell$ is the brightness from Definition~\ref{def:brightness}. To complete our arguments of the hypercube, we show that its brightness is close to the degree. Since degree upper-bounds the distance by Proposition~\ref{proposition:degree} and the brightness approximately lower bounds the distance (up to the sensitivity factor), this will imply that the hypercube distance is at least a constant factor away from its upper bound, and that the greedy decoder succeeds up to this factor. Since the hypercube has cycles of length 4, Lemma~\ref{lemma:3away_is_good} does not apply. Instead, we prove the brightness lower bound directly by studying the structure of the hypercube.

\begin{proposition}[Hypercube brightness] \label{proposition:hypercube_brightness}
Let $\ell$ be the brightness from Definition~\ref{def:brightness}. Then the $m$-dimensional hypercube code from Definition~\ref{def:hypercube} has $\ell = m - 2$.
\end{proposition}
\begin{proof}
Every input is adjacent to $m$ vertices of which one is a pivot and none are inputs.
Thus, the max-illumination of any input vertex is $m-1$.
As the pivots of the hypercube code are 3-spaced by Lemma~\ref{lemma:hypercube_separation}, we know that for any $v\in\CO\cup\CP$, $p(v)$ contains at most one vertex.
Furthermore, $ip(v)$ will share exactly two neighbours with $v$, one of which is $p(v)$ and the other one of which is in $o(v)$.
Thus, $v$ neighbours $m$ vertices of which one might be an input and one might be removed by the symmetric difference of $o(v)$ and $oip(v)$, so $\L_\text{max}(v)\geq m-2$.
Since $\ell$ is the smallest max-illumination, this proves the claim.
\end{proof}

\begin{corollary}
    Let $d(m)$ be the distance of the $m$-dimensional hypercube code where $m \geq 3$. Then $\floor{\frac{m-2}{2+\frac{2}{m-2}}}+1
    =\floor{\frac{m-1}{2}}\leq d(m)\leq m$. Notably, $d(m)\in\Th(m)$. Moreover, Algorithm~\ref{alg:greedy} can correct all Pauli errors of weight $\lfloor \frac{m - 1}{4 + \frac{4}{m-2}} \rfloor - 1 = \lfloor \frac{m - 1}{4} \rfloor - 1$ or less.
\end{corollary}
\begin{proof}
    This is a direct consequence of Proposition~\ref{proposition:hypercube_brightness}, Theorem~\ref{thm:distanceBound}, and Theorem~\ref{thm:greedy_guarantee}.
\end{proof}

This example shows that we can use the graph formalism to construct graphs that admit good algorithms for decoding.

\begin{conjecture}
For the $m$-dimensional hypercube code, as defined in Definition~\ref{def:hypercube}, $d(m) = m$ (where $d(m)$ is the distance of the $m$-dimensional hypercube code) when $m\geq7$ and a greedy decoding algorithm can decode up to $\lfloor\frac{m-1}2\rfloor$ errors.
\end{conjecture}
In the case of the $\llbracket 112, 16\rrbracket$ ($m = 7$) code as well as the $\llbracket 240, 16, 8 \rrbracket$ ($m = 8$) code, we verified numerically that the distance portion of our conjecture holds~\cite{Lu_graphcodes}. For $m<7$, we verified that the conjecture does not hold and believe this to be due to low-dimensional effects.

We next proceed to upper bound the sensitivity. Recall that the girth of a graph $G$ is the length of the shortest cycle in $G$.
Moreover, recall that a subset of nodes is 3-spaced if every path between these nodes has length at least 3. (See Definition~\ref{def:t_spacing}.) Previously, we showed that girth and 3-spacing were sufficient conditions to lower-bound the brightness. As we now show, the same properties are sufficient to upper-bound the sensitivity. Specifically, a larger girth implies that the graph has a lower sensitivity, with there being options to relax the girth requirement at the cost of introducing requirements on the spacing of inputs or pivots. In conjunction with the previous results on brightness upper bounds, we obtain a tight guarantee for efficient greedy decodability up to nearly the information-theoretical limit (given by the distance) for all graphs of sufficient girth and spacing.

\begin{figure}[ht!]
    \centering
    \includegraphics[width=0.65\linewidth]{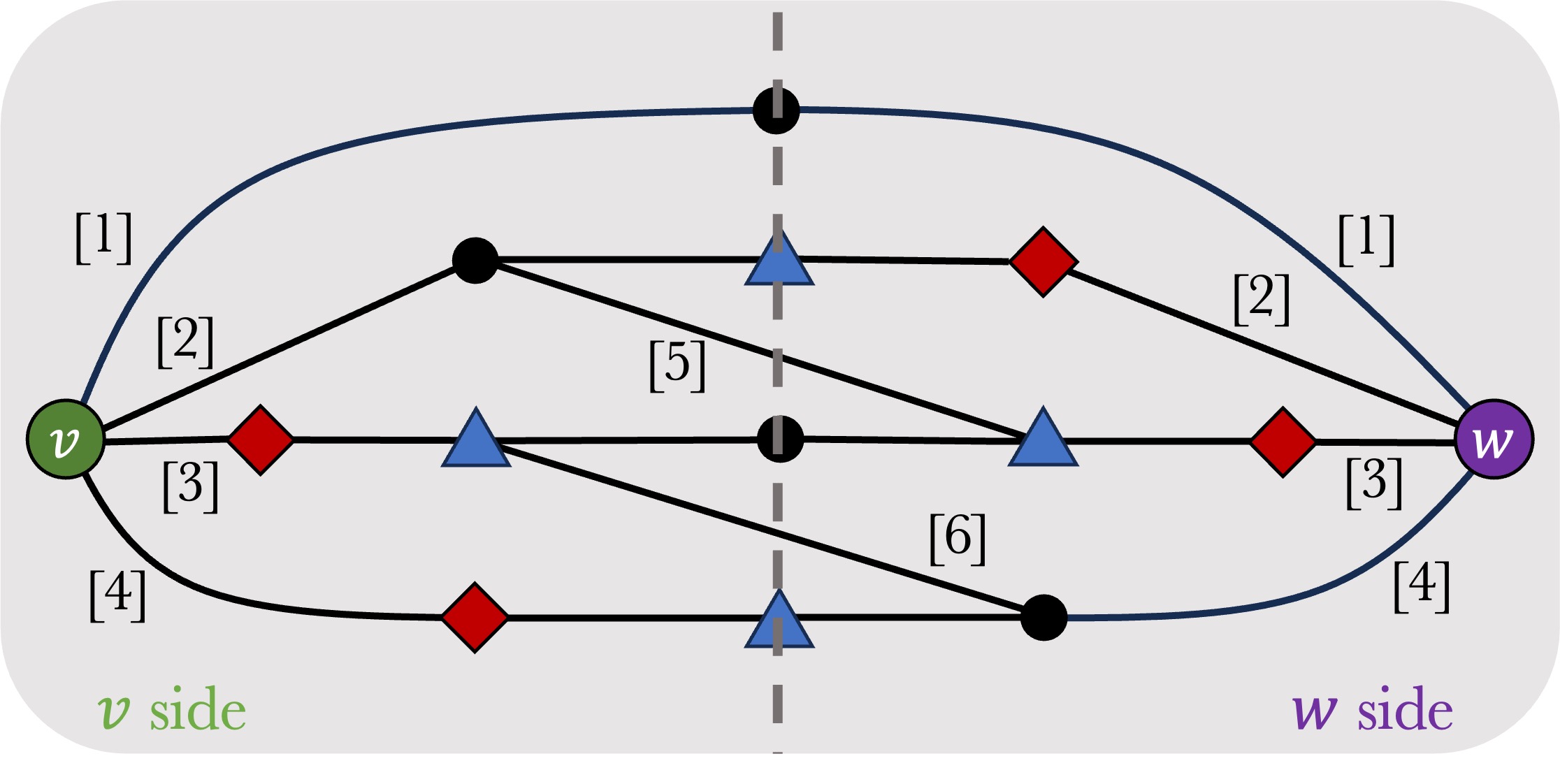}
    \caption{A visualization of the various paths by which flipping a switch at $w$ can affect the illumination of $v$. Blue triangles are inputs and red diamonds are pivots.
    The connecting arcs [1] to [4], correspond to intersections of the pairs $o(v)$ and $o(w)$, $o(v)$ and $oip(w)$, $oip(v)$, and $oip(w)$, and $oip(v)$ and $o(w)$, respectively.
    These are paths through which $w$ can affect the illumination of $v$ and bounds on the girth and spacing of inputs and pivots can affect which and how many of these connections can be present between $v$ and $w$.
    The edges labeled [5] and [6] are used only when necessary to highlight particular possible constructions.}
    \label{fig:girth_diagram}
\end{figure}

\begin{lemma}[Large girth implies sensitivity] \label{lemma:girth_sensitivity}
    Let $G$ be a graph with inputs $\CI$, pivots $\CP$, and sensitivity $B$.
    The following all hold.
    \begin{enumerate}
        \item[(1)] If $\text{girth}(G)>12$, $B=1$.
        \item[(2)] If $\text{girth}(G)>10$ and $G$ is $q$-regular, $B\leq1+\frac{1}{q-1}$.
        \item[(3)] If $\text{girth}(G)>8$ and $\CI$ is 3-spaced, $B=1$.
        \item[(4)] If $\text{girth}(G)>8$ and $\CP$ is 3-spaced, $B=1$.
        \item[(5)] If $\text{girth}(G)>8$ and $G$ is $q$-regular, $B\leq2$.
        \item[(6)] If $\text{girth}(G)>6$ and $\CP$ is 3-spaced, $B\leq2$.
        \item[(7)] If $\text{girth}(G)>6$, $G$ is $q$-regular, and $\CP$ is 3-spaced, $B\leq1+\frac{1}{q-1}$.
        \item[(8)] If $\text{girth}(G)>4$ and $\CP$ is 3-spaced, $B\leq3$.
        \item[(9)] If $\text{girth}(G)>4$, $G$ is $q$-regular, and $\CP$ is 3-spaced, $B\leq2+\frac{2}{q-1}$.
    \end{enumerate}
    Here girth is as in Definition~\ref{def:girth}, sensitivity as in Definition~\ref{def:sensitive}, and 3-spacing as in Definition~\ref{def:t_spacing}.
\end{lemma}

\begin{proof}
    In all cases, the girth of the graph is greater than 4. Therefore if $v\in\CI$ is an input, then for any pivot $p \in \CP$, $oi(p)$ can intersect with $o(v)$ on at most 1 vertex, as otherwise there would be a cycle of length 4. Consequently for all such cases, the illumination of any \textit{input} is not changed by more than 1. It is enough therefore to consider the case of $v\in\CO\cup\CP$.
    We know that for $w$ to affect the illumination of $v$ by more than 1, the operation on $w$ must affect one or both of $o(v)$ and $oip(v)$, and this operation can only do so while toggling the lights in $o(w)$ and $oip(w)$, which gives us four total cases. These cases are illustrated [1]-[4] in Fig.~\ref{fig:girth_diagram}. We refer to each of these cases as \textit{types} of connections. Each type of connection independently contributes to the sensitivity.
    Note that if a $Y$ operation is used instead of an $X$ operation, this also toggles the light at $w$.
    For the purposes of girth, it is always stronger for $w$ to connect to the light in $o(v)$ or $oip(v)$ that it wants to toggle rather than being this light itself, as that is what lengthens the relevant cycles the most.
    Doing both (being a light in one of these sets and being connected to a light in one of these sets) is never an option as that would imply the existence of a cycle of length 3, which is impossible if the girth of $G$ exceeds 4.

    The four cases described above are illustrated in Fig.~\ref{fig:girth_diagram}, which we will refer to repeatedly in this proof, particularly the connections numbered [1]-[4] between $v$ and $w$, as well as edges [5] and [6].
    This figure illustrates all the possible families of connections that are relevant to the conditions in the theorem.
    \begin{enumerate}
        \item[(1)] When $\text{girth}(G)>12$, at most one connection from $v$ to $w$ is possible. If there were two such connections of type [3], we would obtain a cycle of length 12 since each connection has length 6. Thus $B=1$.
        \item[(2)] If $\text{girth}(G)>10$, then we might have many length 6 connections, such as copies of type [3]. These connections come from the intersection of $oip(v)$ and $oip(w)$ which means that paths start with $v$ to a pivot and end with a pivot to $w$.
        If $G$ is $q$-regular, each neighbour set $oi(p)$ for some pivot neighbour $p \in p(v)$ has size $q-1$, and at most $q$ of these length 6 connections will be made, so $B\leq\frac{q}{q-1}=1+\frac1{q-1}$.
        \item[(3)] If $\text{girth}(G)>8$, then we might have a length 4 connection such as [2], a $oip(v)$ and $o(w)$ intersection, or [4], a $o(v)$ and $oip(w)$ intersection. There could also be some length 6 connections as in [3] from the previous case.
        If $\CI$ is 3-spaced, however, [3] cannot exist, because [3] requires two inputs to have a distance of length 2. Moreover, if there were more than 1 connection of type either [2] or [4], there would be a cycle of length 8 (see cycle made of going through [2] and back through [4]). Hence, only one connection of type [2] or [4] is possible, so $B = 1$.
        \item[(4)] As in case (3), the only possible connections are of type [2], [4], or [3]. Since the girth is $>8$, again only 1 connection of type either [2] or [4] is possible. If there were more than 1 connection of type [3], there would exist a pair of pivots with distance 2 from each other. Thus there is at most one connection of type [3] as well, and $B = 1$.
        \item[(5)] As in cases (3) and (4), the only connections allowed is a single one of type [2] or [4], and then potentially many connections of type [3].
        If $G$ is $q$-regular, then as in case (2), each connection of type [3] can contribute at most $\frac1{q-1}$ to the illumination of $v$.
        Since connections of type [2] and [4] can modify the illumination by 1, we have in total $B\leq1+(q-1)\frac1{q-1}=2$.
        \item[(6)] If $\text{girth}(G)>6$, then we might have either connection [1] and copies of [3] (possibly with shared or overlapping central vertices between the copies), or we might have copies of [2], [3], or [4].
        If $\CP$ is 3-spaced, however, we are restricted to a single connection each of type [2], [3], and [4]. Moreover, a connection to $v$ of type either [2] or [4] cannot co-exist with a connection to $v$ of type [3] by pivot 3-spacing, and likewise with $w$ instead of $v$. Thus, we have at most either a single connection each of types [2] and [4], or a single connection each of types [1] and [3].
        In either case, these two connections guarantee $B\leq2$.
        \item[(7)] This case is identical to case (6), with the addition of $q$-regularity as a condition. $q$-regularity implies that the connections that begin with a pivot, namely [3] and [4], can affect the illumination of $v$ by at most $\frac1{q-1}$, so $B\leq1+\frac1{q-1}$.
        \item[(8)] If $\text{girth}(G)>4$, we may have at most one connection of type [1], as two would give a cycle of length 4. The other connections a priori may be plentiful.
        However, since $\CP$ is 3-spaced, there is at most one connection to $v$ each of type [2], [3], and [4]. Further, a connection to $v$ of type [3] cannot coexist with either a connection to $v$ of type [2] or [4]. The same holds for $v$ replaced with $w$.
        By enumerating the possible cases, we find that the structure maximizing connectivity arises from [1], [2] on the (without loss of generality) $v$ side only, [3], [4] on the $w$ side only, and then adding ``staggered edges" given by [5], and [6].
        This connectivity gives two lights in $o(v)$ being toggled and two lights in $oip(v)$ when we operate on $w$.
        As the latter two contribute to the illumination as part of the same pivot node, we have $B \leq 2 + 1 = 3$.
        \item[(9)] This is identical to case (8), but now adding the condition that $G$ is $q$-regular means that the two lights toggled in $oip(v)$ contribute only $\frac2{q-1}$ towards the illumination since they go through the same pivot. Thus, $B\leq2+\frac{2}{q-1}$.
    \end{enumerate}
\end{proof}
While it may at first appear surprising that the entire notion of sensitivity can be simply bounded by the girth and spacing constraints which are both small constants, we note for intuition that the formulation of sensitivity from Definition~\ref{def:sensitive} is highly local.
That is, when an error affects a certain qubit, the stabilizers which can detect that error are no more than a \textit{small, constant} number of edges away from the affected qubit. Thus, if two nodes are separated by a sufficiently large constant minimum path length, then errors on one have no effect on the other. This condition of separation by a long path length is closely related to the girth, since if a node $v$ is in a short cycle, then the path from $v$ to the furthest away node $w$ contained in the cycle is about half the length of the cycle.

\begin{theorem}[Master theorem of graph girth] \label{thm:high_girth_high_dist}
    Let $G$ be a graph with inputs $\CI$, pivots $\CP$, and non-pivot outputs $\CO$.
    Let $d$ be the distance of the code corresponding to $G$ and let $T$ be the number of errors correctable by the decoder in given in Algorithm~\ref{alg:greedy}.
    If one of the following conditions holds, then the brightness of $G$ is 
    \begin{align}
        \ell^* = \min\left(\min\limits_{v\in \CO\cup\CP}\deg(v)-|i(v)|,\min\limits_{v\in\CI}\deg(v)-1\right)
    \end{align}
    and the bounds $d$ and $T$ are as stated. Note that if $G$ is $q$-regular, $\ell\leq q-1$.
    \begin{enumerate}
        \item[(1)] If $\text{girth}(G)>12$, $d\geq\ell^*+1$ and $T\geq\ceil{\frac{\ell^*-1}2}$.
        \item[(2)] If $\text{girth}(G)>10$ and $G$ is $q$-regular, $d\geq\ell^*+1$ and $T\geq\ceil{\frac{(\ell^*+1)(q-1)}{2q}}-1\geq\floor{\frac{\ell^*-1}2}$.
        \item[(3)] If $\text{girth}(G)>8$ and $\CI$ is 3-spaced, $d\geq\ell^*+1$ and $T\geq\ceil{\frac{\ell^*-1}2}$.
        \item[(4)] If $\text{girth}(G)>8$ and $\CP$ is 3-spaced, $d\geq\ell^*+1$ and $T\geq\ceil{\frac{\ell^*-1}2}$.
        \item[(5)] If $\text{girth}(G)>8$ and $G$ is $q$-regular, $d\geq\ceil{\frac{\ell^*}{2}}+1$ and $T\geq\ceil{\frac{\ell^*+1}{4}}-1$.
        \item[(6)] If $\text{girth}(G)>8$, $G$ is $q$-regular, and $\CI$ is 3-spaced, $d=q$ and $T=\floor{\frac{q-1}{2}
        }$.
        \item[(7)] If $\text{girth}(G)>6$ and $\CP$ is 3-spaced, $d\geq\ceil{\frac{\ell^*}{2}}+1$ and $T\geq\ceil{\frac{\ell^*+1}{4}}-1$.
        \item[(8)] If $\text{girth}(G)>6$, $G$ is $q$-regular, and $\CP$ is 3-spaced, $d\geq\ell^*+1$ and $T\geq\ceil{\frac{(\ell^*+1)(q-1)}{2q}}-1\geq\floor{\frac{\ell^*-1}2}$.
        \item[(9)] If $\text{girth}(G)>6$, $G$ is $q$-regular, $\CP$ is 3-spaced, and $\CI$ is 3-spaced, $d = q$ and $T\geq\floor{\frac{q}{2}}-1$.
        \item[(10)] If $\text{girth}(G)>4$ and $\CP$ is 3-spaced, $d\geq\ceil{\frac{\ell^*}{3}}+1$ and $T\geq\ceil{\frac{\ell^*+1}{6}}-1$.
        \item[(11)] If $\text{girth}(G)>4$, $G$ is $q$-regular, and $\CP$ is 3-spaced, $d\geq\ceil{\frac{\ell^*}{2}}+1$ and $T\geq\ceil{\frac{(\ell^*+1)(q-1)}{4q}}-1\geq\floor{\frac{\ell^*-1}{4}}$.
        \item[(12)] If $\text{girth}(G)>4$, $G$ is $q$-regular, $\CP$ is 3-spaced, and $\CI$ is 3-spaced, $d\geq\floor{\frac{q}{2}}+1$ and $T\geq\ceil{\frac{q-1}{4}}-1$.
    \end{enumerate}
\end{theorem}
\begin{proof}
    The claim is a direct consequence of Lemma~\ref{lemma:girth_sensitivity}, Lemma~\ref{lemma:3away_is_good}, Corollary~\ref{corollary:max-bright}, Theorem~\ref{thm:distanceBound}, and Theorem~\ref{thm:greedy_guarantee}.
\end{proof}

As long as any of the various conditions in Theorem~\ref{thm:high_girth_high_dist} are satisfied, we obtain a code that has distance $\W(\ell^*)$, which is a simple function of the degree as we had originally desired. As long as no node has more than a constant fraction of neighbours as inputs, $\ell^* = \W(\CD_{\min})$ (where $\CD_{\min} = \min_{v \in V} \deg(v)$) and so we can correct $\Omega(\CD_{\min})$ errors, matching the upper bound from Proposition~\ref{proposition:degree} up to constants.

We remark that while Theorem~\ref{thm:high_girth_high_dist} provides several sufficient conditions for which the distance has a lower bound near the upper bound and the greedy decoder decodes a number of errors also near the upper bound, it is not necessary. If one wishes to construct graphs with girth $\leq 4$, the corresponding code is not necessarily immediately low distance or not greedily decodable. Rather, the graph must instead be carefully designed to still have high brightness and low sensitivity. The most direct example of this phenomenon is the hypercube code. The hypercube is full of length 4 cycles, e.g. $00\dots \to 01\dots \to 11\dots \to 10\dots \to 00\dots$, and yet by carefully studying its symmetries, we were able to show in Theorem~\ref{proposition:hypercube_brightness} and Theorem~\ref{thm:hypercube_sensitivity} that the family of hypercube codes indeed has near-maximal distance and greedy decodability.

For the remainder of this work, however, we will use girth to our advantage. Theorem~\ref{thm:high_girth_high_dist} suggests a completely graph-theoretic recipe for finding good codes optimally decodable by the greedy strategy. Specifically, we seek families of graphs which satisfy \begin{enumerate}
    \item any of the conditions in Theorem~\ref{thm:high_girth_high_dist} hold for $G$, for example having $\text{girth}(G)>10$ and being $m$-regular, and
    \item the minimum degree of $G$ is as large as possible, ideally at least $\w(\log n)$ to beat the distance of the hypercube code, where $\w$ is little-omega notation. That is, a function $f(n)$ is $\omega(g(n))$ if $\lim_{n \to \infty} \frac{f(n)}{g(n)} = \infty$.
\end{enumerate}
These two simple conditions invite the application of extremal graph theory for code construction. To that end, we consider the following seminal result of extremal graph theory.

\begin{theorem}[\citet{benson1966minimal}] \label{thm:Benson_graphs}
    There exists an explicit family of bipartite graphs $G_q$, where $q$ is any power of an odd prime, which satisfies \begin{enumerate}
        \item[(a) ] $G_q$ is $(q+1)$-regular, 
        \item[(b) ] $G_q$ has $2\sum\limits_{k=0}^5 q^k$nodes, and 
        \item[(c) ] $G_q$ has girth at least 12.
    \end{enumerate}
\end{theorem}

We refer to such graphs as Benson graphs. The only remaining obstacle between a Benson graph and a greedy-optimal graph code is the choice of input and pivots. Although the maximum packing of inputs is generically hard, we show that it is still possible to do so up to a good approximation ratio. We give two techniques for such an input packing, both of which are completely generic and have no dependence on the Benson graph. They thus apply equally well towards transforming any other graph into a code. The first approach is very simple but only gives a guarantee $k = \W(n/\CD_{\text{max}}^2)$ in the worst case, where $\CD_{\text{max}} = \max_{v \in V} \deg(v)$. In Theorem~\ref{thm:Benson_graphs}, $q=\CD_\text{max}$ The second approach is more sophisticated and achieves the guarantee $k = \W(n/\CD_{\text{max}})$. Recall that for a node $v \in V$, where $V$ is the vertex set of a graph $G$, $N(v) = \set{w \in V \,:\, (v, w) \in E}$ for $E$ the edge set of $G$.

\begin{algorithm}\caption{Simple input-pivot assigner} \label{alg:simple_input}
\DontPrintSemicolon
  \SetKwFunction{FMain}{SimpleInputPivot}
  \SetKwProg{Fn}{Function}{:}{}
  \Fn{\FMain{$G$}}{
        \For{$v \in V$}{
            $\text{marked}_{v} \leftarrow \text{\texttt{False}}$\;
        }
        \While{$\exists v, w \in V \,:\, (v, w) \in E \text{\texttt{ and }} \text{marked}_{v} = \text{marked}_{w} = \text{\texttt{False}}$}{
            $\text{marked}_{v} \leftarrow \text{\texttt{True}}$\; 
            $\text{marked}_{w} \leftarrow \text{\texttt{True}}$\;
            $\text{\texttt{set }} v \text{\texttt{ as }input}$\;
            $\text{\texttt{set }} w \text{\texttt{ as }pivot}(v)$\;
            \For{$u \in N(N(v) \cup N(w))$}{
                $\text{marked}_{u} \leftarrow \text{\texttt{True}}$\;
            }
        }
        \KwRet
}
\end{algorithm}

\begin{figure}[ht!]
    \centering
    \includegraphics[width=0.8\linewidth]{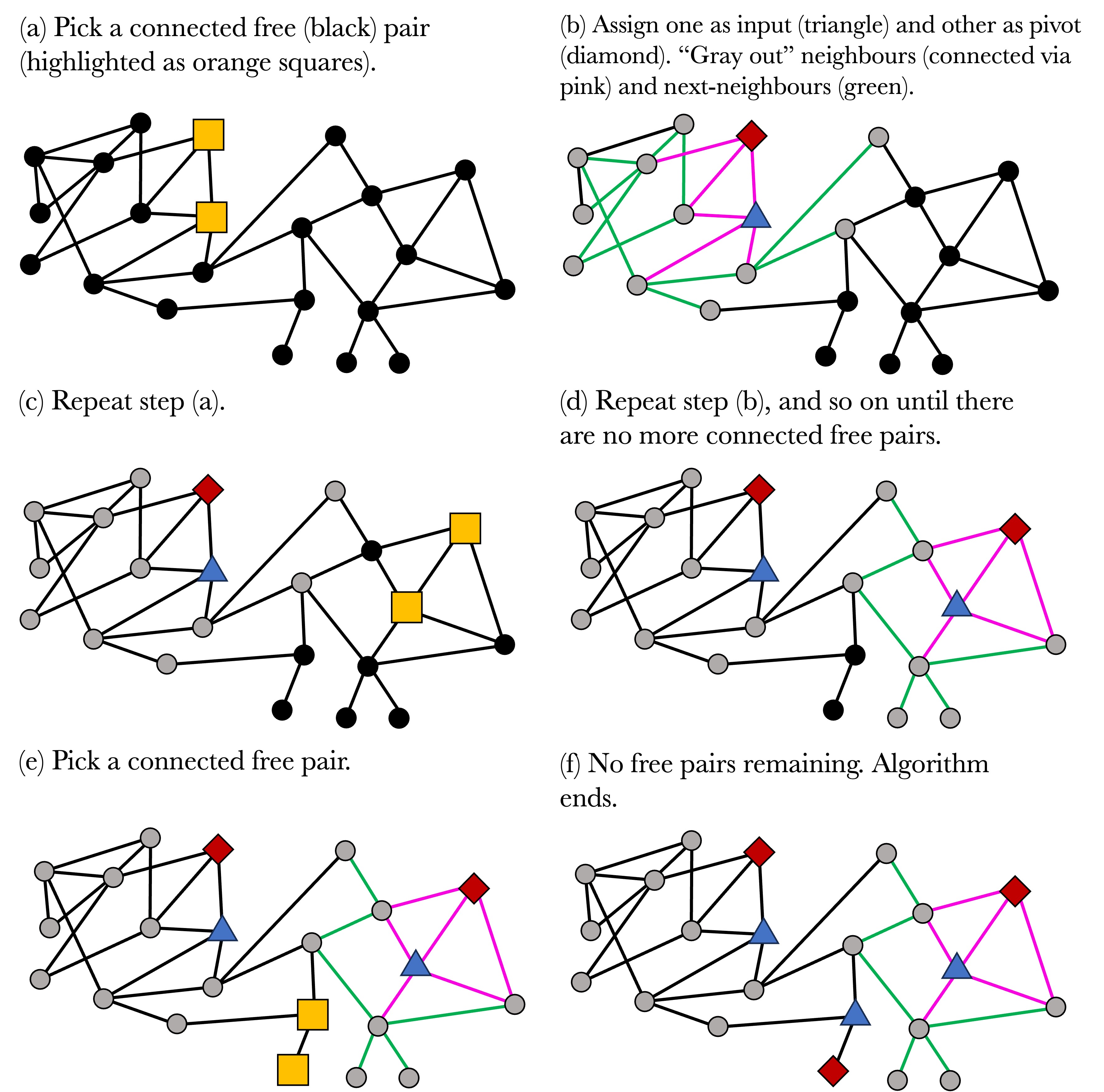}
    \caption{Example run of the simple input-pivot assignment process given by Algorithm~\ref{alg:simple_input}. Initially, all nodes are free, depicted by black circles. We begin in (a) by picking a connected free pair (orange squares). One is assigned as input (blue triangle) and the other as pivot (red diamond). Both the neighbours of the pair and the neighbours of those neighbours are grayed out in (b), making them no longer free. We repeat this process in (c)--(e) until there are no free pairs. The algorithm then terminates with the final assignment as in (f).}
    \label{fig:AssignmentAlgoSimple}
\end{figure}

For concreteness, we depict an example run of Algorithm~\ref{alg:simple_input} in Fig.~\ref{fig:AssignmentAlgoSimple}.

\begin{claim}\label{claim:bad_input_packing_algo}
    Let $G$ be a graph with $m$ nodes.
    Algorithm~\ref{alg:simple_input} gives a valid assignment of $k$ inputs, $k$ pivots, and $n-k$ non-pivot outputs such that $n+k = m$ and $k = \W\left(\frac{n}{\CD_{\text{max}}^2}\right)$. By valid, we mean that every pivot is connected to exactly one input. Moreover, the set of inputs $\CI$ and the set of pivots $\CP$ are both 3-spaced, as in Definition~\ref{def:t_spacing}.
\end{claim}
\begin{proof}
    At each step of the loop, all nodes within a distance $\leq 2$ of the newly assigned input and pivot are marked, and therefore cannot be later made inputs or pivots. Hence, all inputs and pivots are separated by paths of length at least 3. In addition, each input has a uniquely connected pivot, giving a valid assignment. Moreover, at each step at most $2 \CD_{\text{max}}^2$ nodes are marked, so the loop runs $\W\left(\frac{m}{\CD_{\text{max}}^2}\right)$ times. 
\end{proof}

Informally, Algorithm~\ref{alg:input} takes advantage of the observation that Algorithm~\ref{alg:simple_input} is maximally conservative at packing inputs. It is not strictly necessary for all inputs and pivots to be separated by distance at least 3, so long as we are careful.
Thus, Algorithm~\ref{alg:input} tries to pack input nodes as closely as possible without packing them ``too closely''.
Specifically, if we desire that the brightness of the graph remains within a constant fraction of the minimum degree of the graph, then we can have at most a constant fraction of the nodes around $v$ be inputs or pivots.
More generally, we define a parameter $\gamma>1$ which quantifies how closely we are willing to pack our inputs. To be more precise, we construct our algorithm so that no output node can have more than $\CD_{\text{min}}/\gamma$ neighbouring inputs or pivots.
A convenient technique to enforce this constraint is to give each node an \textit{availability}, initialized to $\CD_{\text{min}} / \g$, which indicates how many more inputs or pivots can be placed around that node. When the availability of a node $v$ drops to zero, all nodes in $N(v)$ become ineligible to be chosen as the next input or pivot. Algorithm~\ref{alg:simple_input} is the special case of this new approach when $\gamma = \CD_{\text{min}}$. As $\g$ decreases the rate increases while the distance decreases, but not necessarily at the same rate. We can therefore choose $\g$ carefully to improve the rate significantly without degrading the distance by too much.
In fact, we show that, up to constant factors, the following algorithm improves strictly upon Algorithm~\ref{alg:simple_input} in terms of distance and rate. The cost, however, is that the dense packing of inputs causes our bound on the check weight to go up from a linear to a quadratic bound, as discussed in Proposition~\ref{proposition:stab_weight}.

\begin{algorithm}\caption{Clever input-pivot assigner} \label{alg:input}
\DontPrintSemicolon
  \SetKwFunction{FMain}{CleverInputPivot}
  \SetKwProg{Fn}{Function}{:}{}
  \Fn{\FMain{$G, \g$}}{
        $\CD_{\text{min}} \leftarrow \min_{v \in V} \deg(v)$\;
        \For{$v \in V$}{
            $\text{availability}_{v} \leftarrow \lfloor \CD_{\text{min}}/\g \rfloor$\;
            $\text{marked}_{v} \leftarrow \text{\texttt{False}}$\;
        }

        \While{$\exists v_1, v_2 \in V \,:\, (v_1, v_2) \in E \text{\texttt{ and }} \text{marked}_{v_1} = \text{marked}_{v_2} = \text{\texttt{False}}$}{
            $\text{\texttt{set }} v_1 \text{\texttt{ as }input}$\;
            $\text{\texttt{set }} v_2 \text{\texttt{ as }pivot}(v_1)$\;
            $\text{marked}_{v_1} \leftarrow \text{\texttt{True}},\, \text{marked}_{v_2} \leftarrow \text{\texttt{True}}$\;
            \For{$w \in (N(v_1) \cup N(v_2)) \setminus \set{v_1, v_2}$}{
                $\text{marked}_w \leftarrow \text{\texttt{True}}$\;
                $\text{availability}_{w} \leftarrow \text{availability}_{w} - 1$\;
                \If{$w \in N(v_1) \cap N(v_2) \text{\texttt{ and }} \text{availability}_{w} > 0$}{
                    $\text{availability}_{w} \leftarrow \text{availability}_{w} - 1$\; \tcp{Availability decrements once for \textit{each} time it appears in a neighbouring set, unless already 0}
                }
                \If{$\text{availability}_{w} = 0$}{
                    \For{$u \in N(w)$}{
                        $\text{marked}_u \leftarrow \text{\texttt{True}}$\;
                    }
                }
            }
        }
        \KwRet
  }
\end{algorithm}

\begin{figure}[ht!]
    \centering
    \includegraphics[width=0.9\linewidth]{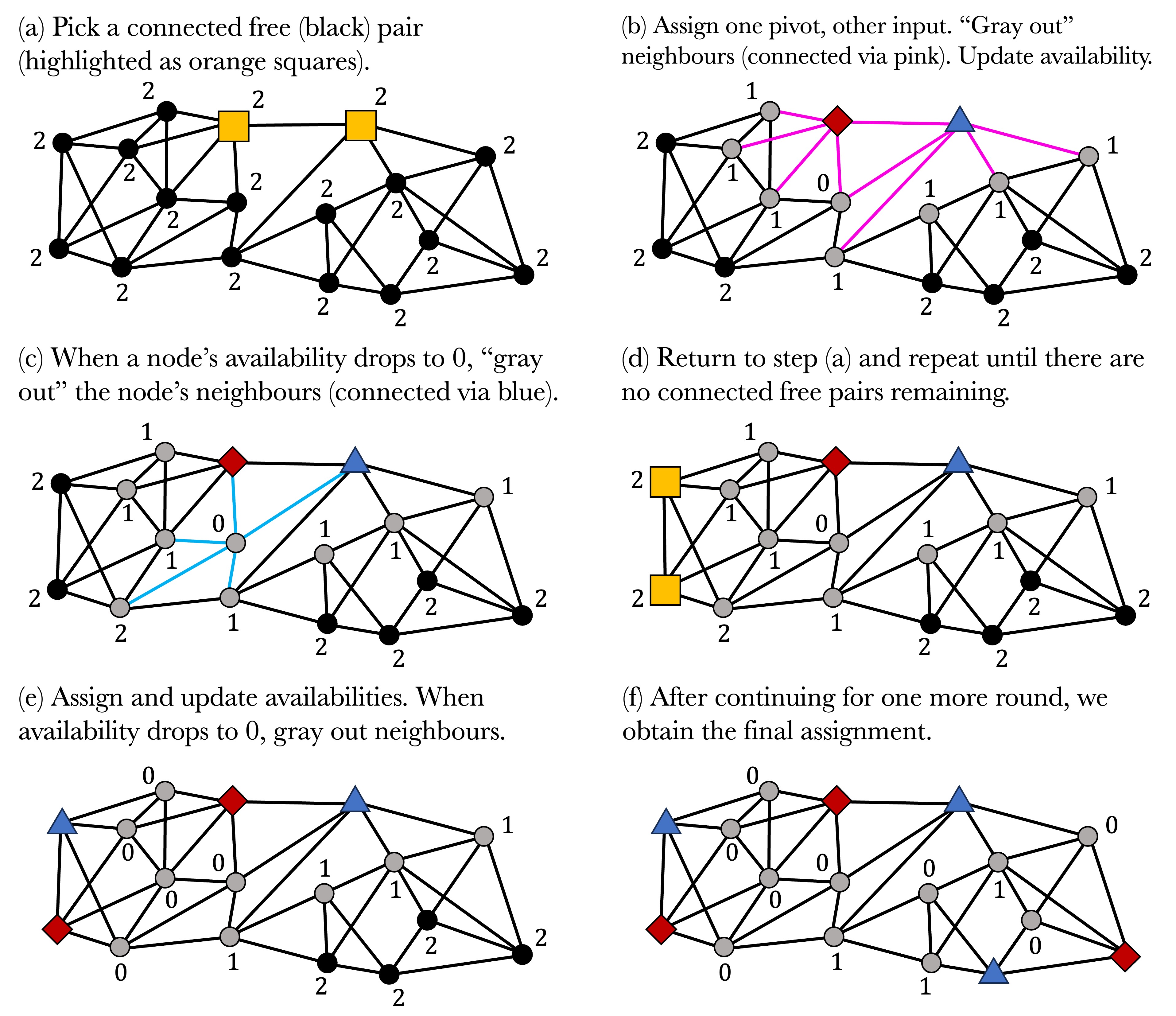}
    \caption{Example run of the clever input-pivot assignment process given by Algorithm~\ref{alg:input}. The graph shown has minimum degree $\CD_{\min} = 4$ and we choose the parameter $\gamma = 2$, so that the availability is $\CD_{\min} / \gamma = 4/2 = 2$. The graph begins in (a) with all nodes free (shown as black circles) with availability 2 on every node. We choose some connected free pair (orange squares) and assign one as input (blue triangle) and the other as pivot (red diamond). The neighbours are ``grayed out", i.e. marked as not free, and their availability is decreased in (b). The availability of each neighbour drops by 1 if it neighbours only the pivot or only the input, or 2 if it neighbours both the input and the pivot. In (c), all nodes whose availabilities drop to 0 have their own neighbours grayed out. This process is repeated in (d) and (e), and so on, until the final assignment is reached in (f).}
    \label{fig:AssignmentAlgoClever}
\end{figure}

We depict also an example run of Algorithm~\ref{alg:input} in Fig.~\ref{fig:AssignmentAlgoClever}.

\begin{claim} \label{claim:input_packing_algo}
    Let $G$ be a graph with $m$ nodes and minimum (maximum) degree $\CD_{\min}$ ($\CD_{\max}$).
    Let $\g > 2$ be a constant and suppose $\CD_{\min} \geq \gamma$.
    Then Algorithm~\ref{alg:input} gives a valid assignment of $k$ inputs, $k$ pivots, and $n-k$ non-pivot outputs such that $n+k = m$ and $k = \W\left(\frac{n \CD_{\min}^2}{ (\g + 1) \CD_{\max}^3}\right)$. By valid, we mean that every pivot is connected to exactly one input.
\end{claim}

\begin{proof}
    First, by construction each input has a unique pivot neighbour. Moreover, since all neighbours of the newly assigned input-pivot pair are marked at each step, no two inputs or pivots can be neighbours. The assignment is thus valid. 

    Next, at each step, $v_1, v_2$, and $N(v_1) \cup N(v_2) \setminus \set{v_1, v_2}$ are marked explicitly (ignoring for the moment the markings due to the change of availability). Thus, each step marks $\leq 2 (\CD_{\max} - 1) + 2 = 2 \CD_{\max}$ nodes explicitly.
    We analyze the effect of the markings due to a change in availability by appealing to the pigeonhole principle.
    When the availability of some node drops to zero, $\leq \CD_{\text{max}}$ nodes are marked. 
    However, in each step, the total availability summed across all nodes drops by $\leq 2(\CD_{\max} - 1)$. Therefore, after $s$ steps, the total availability across all nodes has changed by $\leq 2 s \CD_{\text{max}}$, so at most $\frac{2 s \CD_{\text{max}}}{\CD_{\text{min}}/\g}$ nodes have reached availability 0.
    Hence, at most $\frac{2 s \CD_{\text{max}}} {\CD_{\text{min}}/\g} \cdot \CD_{\text{max}}$ nodes are marked at after $s$ steps due to a change in availability.
    The total number of marked nodes after $s$ steps is therefore at most $2 s \CD_{\text{max}} + 2 \g s \CD_{\text{max}}^2 / \CD_{\text{min}} \leq 2 (\g + 1) s \CD_{\text{max}}^2 / \CD_{\text{min}}$. Consequently, it takes at least $m \CD_{\text{min}} / 2 (\g+1) \CD_{\text{max}}^2$ steps before all nodes are marked and therefore no more input-pivot pairs can be added.

    To complete the proof, we show that for $\W\left(\frac{m \CD_{\text{min}}^2}{(\g + 1) \CD_{\text{max}}^3}\right)$ steps, there exists a node which is unmarked and has an unmarked neighbour, so that the loop runs for $\W\left(\frac{m \CD_{\text{min}}^2}{ (\g + 1) \CD_{\text{max}}^3}\right)$ steps and thus assigns $k = \W\left(\frac{m \CD_{\text{min}}^2}{ (\g + 1) \CD_{\text{max}}^3}\right)=\W\left(\frac{n \CD_{\text{min}}^2}{ (\g + 1) \CD_{\text{max}}^3}\right)$ inputs. The condition of the existence of an unmarked node with an unmarked neighbour is equivalent to the existence of an edge in the subgraph induced by the unmarked nodes. Each time a node is marked, it removes at most $\CD_{\text{max}}$ edges from the unmarked subgraph. After $s$ steps, at most $2 (\g + 1) s \CD_{\text{max}}^2 / \CD_{\text{min}}$ nodes are marked, which in turn remove at most $2 (\g + 1) s \CD_{\text{max}}^3 / \CD_{\text{min}}$ edges. There are, to begin with, at least $m \CD_{\text{min}}/2$ edges. Therefore, for $s \leq m \CD_{\text{min}}^2 / 4 (\g+1) \CD_{\text{max}}^3$ steps, the unmarked subgraph has at least one edge.
\end{proof}

Although we have for the sake of generality included $\gamma$ in the asymptotic scaling, we will typically choose $\gamma$ to be a small constant greater than $1$, such as $2$. This choice is sufficient for all constructive purposes which might involve Algorithm~\ref{alg:input}. Combining Claim~\ref{claim:bad_input_packing_algo}, Claim~\ref{claim:input_packing_algo}, Theorem~\ref{thm:distanceBound}, Theorem~\ref{thm:greedy_guarantee}, Lemma~\ref{lemma:3away_is_good}, and Corollary~\ref{corollary:max-bright}, we obtain the following general results.

\begin{theorem}[Sensitive girth-5 graphs as codes]
\label{thm:general-code-parameters}

If $G$ is a $B$-sensitive graph with $m = n+k$ nodes, $\text{girth}(G)>4$, maximum degree $\CD_{\max}$, and minimum degree $\CD_{\min}$, Algorithm~\ref{alg:simple_input} describes an input packing which yields a $\llbracket n, k, d \rrbracket$ code, where
\begin{align}
    n = \Th(m),\quad k = \W\br{\frac{m}{\CD_{\max}^2}} ,\quad d = \W\br{\frac{\CD_{\min}}{B}}.
\end{align}
The greedy decoder can (efficiently) correct $\ceil{\frac{\CD_{\min}}{2B}}-1$ errors.
\end{theorem}
\begin{proof}
Since each node has at most 1 input and 1 pivot around it and since $G$ has $\text{girth}(G)>4$, this guarantees that $\ell=\CD_{\min}-1$ by Lemma~\ref{lemma:3away_is_good}, where $\ell$ is the brightness of $G$.
Since the graph is $B$-sensitive, the code's distance will be at least $\ceil{\frac{\ell}B}+1=\ceil{\frac{\CD_{\min}-1}B}+1$ by Theorem~\ref{thm:distanceBound}.
The efficient decoding of the code is guaranteed by Theorem~\ref{thm:greedy_guarantee}.
The quantities $n$ and $k$ are produced by Claim~\ref{claim:bad_input_packing_algo}.
\end{proof}

\begin{theorem}[Sensitive girth-7 graphs as codes]
\label{thm:general-code-parameters-clever}

If $G$ is a $B$-sensitive graph with $m = n+k$ nodes, $\text{girth}(G)>6$, maximum degree $\CD_{\max}$, and minimum degree $\CD_{\min}$, a choice of $\g$ for Algorithm~\ref{alg:input} describes an input packing which yields a $\llbracket n, k, d \rrbracket$ code, where \begin{align}
    n = \Th(m),\quad k = \W\br{\frac{m \CD_{\min}^2}{(\g + 1) \CD_{\max}^3}} ,\quad d = \W\br{\frac{\CD_{\min}(1 - 1/\g)}{B}}.
\end{align}
The greedy decoder can (efficiently) correct $\ceil{\frac{\CD_{\min} (1 - 1/\g)+1}{2B}}-1$ errors.
\end{theorem}
\begin{proof}
Since each node has at most $\CD_{\min}/\g$ inputs or pivots around it and since $G$ has $\text{girth}(G)>6$, this guarantees that $\ell\geq \CD_{\min}(1-1/\g)$ by Lemma~\ref{lemma:3away_is_good}, where $\ell$ is the brightness of $G$.
Since the graph is $B$-sensitive, the code's distance will be at least $\ceil{\frac{\ell}B}+1\geq \ceil{\frac{\CD_{\min}}{B}(1-1/\g)}+1$ by Theorem~\ref{thm:distanceBound}.
The efficient decoding of the code is guaranteed by Theorem~\ref{thm:greedy_guarantee}.
The quantities $n$ and $k$ are produced by Claim~\ref{claim:input_packing_algo}.
\end{proof}

With the input packing considered, we return to Theorem~\ref{thm:Benson_graphs} and its construction of a $(q+1)$-regular graph with girth at least $12$, where $q$ is any power of an odd prime. We may apply Algorithm~\ref{alg:input}, choosing $\g = \Th(1)$, e.g. $\g = 2$, to obtain an explicit construction of a code by packing input-pivot pairs into Benson graphs.

\begin{theorem}[A greedy-optimal code with high distance] \label{thm:girthy_code}
    There exists an explicit family of CSS codes with parameters $\llbracket n, \W(n^{4/5}), \Th(n^{1/5}) \rrbracket$, based on Benson graphs, which are efficiently decodable by the greedy algorithm. The corresponding stabilizers in $\CS(G)$, by Proposition~\ref{proposition:stab_weight}, have weight bounded above by $O(n^{2/5})$.
\end{theorem}
By using Algorithm~\ref{alg:simple_input} instead of Algorithm~\ref{alg:input}, we may trade off the check weight bound and rate to construct a $\llbracket n, \W(n^{3/5}), \Th(n^{1/5}) \rrbracket$ with check weight $O(n^{1/5})$. These tradeoffs are, however, generic in that they do not depend on the structure of Benson graphs at all. It is possible that a more careful analysis into the geometry of the Benson graph (beyond its girth and degree properties) will reveal a more clever assignment of input-pivot pairs which ensures that inputs and pivots are separated by distance at least 3, giving a code with rate $\W(n^{4/5})$ and check weight bound $O(n^{1/5})$.

Our primary purpose for constructing codes from Benson graphs in Theorem~\ref{thm:girthy_code} is to demonstrate concretely that the technique of using girth can readily give greedily decodable code constructions whose distance scales relatively well with $n$, i.e. as a power law $n^{\tau}$ for constant $\tau$ rather than something constant or logarithmic. The Benson graph code, however, is not immediately practical in the sense of fault tolerance because the check weight will scale as around the distance. 
Nonetheless, we believe that further investigation into graph constructions may reveal improved results, including larger distance codes as well as ways to find better choices of stabilizer generators by devising QLO strategies of minimum weight generator selection as in Theorem~\ref{thm:MWGSQLO}.

As a final example to show the flexibility of Theorem~\ref{thm:high_girth_high_dist}, we construct a code with a girth smaller than 12. Specifically, a construction due to \citet{benson1966minimal} and independently \citet{singleton1966minimal} gives a $(q+1)$-regular graph (for $q$ a prime power) with girth 8 and $n = \Theta(q^3)$ vertices. We refer to this family as the \textit{Benson-Singleton} graphs. Theorem~\ref{thm:high_girth_high_dist} requires for girth 8 that the pivots be 3-spaced (i.e. all pivot nodes have distance at least 3 from each other on the graph). Therefore, we will assign inputs and pivots according to Algorithm~\ref{alg:simple_input}. This ensures that inputs and pivots are both 3-spaced, with a logical dimension guarantee $k = \W(n/(q+1)^2) = \W(n^{1/3})$. Along with the regularity of the Benson-Singleton graph, allows us to use case (9) of Theorem~\ref{thm:high_girth_high_dist}, so that the distance of the code is $q+1$ and the greedy decoder can correct at least $\lfloor (q+1)/2 \rfloor - 1$ errors. We have thereby proven the following result.

\begin{theorem}[Benson-Singleton graph codes]
    There is an explicit family of $\llbracket n, \W(n^{1/3}), d=\Th(n^{1/3}) \rrbracket$ codes, based on Benson-Singleton graphs, which are efficiently decodable up to $\lfloor d/2 \rfloor - 1$ errors via Algorithm~\ref{alg:greedy}. The check weight of the code is at most $O(n^{1/3})$, by Proposition~\ref{proposition:stab_weight}.
\end{theorem}
As with the Benson graph codes, this check weight is too large to be practical for fault tolerance purposes, but further improvements on techniques to reduce check weight it is plausible that constructive techniques of this form may prove useful for practical purposes.

\subsection{\label{subsec:randcodes}Random codes with reduced stabilizer weights}
One direct technique for which a graph formalism enables more controlled analysis is random codes. Under the stabilizer tableau representation, the quantum Gilbert-Varshamov-bound gives an asymptotic (large $n$) relation between code parameters $n$, $k$, and $d$ that, if satisfied, guarantees the existence of a $\llbracket  n, k, d\rrbracket$ code. \begin{theorem}[Quantum Gilbert-Varshamov] \label{thm:quantum_gv}
    For parameters $d \leq \frac{n}{2},\, k$ which depend implicitly on $n$, as $n \to \infty$ there exists a $\llbracket  n, k, d\rrbracket$ code if $n H(d/n) + d \log 3 < n - k$, where $H(p) := -p \log p - (1-p) \log(1-p)$ is the binary entropy function of $p \in [0, 1]$.
\end{theorem}
We give a self-contained proof of Theorem~\ref{thm:quantum_gv} in Appendix~\ref{app:sec:QGV}.
The proof proceeds probabilistically, namely by generating a uniformly random stabilizer tableau and arguing that such a tableau has a nonzero probability of having distance $d$ if the bound is satisfied. However, because the expected weight (i.e. number of non-identity elements) of a random Pauli string $P_{1} \otimes \cdots \otimes P_{n}$, where $P_i \in \set{I, X, Y, Z}$, is $O(n)$, the random codes generated in the Gilbert-Varshamov fashion will have linear stabilizer weight. A natural question is whether the average weight can be decreased by a stochastic algorithm that is more fine-grained than uniform randomness. Such a task has proven difficult in the stabilizer tableau representation due to the challenge of generating Paulis that both commute and are not uniformly distributed. On the other hand, we can easily generate random graphs of a certain structure by defining first the possible edges that may be included, and then randomly choosing edges to include in the graph. This insight enables us to extend the result of the quantum Gilbert-Varshamov bound by way of the following theorem. Recall that a function $f(n)$ is $\w(g(n))$ if $\lim_{n \to \infty} \frac{f(n)}{g(n)} = \infty$.

\begin{theorem} \label{thm:random_graphs}
    Let $n$ be the number of physical qubits and define $R(n) < 1$ and $d(n) \in [\w(1), \frac{n}{\log n}]$. There exists an efficiently sampleable family of distributions $\CS_n$, parameterized by $n$, over stabilizer codes that as $n \to \infty$ satisfy three properties. \begin{enumerate}
        \item[(a) ] The stabilizer weights are \begin{align}
            |S_i| \leq \min\left(n,\br{4 + 2\frac{R}{1 - R}} d \log n\right),
        \end{align}
        where $S_i$ is a stabilizer from a tableau sampled from $\CS_n$, 
        \item[(b) ] the rate ratio of logical to physical qubits) is $R(n)$, and
        \item[(c) ] with probability at least $1 - n^{-d + o(d)}$, a random code from $\CS_n$ will have distance $d(n)$.
    \end{enumerate}
\end{theorem}

Theorem~\ref{thm:random_graphs} is positive in that it provides a nontrivial bound on the stabilizer weight of random codes for many choices of $R(n)$ and $d(n)$, and it gives a strong probabilistic concentration of $1 - n^{-d + o(d)}$. For any $d \in [\w(1), O(\frac{n}{\log n})]$, this concentration implies that almost all codes in $\CS_n$ have the desired properties. On the other hand, Theorem~\ref{thm:random_graphs} complements---as opposed to subsuming---the Gilbert-Varshamov bound because it provides a nontrivial weight reduction at substantially sublinear distances. It also forces the stabilizer weight---at least, for the generating set we construct---to be at least as large as the distance, which is an artifact of our proof technique. However, the result nonetheless remains interesting because of the guarantees it is able to provide and the flexibility that it offers. Indeed, practically useful codes are seldom asymptotically good. Since the distance and the rate are both determining factors for the stabilizer weight, we obtain a three-way distance-rate-weight trade-off of random codes that can be leveraged to study sublinear-distance codes in which we care much less about one of the three properties than the others. As with the other claims in this paper, this result is thus complementary of LDPC constructions that are asymptotically good but whose utility lies primarily in complexity theory rather than practical usage, due to, e.g. the smallest-$n$ instance of such quantum LDPC families being far larger than a practical block length. Instead, we interpret this result as evidence that for moderately large codes---e.g. $n \sim 50$ at a scale where we expect little-$o$ asymptotics to take effect---there is a great deal of flexibility in the graph construction to build codes of certain ranges of $R$ and $d$ which also may have a relatively small stabilizer weight.

\begin{figure*}[ht!]
    \centering
    \includegraphics[width=0.67\linewidth]{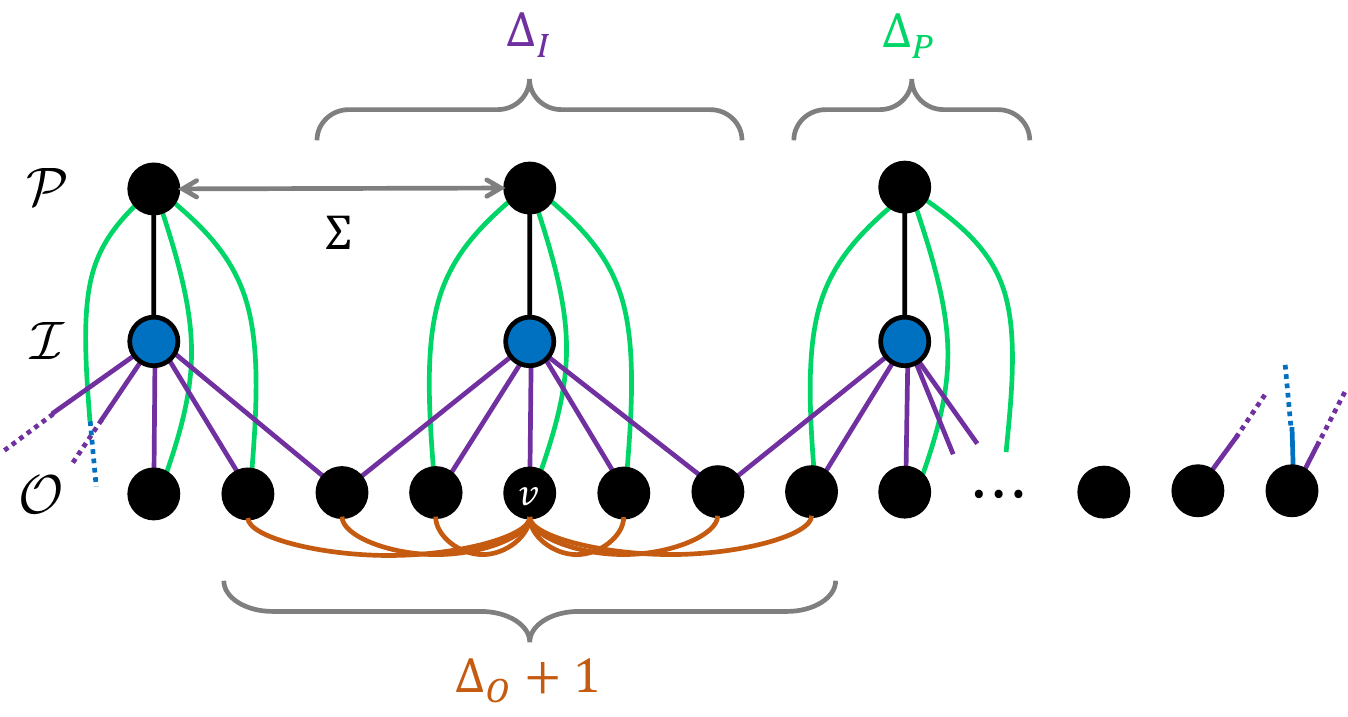}
    \caption{General localized random graph construction. Nodes are separated into $k$ input nodes $i \in \CI$, (which are spaced apart by $\Sigma$, depicted as 4 in the figure), $k$ pivot nodes $p \in \CP$, and $n-k$ non-pivot output nodes $o \in \CO$. Each pivot $p$ (input $i$) can connect to output nodes within a diameter $\D_{P}$ of $p$ ($\D_I$ of $i$); here, $\D_P = 3$ ($\D_I = 5$) and the \textit{possible} lines are green (purple). We emphasize that the edges drawn represent edges which appear with nonzero probability in the random graph construction, but the actual randomly drawn graph will not include all drawn edges. Similarly, output nodes $o$ may connect to outputs in a diameter $\D_O$ of $o$. For clarity, only the possible connections for a single output $v$ is shown, and here $\D_O = 6$.}
    \label{fig:localgraph}
\end{figure*}

We proceed to the proof of Theorem~\ref{thm:random_graphs}.
\begin{proof}
    We construct elements of $\CS_n$ graphically as shown in  Fig.~\ref{fig:localgraph}; denote this graph $G$. In particular, we begin with an encoder-respecting form, separating the inputs $\CI$, pivots $\CP$, and (non-pivot) outputs $\CO$. We choose a spacing $\Sigma$ so that every $\Sigma$ output nodes, we place an input-pivot pair. The rate satisfies $R = 1/(\Sigma + 1)$, since for every $\Sigma + 1$ outputs (including pivots) there is 1 input. Next, we assign each set of nodes a \textit{locality diameter}, respectively $\D_I, \D_P, \D_O+1$ (the $+1$ since the output node is included in its diameter but does not have a self-edge). The boundary conditions are periodic. Each node may only connect to output nodes within their locality diameter. If two nodes may be connected, we say one is \textit{reachable} from the other.
    We sample from $\CS_n$ by iterating through each connectable edge and randomly including or excluding it from the graph with probability $1/2$; the resulting stabilizer tableau is the one defined by the collection $S := \set{S_v := X(v) Z(N_o(v)) X(p(i(v))) Z(N_o(p(i(v))))}_{v \in \CO}$ as described in Section~\ref{sec:inversion}. For notation, we say that $S_v$ contains a Pauli $X_i$ ($Z_i$) if the $i$th element of $S_v$ is $X_i$ or $Y_i$ ($Z_i$ or $Y_i$). Our analysis of this stochastic algorithm adapts the techniques in the analysis of Theorem~\ref{thm:quantum_gv} to the graph picture. We fix a non-identity Pauli $P$ and consider four cases. \begin{enumerate}
        \item[(1) ] $P = Z^{\otimes T}$ for $T \subseteq [n]$ and $T$ does not contain any pivots. Pick $i\in T$ and consider $v_i \in G$. Then $S_{v_i}$ must anticommute with $P$ because $S_{v_i}$ contains $X(v_i)X(p(i(v_i)))$ which only overlaps $P$ in one position. Hence, $P$ does not commute with $S$.

        \item[(2) ] $P = Z^{\otimes T}$ for $T \subseteq [n]$ and $T$ contains at least one pivot $p \in \CP$. Let $i \in \CI$ be the unique input connected to $p$. Each of the $\D_I$ output nodes $v$ within the diameter $\D_I$ of $i$ have probability $1/2$ of connecting to $i$. $S_{v}$ contains $X(p)$ if and only if $i$ and $v$ are connected. Since the (anti-)commutativity of $P$ and $S_v$ are determined only by the parity of the number of qubits on which $X$'s in $S_v$ overlap with $Z$'s in $P$, the probability that $S_v$ and $P$ commute is $1/2$ regardless of the other Paulis in $S_v$ and $P$. Furthermore, consider any other $S_w$ where $w$ is also in the diameter of $i$. Because $w$ has an edge connecting to $i$ distinct from that of $v$, the probability of $S_w$ and $P$ commuting is independently $1/2$. This argument is general to all output nodes in the diameter of $i$, and remaining stabilizer act on disjoint subspaces. Hence the probability that $P$ commutes with $S$ is $2^{-\D_I}$. 

         \item[(3) ] $P$ contains at least one $X(p)$ for $p \in \CP$. Analogously, $\Pr[P \text{ commutes with } S] = 2^{-\D_P}$ by considering the $\D_P$ nodes in $\CO$ which each independently have probability $1/2$ of having a $Z(p)$.

        \item[(4) ] $P$ contains at least one $X(v)$ for $v \in \CO$. $\Pr[P \text{ commutes with } S] \leq 2^{-\D_O}$ by an analogous argument. We neglect some stabilizers which could anticommute with $P$ by acting through neighbouring pivots, i.e. the $Z(N_o(p(i(v))))$ component, since that can only decrease the probability and complicates the expression.
    \end{enumerate}
    Set $\D_O = \D_P = \D_I \coloneqq \D$.
    We proceed to bound the probability of the event $A(S)$ that there exists non-identity $P$ with weight $\leq d$ which commutes with $S$. As shown in Appendix~\ref{app:sec:QGV}, there are $2^{n H(d/n) + d \log 3 + O(\log n)}$ such Paulis, where $H(p) = -p\log p - (1-p) \log(1-p)$ is the binary entropy. By the union bound, \begin{align}
        \Pr[A(S)] \leq 2^{n H(d/n) + d \log 3 + O(\log n) - \D} \leq \e ,
    \end{align}
    where $\e$ bounds the fraction of codes which do not have distance at least $d$. We therefore solve the equation \begin{align}
        \D \geq \log \frac{1}{\e} + n H\left(\frac{d}{n}\right) + d \log 3 + O(\log n) .
    \end{align}
    Since $\frac{d}{n} \leq \frac{1}{\log n}$, we can asymptotically Taylor expand the binary entropy as $H(d/n) \to (\frac{1}{\ln 2} - \log \frac{d}{n}) \frac{d}{n}$, giving \begin{align}
        \D \geq \log \frac{1}{\e} + d \log n + o(d \log n) .
    \end{align}
    Here we have used our assumption that $d = \w(1)$, so that $O(\log n) = o(d \log n)$.
    Let $\e = n^{-d + o(d)}$ (where the $o(d)$ absorbs the $o(d \log n)$ term), so that if $\D = 2 d \log n$ at least $1 - n^{-d + o(d)}$ of codes asymptotically have distance $d$. Lastly, we calculate the stabilizer weights.
    
    For $v \in \CO$, we consider the weight of $S_v = Z(N_o(v))X(p(i(v))) Z(N_o(p(i(v))))$ which has four distinct contributions. The first term has weight 1. The second and fourth terms are all in $\CO$, and collectively may only have weight at most $2\D-1$.  This is because the number of nodes in $\CO$ that can be affected by this is at most $2\D-1$, as the locality diameter dictates that the last term can reach $\D$ different nodes from a given input on $v$'s left-hand side, including $v$. An additional $\D$ nodes can be reached on the right-hand side, again including $v$. But we double-counted $v$, so we have in total $2\D-1$ distinct nodes that the fourth term can reach. The fourth term's contribution subsumes that of the first and second. Hence, the total contribution of the first, second, and fourth terms is thus $2\D - 1$. As for the third term, we count the number of reachable input nodes, which is $ \leq \frac\D\Sigma+1$. This gives us a total stabilizer weight of at most $(2+\frac1\Sigma)\D$. Since $\Sigma = \frac{1}{R} - 1$, we have \begin{align}
        |S_v| & \leq \left(2+\frac1\Sigma\right)\D = \br{2 + \frac{1}{1/R - 1}} \Delta \\
        & = \br{2 + \frac{R}{1 - R}} \Delta = \br{2 + \frac{R}{1 - R}} 2 d \log n \\
        & = \br{4 + 2\frac{R}{1 - R}} d \log n .
    \end{align}
\end{proof}
    Obviously, $|S_v| \leq n$, which gives the claimed result.

\section{\label{sec:conclusion}Conclusion \& Outlook}

We introduced a graph structure which universally represents all stabilizer codes. Stabilizer tableaus can be efficiently compiled into such graphs, and vice versa. Our primary motivation for such a representation was to gain access to natural graph properties and notions---degree, geometry, connectivity---and then leverage them to improve code construction and analysis. As first steps in this direction, we chose several geometric shapes discretized into graphs and analyzed them as code representations. In doing so, we found a number of constant-size codes with desirable rates and reasonably large numerical distances, as well as a family of hypercube codes that have near-linear rate, logarithmic distance, and good encoding/decoding properties. In similar spirit, we also constructed a class of codes for which, given a desired rate $R$ and distance $d$, a random code drawn from the class has high probability of having rate $R$, distance $d$, and, so long as $R$ and $d$ were not both too large, nontrivially bounded stabilizer weights. This analysis extends the result of the quantum Gilbert-Varshamov bound, which has a distance-rate trade-off but no bound on stabilizer weight due to its coarse-grained analysis.

More generally, we showed that graph properties are intimately tied to code properties. Key code properties (weight, distance, encoding and logical operator circuit depths) are all bounded by small linear functions of graph degrees, and key coding algorithms (distance approximation, minimum weight generator selection, decoding) are all strategies on various instances of quantum lights out (QLO) games. The former observation allowed us to leverage graph algorithms to efficiently construct encoding circuits with controlled depth upper bounds as well as build logical operation circuits whose depths, in the case of diagonal or certain Clifford gates, are conceptually distinct and quantitatively better than simply unencoding, applying, and re-encoding. The latter result led us to design an efficient greedy QLO strategy for decoding, and prove that it succeeds up to a small constant factor on at least a certain class of graphs which include the hypercube code. We believe that further study in graph-based decoding algorithms will reveal efficient decoders for many other families of graphs. In sum, the universal graph representation enables insights and improvements that are general across all or large classes of stabilizer codes, by providing avenues of construction and analysis that are substantially less obvious via conventional representations of additive quantum codes.

There are several directions in which to explore from this point. The first is general. Classically, graph representations have shown a strong correspondence between graph properties and code properties beyond simple objects like the degree. Notably, spectral expander graphs have led to good classical codes. A characterization effort connecting graph properties such as spectrum, expansion, cycle sizes, density, geometry, and topology to distance, decoding, stabilizer weight, etc. could significantly extend this work's insight into stabilizer codes. 

A second pressing direction involves more deeply studying the quantum lights out game, as insights onto their strategies have equivalent interpretations on coding algorithm performance.
This paper studied the simplest QLO strategy and derived performance guarantees for graphs of sufficient girth. Such a connection is reminiscent of belief propagation insights from classical coding theory, which led to significant progress for classical error correction. By devising more clever QLO strategies, could we weaken the constraints on graphs decoded by the strategy, and therefore construct efficiently QLO-decodable codes with distance large enough to be practical?

A third direction is directly constructive. Given particular architectures or algorithms of experimental interest, is it possible to systematically design graphs whose codes are particularly suitable for such devices or algorithms in terms of locality of connectivity, geometry, rate, and distance? As a starting point, one can conduct a more thorough investigation into extremal graph theory to obtain the best explicit constructions of graphs with a particular girth (as dictated by the Master Theorem of girth, Theorem~\ref{thm:high_girth_high_dist}). Such graphs come with guaranteed efficient decodability and distance lower bounds nearly matching the upper bound. Even further, one can devise nontrivial strategies for selecting a better choice of stabilizer generators under certain structural conditions, in hopes of chaining these ideas together in order to create high distance, low check weight, and efficiently decodable stabilizer codes. It is plausible that such approaches involve modifying the representation itself. Representations that are universal are doomed in certain aspects of coding because their generality implies that properties scale with distance. If there exists a collection of graphs that, in exchange for representing a restricted subset of stabilizer codes, satisfies low-weight stabilizers, it may open new paths forward for practical code constructions and corresponding algorithms that are fundamentally not based on topological or classical (i.e. CSS-type product) reasoning. Such a representation would essentially lie in an optimal middle ground between the approach of surface-type code constructions and the universal graph representation.

\section*{\label{sec:acknolwedgements}Acknowledgments}
The authors are grateful to Zhiyang He (Sunny) for fruitful discussions and insightful observations. We also thank Adam Hesterberg for suggesting a closer investigation into the girths of graphs.

ABK was supported by the National Science Foundation (NSF) under Grant No. CCF-1729369. JZL was supported by a National Defense Science and Engeineering Graduate (NDSEG) Fellowship. PWS was supported by the NSF under Grant No. CCF-1729369, by the NSF Science and Technology Center for
Science of Information under Grant No. CCF-0939370, by the U.S. Department of Energy, Office of Science, National Quantum Information Science Research Centers, Co-design Center for Quantum Advantage (C2QA) under contract number DE-SC0012704., and by NTT Research Award AGMT DTD 9.24.20.

\sloppy
\RaggedRight
\bibliography{bib}

\clearpage
\newpage
\appendix

\section{Tableau to Graph Proofs} \label{app:sec:compiler}
In the main text, we sketched briefly the algorithm that maps encoders to ZXCFs. Here, we give a detailed explication. The first part of our algorithm maps the encoder into \textit{some} ZX diagram; we can then transform it via a sequence of equivalence rules into a ZXCF. To accomplish a transformation into ZX diagrams, we first observe prior work by \citet{hu2022improved} for stabilizer states.
\begin{theorem}[\citet{hu2022improved}]
    There exists a ZXCF for stabilizer quantum states, which we call the HK form.
\end{theorem}
Although Hu and Khesin did not use ZX calculus in their work, their construction uses graph states decorated with single-qubit operators, which can be mapped directly to ZX calculus by a manner we will briefly describe. A HK diagram is constructed by starting with a graph state---a state corresponding to an undirected graph wherein the nodes become $\ket{+}$ qubits and the edges become controlled-$Z$ gates. Each node is then endowed with a single \textit{free edge}, which does not connect to any other node. Local Clifford operations in $\langle H,S,Z\rangle$ are then applied to the qubits.
Such a construction has the capacity to express any stabilizer state. A direct mapping enables the presentation of a HK diagram into the ZX calculus.
\begin{enumerate}
    \item[(1) ] HK vertices become $Z$ nodes, as a vertex in a graph state begins in the state $\ket+$, which is a $Z$ node with a single output.
    \item[(2) ] HK (non-free) edges become (non-free) edges with Hadamard gates, as both of these correspond to $CZ$ gates.
    \item[(3) ] HK local operations of $S$, $Z$, or $SZ$ become phases on the corresponding node of $\frac\pi2$, $\pi$, and $\frac{3\pi}2$, respectively. The local operation $H$ becomes a Hadamard gate on the node's free edge.
\end{enumerate}
Thus, we will without loss consider HK forms to be in the ZX calculus representation and denote them \textit{ZX-HK forms}. We remark that
transformations of similar decorated graph state families into ZX calculus presentations have been considered in other contexts, such as in \citet{backens2014zx}.

The work of \citet{hu2022improved} is limited to stabilizer states. However, by vector space duality one can equivalently transform operators into states and vice versa. (The formal statement of this duality in quantum information theory is given by the Choi-Jamio\l{}kowski isomorphism.) The composition of such a transformation with the result above gives the following lemma.
\begin{lemma}
    There exists an efficient transformation of a ZX encoder diagram with $k$ input edges and $n$ output edges into a corresponding ZX presentation of a stabilizer state in ZX-HK form, with $n+k$ free edges.
    \label{lemma:ZXencoder_to_HK}
\end{lemma}
\begin{proof}
We will turn the encoder diagram into a Clifford circuit, and then map the output of that circuit (with input the all-$\ket{+}$ state) to HK form. 
Any $X$ (red) nodes in the ZX encoder diagram can be transformed into $Z$ (green) nodes with the same phase surrounded by Hadamard gates. Now, we can interpret the ZX encoder diagram as a state by treating the input edges as output edges (the ZX version of the Choi-Jamio\l{}kowski isomorphism).

Our ZX diagram now computes a state, so we can express it entirely in terms of the following elementary operations and their circuit analogs.
First, we could have a $Z$ node with only one output or only one input. In a circuit, this is a $\ket+$ qubit or a post-selected $\bra+$ measurement, respectively.
We can also have a Hadamarded edge or a node with $\frac\pi2$ phase, which are represented as $H$ or $S$ gates in circuits, respectively.
Lastly, we can have a $Z$ node of degree 3, with either 2 inputs and 1 output or 1 input and 2 outputs that acts as a merge or a split.
This is equivalent to applying a $CX$ operation between two qubits where the second qubit is either initialized to $\ket0$ before the $CX$ gate or post-selected by the measurement $\bra0$ after the $CX$ gate, respectively.

Any ZX diagram can be expressed in terms of these operations \cite{backens2014zx}.
Furthermore, as we apply each of these operations we can keep track of the current HK form for our state. (The \citet{hu2022improved} method shows us how to keep a diagram in HK form as each operation is applied.)
All of these steps can be done efficiently \cite{hu2022improved}, so this gives us an efficient procedure for turning a ZX diagram into a corresponding stabilizer state in HK form.
\end{proof}

Application of Lemma \ref{lemma:ZXencoder_to_HK} results in a ZX-HK form. That is, there is one $Z$ node per vertex of the graph in HK form, with any internal edge having a Hadamard gate on it.
Free edges can only have Hadamard gates on them if their phase is a multiple of $\pi$, corresponding to the local Clifford gates $H$ and $HZ$ in HK form, and if the associated node is not connected to any lower-numbered nodes.
Nodes whose free edges have no Hadamard gate are free to have any multiple of $\frac\pi2$ as a phase, corresponding to local Clifford operations $I$, $S$, $Z$, or $SZ$, respectively.

We are now equipped with a ZX-HK diagram that represents a state. This diagram has only output edges. To return it into an encoder diagram, we partition the $n+k$ free edges into $k$ input edges and $n$ output edges. In the circuit representation, this is equivalent to turning bra's into ket's and vice versa to map between an operator and a state. For example, $\ketbra{00}{1} - \ketbra{11}{0} \leftrightarrow \ket{001} - \ket{110}$. Now, if there are any edges between input nodes (those in $\CI$), we can simply remove them. They correspond to controlled-$Z$ operations, and we can take off any unitary operation on the input by the encoder definition. The same goes for local Cliffords on $\CI$. 

At this stage, the obtained ZX-HK form is in encoder-respecting form. It also satisfies the Edge and Hadamard rules---there are no input edges in ZX-HK form so the analogous rule is enforced only on lowered-numbered nodes, but if we number the nodes from the beginning such that the input nodes are lowered-numbered than all output nodes, then the transformed ZX-HK form will satisfy our ZXCF Hadamard rule. So, all that remains is to simplify the diagram to obey the RREF and Clifford rules.

\begin{lemma}
    An encoder diagram in ZX-HK form can be efficiently transformed to satisfy the RREF rule, while continuing to satisfy the Edge and Hadamard rules.
    \label{theorem:rref}
\end{lemma}

\begin{proof}
This proof relies on two basic properties of ZX calculus.
The first is that a $CX$ gate can be expressed on a pair of qubits by applying a pair of spiders connected by an edge, with a $Z$ spider on the control and an $X$ spider on the target.
The second fact is a rewrite rule called the \textit{bialgebra rule} \cite{backens2014zx}, which allows us to rewrite a green node connected to a red node by the following sequence of steps.
First, remove the two spiders and the edge between them.
Next, add a red node on each edge previously leading into the green node; do the same for the red node by adding green nodes.
Lastly, connect each red node to each green node.

Applying a $CX$ gate to a pair of inputs is a unitary operation and does not change our code.
The control spider of the $CX$ gate fuses with the first input.
The target spider is used to perform the bialgebra rule with the second input.
This results in two green nodes where the $CX$ target used to be.
The first is merged with the first input while the second takes the place of the recently-destroyed second input.
Meanwhile, the red nodes on each input-output edge previously connected to the second input now find themselves connected to both the first and second input, and turn green after commuting through the Hadamard gate on the input-output edge, merging with the output on the other side.
Any doubled edges are removed by the Hopf rule \cite{backens2014zx}.
The result is that we added the row of $M_{\mathcal{D}}$ corresponding to the first input the the row of the second.
This allows to do arbitrary row operations since we are working in a field of characteristic 2.
This concludes the proof of the fact that $M_{\mathcal{D}}$ can be turned into RREF without affecting the Edge and Hadamard rules.

\end{proof}

The only remaining task is to enforce the Clifford rule. For any pivot node $p \in \CP$ with a non-zero phase, denote $v_{\text{in}}$ its associated input node. We apply a local complementation about $v_{\text{in}}$, which notably does not change the entries of $M_\mathcal{D}$.
However, this operation also increases the phase of each neighbours of $v_{\text{in}}$ by $\frac\pi2$ (due to multiplication by $S$) so we repeat this process until the phase of that pivot vanishes. The local complementation transformation is a ZX equivalence identity, and therefore is a valid operation on our diagram~\cite{backens2014zx}.

Lastly, if there are any edges between pivots $p_1$ and $p_2$, we can remove them by applying a different transformation $\phi$ to their pair of associated input nodes $v_1$ and $v_2$. Let $N_1$ be the set of the neighbours of $v_1$ as well as $v_1$ itself and let $N_2$ be defined respectively.
The effect of $\phi$ is to toggle all edges in the set $N_1\times N_2$, including multiplicity. This means that edges that get toggled twice are not affected.
Any self-edge of the form $(v,v)$ is treated as a $Z$ operation on vertex $v$.
In addition, $\phi$ also applies an additional local operation of $HZ$ to each of $v_1$ and $v_2$, but this can be removed as $v_1$ and $v_2$ are input vertices.
As a result of this, we have swapped the neighbours of $v_1$ and $v_2$ as well as toggled the one pivot-pivot edge we wish to flip, $(p_1,p_2)$.
Note that $\phi$ does not violate the RREF rule because once this process is complete we can simply swap the neighbours back (without any toggling) via a row operation. This also does not add any local operations to the nodes $p_1$ and $p_2$, as each is only in one of $N_1$ and $N_2$, and thus does not receive a $Z$ operation.
We repeat until no pivot-pivot edges remain. $\phi$ can be shown to be a valid ZX equivalence identity from Eqn.~(101) in~\citet{van2020zx}.

With that step, the transformation from Clifford encoder to ZXCF is complete. Although this process may seem lengthy, all of these steps can be done systematically in an efficient manner, without having to go back to fix earlier rules.
Specifically, this algorithm takes $O(n^3)$ time. Creating the HK diagram for the encoder requires us to apply $O(n)$ Pauli projections, each of which can be applied in $O(n^2)$ time. After the HK form is created, operations such as row-reducing a matrix also take $O(n^3)$ time. We suspect that the time complexity cannot be improved in the worst case, but that there is a lot of room for heuristic runtime improvements and optimizations.

\section{Derivation of the ZXCF Counting Recursion} \label{app:recursion}

We conclude this section with the derivation of the recursive counting formula of ZXCF diagrams, given in Eq.~(\ref{eq:zxcf-count}). To derive $f$, imagine that, starting with two empty bins, we must assign the $\bar k\coloneqq n-k$ output nodes to be pivots (case $A$) or non-pivots (case $B$), where pivots need to be matched with input nodes.
The current number of pivots is tracked by $p$ and the number of non-pivot outputs is tracked by $o$.

Suppose we want the next output node to be a pivot (in $\CP$).
Since there is a one-to-one correspondence between pivots and inputs, we can add pivots only if $n > k$, and thus $\bar k>0$.
The matching between pivots and input nodes is fully constrained by the RREF rule, which sorts the inputs and pivots together.
The Clifford rule says that no pivot nodes may have local Clifford operations, so we just need to choose the edges connecting them to nodes we have already assigned. Since there are no pivot-pivot edges and the pivot connects to only one input, we have exactly $2^o$ possibilities.
Having made an assignment, $p$ increases by 1, and $n$ decreases by 1 (both input and output are decremented since the pivot matches with an input).

Suppose instead we want the next output node to be a non-pivot (in $\CO$).
Then we have to choose the local Clifford operation as well as its edges to previously assigned vertices.
If we choose to connect the node to none of the $p$ assigned inputs, $p$ assigned pivots, or $o$ assigned vertices, then we are allowed to place any of the 6 local Cliffords on the node.
If any of those edges are present, however, we cannot apply a Hadamard to the output edge due to the Hadamard rule. Hence, 4 choices remain for the local Clifford operation.
This works out to a total of $4(2^{2p+o}-1)+6=2^{2p+o+2}+2$ possibilities.
To finish, we decrement the number of output qubits without changing the number of inputs.

\section{Quantum Gilbert-Varshamov} \label{app:sec:QGV}
The quantum Gilbert-Varshamov bound, Theorem~\ref{thm:quantum_gv}, is a somewhat folklore result but is discussed in, e.g., \citet{nielsen2002quantum}.
We here provide a self-contained proof. \begin{proof}
    We proceed via a probabilistic method. That is, we compute the probability that a random code has parameters $\llbracket n, k, d\rrbracket$ with $n$ large. To construct a random stabilizer code, we begin with $Z_1, \dots, Z_{n-k}$ as stabilizers, and then choose a random Clifford operator $U$ on $n$ qubits. The random stabilizer code is given by $S_1 := U Z_1 U^\dag, \dots S_{n-l} := U Z_{n-k} U^\dag$. Such a code is a uniformly random stabilizer code because (a) all generators commute and (b) a random Clifford takes distinct non-Identity Paulis to uniformly random non-Identity Paulis.

    Fix a $n$-qubit non-Identity Pauli $P$ and note that $P' := U P U^\dag$ is a fixed uniformly random non-Identity Pauli. Let $p = \Pr[[ P, S_1] = 0, \dots, [P, S_{n-k}] = 0] = \Pr[[ P', Z_1] = 0, \dots, [P', Z_{n-k}] = 0]$ be the probability that $P$ commutes with the stabilizer tableau. If $P'$ commutes with $Z_i$, then the Pauli on the $i$th qubit must be either $I$ or $Z$. There are $2^{n-k} 4^k - 1$ non-identity Paulis which have $I$ or $Z$ on the first $n-k$ qubits, and $4^{n} - 1$ total non-identity Paulis. For large $n$ then, \begin{align}
        p = \frac{2^{n-k} 4^k - 1}{4^n - 1} \approx \frac{2^{n+k}}{4^n} = \frac{1}{2^{n-k}} .
    \end{align}
    On the other hand, the total number of non-identity Paulis with weight at most $d$ is \begin{align}
        N = \sum_{j = 1}^{d} 3^j \binom{n}{j} \leq d 3^d \binom{n}{d} ,
    \end{align}
    using the assumption that $d \leq \frac{n}{2}$ so that the last term is the greatest. By Stirling's approximation, \begin{align}
        N \longrightarrow 2^{n H(d/n) + d \log 3 + O(\log n)} .
    \end{align}
    To complete the proof, we apply the union bound. Let $A(S, P)$ be the event that $P$ commutes with $S_1, \dots, S_{n-k}$, and let $A(S) = \bigcup_{P \neq I} A(S, P)$. Then \begin{align}
    \begin{aligned} \label{eq:app:QGV}
        \Pr[A(S)] & \leq \sum_{P \neq I} \Pr[A(S, P)] \\
        & \leq 2^{n H(d/n) + d \log 3 + O(\log n)} 2^{k - n} .
    \end{aligned}
    \end{align}
    A $\llbracket n, k, d\rrbracket$ code exists if and only if $\Pr[A(S)] < 1$. By setting the right-hand side of Eqn.~(\ref{eq:app:QGV}) to $<1$ and solving, we obtain the quantum Gilbert-Varshamov bound.
\end{proof}
\end{document}